\documentclass[a4paper,fleqn,usenatbib]{mnras}

\usepackage{newtxtext,newtxmath}

\usepackage[T1]{fontenc}
\usepackage{ae,aecompl}

\usepackage{color,soul}
\usepackage{url}
\usepackage{multirow}
\usepackage{pdflscape}
\usepackage{rotating}

\definecolor{myred}{RGB}{228,26,28}
\definecolor{myblue}{RGB}{55,126,184}
\definecolor{mygreen}{RGB}{77,175,74}
\definecolor{mypurple}{RGB}{152,78,163}

\usepackage{graphicx}	
\usepackage{amsmath}	
\usepackage{amssymb}	

\newcommand{\myemail}{\textcolor{blue}{benjamindavis@swin.edu.au}}

\title[Short title, max. 45 characters]{MNRAS \LaTeXe\ template -- title goes here}
\title[The $M_{\rm BH}$--$\phi$ relation]{Updating the (supermassive black hole mass)--(spiral arm pitch angle) relation: a strong correlation for galaxies with pseudobulges}

\author[B. L. Davis, A. W. Graham and M. S. Seigar]{
Benjamin L. Davis,$^{1}$\thanks{E-mail: \myemail}
Alister W. Graham$^{1}$
and Marc S. Seigar$^{2}$
\\
$^{1}$Centre for Astrophysics and Supercomputing, Swinburne University of Technology, Hawthorn, Victoria 3122, Australia\\
$^{2}$Department of Physics and Astronomy, University of Minnesota Duluth, Duluth, MN 55812, USA
}

\date{Accepted 2017 July 13. Received 2017 June 27; in original form 2017 May 24}

\pubyear{2017}

\begin{document}
\label{firstpage}
\pagerange{\pageref{firstpage}--\pageref{lastpage}}
\maketitle

\begin{abstract}
We have conducted an image analysis of the (current) full sample of 44 spiral galaxies with directly measured supermassive black hole (SMBH) masses, $M_{\rm BH}$, to determine each galaxy's logarithmic spiral arm pitch angle, $\phi$. For predicting black hole masses, we have derived the relation: $\log({M_{\rm BH}/{\rm M_{\sun}}}) = (7.01\pm0.07) - (0.171\pm0.017)\left[|\phi|-15\degr\right]$. The total root mean square scatter associated with this relation is 0.43~dex in the $\log{M_{\rm BH}}$ direction, with an intrinsic scatter of $0.33\pm0.08$~dex. The $M_{\rm BH}$--$\phi$ relation is therefore at least as accurate at predicting SMBH masses in spiral galaxies as the other known relations. By definition, the existence of an $M_{\rm BH}$--$\phi$ relation demands that the SMBH mass must correlate with the galaxy discs in some manner. Moreover, with the majority of our sample (37 of 44) classified in the literature as having a pseudobulge morphology, we additionally reveal that the SMBH mass correlates with the large-scale spiral pattern and thus the discs of galaxies hosting pseudobulges. Furthermore, given that the $M_{\rm BH}$--$\phi$ relation is capable of estimating black hole masses in bulge-less spiral galaxies, it therefore has great promise for predicting which galaxies may harbour intermediate-mass black holes (IMBHs, $M_{\rm BH}<10^5$~${\rm M_{\sun}}$). Extrapolating from the current relation, we predict that galaxies with $|\phi| \geq 26\fdg7$ should possess IMBHs. 
\end{abstract}




\section{Introduction}

Qualitatively, pitch angle is fairly easy to determine by eye. This endeavour famously began with the creation of the Hubble--Jeans sequence of galaxies \citep{Jeans:1919,Jeans:1928,Hubble:1926,Hubble:1936}. In modern times, this legacy has been continued on a grand scale by the Galaxy Zoo project \citep{Lintott:2008}, which has been utilizing citizen scientist volunteers to visually classify spiral structure in galaxies on their website.\footnote{\url{https://www.galaxyzoo.org/}} If one is familiar with the concept of pitch angle, and presented with high-resolution imaging of grand design spiral galaxies, pitch angles accurate to $\pm 5\degr$ could reasonably be determined by visual inspection and comparison with reference spirals. Fortunately, instead of relying on only the human eye, several astronomical software routines now exist to measure pitch angle more precisely, particularly helpful with lesser quality imaging, and in galaxies with more ambiguous spiral structure (i.e. flocculent spiral galaxies).

Not surprisingly, pitch angle is intimately related to Hubble type, although it can at times be a poor indicator due to misclassification and asymmetric spiral arms \citep{Ringermacher:2010}. The Hubble type is also known to correlate with the bulge mass \citep[e.g.][and references therein]{Yoshizawa:1975,Graham:Worley:2008}, and more luminous bulges are associated with more tightly wound spiral arms \citep{Savchenko:2013}. Additionally, \citet{Davis:2015} present observational evidence for the spiral density wave theory's (bulge mass)--(disc density)--(pitch angle) Fundamental Plane relation for spiral galaxies.

It has been established that bulge mass correlates well with supermassive black hole (SMBH) mass \citep{Dressler:1989,Kormendy:Richstone:1995,Magorrian:1998,Marconi:Hunt:2003}. A connection between pitch angle and SMBH mass is therefore expected given the relations mentioned above. Moreover, pitch angle has been demonstrated to be connected to the shear rate in galactic discs, which is itself an indicator of the central mass distribution contained within a given galactocentric radius \citep{Seigar:2005,Seigar:2006}. In fact, it is possible to derive an indirect relationship between spiral arm pitch angle and SMBH mass through a chain of relations from (spiral arm pitch angle) $\rightarrow$ (shear) $\rightarrow$ (bulge mass) $\rightarrow$ (SMBH mass), $|\phi| \rightarrow \Gamma \rightarrow M_{\rm Bulge} \rightarrow M_{\rm BH}.$ From an analysis of the simulations by \citet{Grand:2013}\footnote{$\Gamma \approx 1.70-0.03|\phi|$ and $\log({M_{\rm Bulge}/{\rm M_{\sun}}}) \approx 1.17\Gamma +9.42$.} and application of the $M_{\rm BH}$--$M_{\rm Bulge}$ relation of \citet{Marconi:Hunt:2003}, the following (black hole mass)--(spiral arm pitch angle), $M_{\rm BH}$--$\phi$, relation estimation is obtained:
\begin{equation}
\log({M_{\rm BH}/{\rm M_{\sun}}}) \approx 8.18 - 0.041\left[|\phi|-15\degr\right].
\label{indirect}
\end{equation}

In this paper we explore and expand upon the established $M_{\rm BH}$--$\phi$ relation \citep{Seigar:2008,Berrier:2013}, which revealed that $M_{\rm BH}$ decreases as the spiral arm pitch angle increases. One major reason to pursue such a relation is the potential of using pitch angle to predict which galaxies might harbour intermediate-mass black holes (IMBHs, $M_{\rm BH}<10^5$~${\rm M_{\sun}}$). The $M_{\rm BH}$--$\phi$ relation will additionally enable one to probe bulge-less galaxies where the (black hole mass)--(bulge mass), $M_{\rm BH}$--$M_{\rm Bulge}$, and (black hole mass)--(S\'ersic index), $M_{\rm BH}$--$n$, relations can no longer be applied. Furthermore, pitch angles can be determined from images without calibrated photometry and do not require carefully determined sky backgrounds. The pitch angle is also independent of distance. Pitch angles have been measured for galaxies as distant as $z > 2$ \citep{Davis:2012}.

We present the mathematical formulae governing logarithmic spirals in Section \ref{theory}. We describe our sample selection of all currently known spiral galaxies with directly measured black hole masses and discuss our pitch angle measurement methodology in Section \ref{DM}. In Section \ref{AR}, we present our determination of the $M_{\rm BH}$--$\phi$ relation, including additional tests upon its efficacy, and division into subsamples segregated by barred/unbarred and literature-assigned pseudo/classical bulge morphologies; revealing a strong $M_{\rm BH}$--$\phi$ relation amongst the pseudobulge subsample. Finally, we comment on the meaning and implications of our findings in Section \ref{DI}, including a discussion of pitch angle stability, longevity and interestingly, connections with tropical cyclones and eddies in general.

We adopt a spatially flat lambda cold dark matter ($\Lambda$CDM) cosmology with the best-fitting Planck TT+lowP+lensing cosmographic parameters estimated by the Planck mission \citep{Planck:2015}: $\Omega_{\rm M} = 0.308$, $\Omega_\Lambda = 0.692$ and $h_{67.81} = h/0.6781 = H_0/(67.81$~km~s$^{-1}$~Mpc$^{-1}) \equiv 1$. Throughout this paper, all printed errors and plotted error bars represent $1\sigma$ ($\approx 68.3$~per~cent) confidence levels.

\section{Theory}\label{theory}

Logarithmic spirals are ubiquitous throughout nature, manifesting themselves as optimum rates of radial growth for azimuthal winding in numerous structures such as mollusc shells, tropical cyclones and the arms of spiral galaxies. Additional astrophysical examples of the manifestation of logarithmic spirals include protoplanetary discs \citep{Perez:2016,Rafikov:2016}, circumbinary discs surrounding merging black holes \citep{Zanotti:2010,Giacomazzo:2012} and the geometrically thick disc surrounding active galactic nuclei (AGN) central black holes \citep{Wada:2016}. This sort of expansion allows for radial growth without changing shape. One such special case of a logarithmic spiral is the golden spiral ($|\phi| \approx 17\fdg0$), which widens by a factor of the golden ratio ($\approx1.618$) every quarter turn, itself closely approximated by the Fibonacci spiral \citep[see appendix A of][]{Davis:2014}.

\begin{figure}
\includegraphics[clip=true,trim= 0mm 0mm 0mm 0mm,width=\columnwidth]{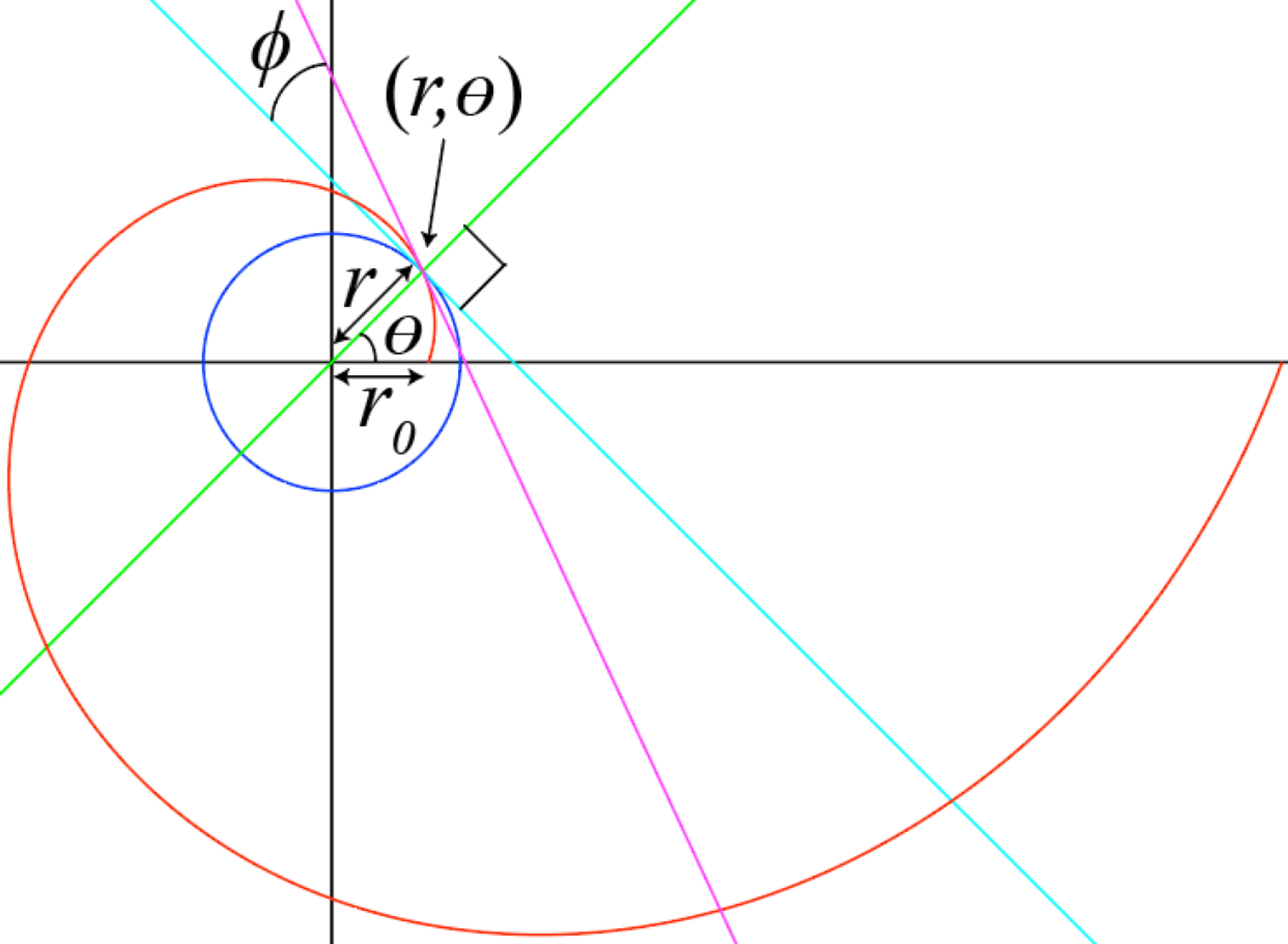}
\caption{A logarithmic spiral (red), circle (blue), line tangent to the logarithmic spiral (magenta) at point ($r,\theta$), line tangent to circle (cyan) at point ($r,\theta$) and radial line (green) passing through the origin and point ($r,\theta$). Included are the length of $r_0$, the angle $\phi$ and the location of the reference point ($r,\theta$). For this example, $r_0 = 1$, $|\phi| = 20\degr$ and point $(r,\theta) = (\approx1.33,{\rm \pi}/4)$; making the circle radius $\approx1.33$. Note that the spiral (red) radius can continue outward towards infinity as $\theta\rightarrow+\infty$ and continue inward towards zero as $\theta\rightarrow-\infty$.}
\label{example}
\end{figure}

One can define the radius from the origin to a point along a logarithmic spiral at $(r, \theta)$ as
\begin{equation}
r = r_0e^{\tau\theta},
\label{spiral_equation}
\end{equation}
where $r_0$ is an arbitrary real positive constant representing the radius when the azimuthal angle $\theta = 0$, and $\tau$ is an arbitrary real constant (see Fig.~\ref{example}).
If $\tau = 0$, then one obtains a circle at constant radius $r = r_0$, while if $|\tau| = \infty$, one obtains a radial ray from the origin to infinity. The parameter $\tau$ therefore quantifies the tightness of the spiral pattern.

A logarithmic spiral is self-similar and thus always appears the same regardless of scale. Every successive $2{\rm \pi}$ revolution of a logarithmic spiral grows the radius at a rate of
\begin{equation}
\frac{r_{n+1}}{r_n}=e^{2{\rm \pi}\tau},
\label{growth}
\end{equation}
where $r_n$ is any arbitrary radius between the origin and the point $(r_n,\theta)$ and $r_{n+1}$ is the radius of the spiral after one complete revolution, such that $r_{n+1}$ is the radius between the origin and the point $(r_{n+1},\theta+2{\rm \pi})$.

The rate of growth (of the spiral radius as a function of azimuthal angle) of such a logarithmic spiral can be defined using the derivative of equation~(\ref{spiral_equation}), such that
\begin{equation}
\frac{dr}{d\theta} = r_0\tau e^{\tau\theta} = \tau r.
\label{spiral_derivative}
\end{equation}
Notice that $\tau = \frac{dr}{d\theta}/r = 0$ generates a circle and $\tau = \frac{dr}{d\theta}/r = \infty$ generates a radial ray. Given these two extremes, one can more conveniently quantify the tightness of logarithmic spirals via an inverse tangent function. Specifically,
\begin{equation}
\tan^{-1}\tau = \tan^{-1}\left(\frac{\frac{dr}{d\theta}}{r}\right) = \phi,
\label{pitch}
\end{equation}
with $\phi$ being referred to as the `pitch angle' of the logarithmic spiral. In general terms, it is the angle between a line tangent to a logarithmic spiral and a line tangent to a circle of radius $r$ that are constructed from and intersect both at $(r,\theta)$, the reference point (see Fig.~\ref{example}). Rearrangement of equation~(\ref{pitch}) implies that $\tau = \tan\phi$. Therefore, in terms of pitch angle, equation~(\ref{spiral_equation}) becomes
\begin{equation}
r = r_0e^{\theta\tan{\phi}},
\label{pitch1}
\end{equation}
and equation~(\ref{spiral_derivative}) becomes\footnote{Equation~(\ref{pitch2}), without the preceding derivation, is provided in equation~6-2 of \citet{Binney:Tremaine:1987}.}
\begin{equation}
\cot{\phi}=r\frac{d\theta}{dr},
\label{pitch2}
\end{equation}
with $|\phi| \leq \frac{{\rm \pi}}{2}$. Therefore, as $\phi\rightarrow0$, the spiral approaches a circle and as $|\phi|\rightarrow\frac{{\rm \pi}}{2}$, the spiral approaches a radial ray.

The sign of $\phi$ indicates the chirality of winding, with positive values representing a clockwise direction of winding with increasing radius (`S-wise') and negative values representing a counterclockwise direction of winding with increasing radius (`Z-wise') for our convention. For galaxies, this merely indicates the chance orientation of a galaxy based on our line of sight to that galaxy. \citet{Hayes:2016} demonstrates through analysis of $458~012$ Sloan Digital Sky Survey \citep[SDSS,][]{York:2000} galaxies contained within the Galaxy Zoo 1 catalogue \citep{Lintott:2008,Lintott:2011} with the \textsc{sparcfire} software \citep{Davis:Hayes:2014}, that the winding direction of arms in spiral galaxies, as viewed from Earth, is consistent with the flip of a fair coin.\footnote{\citet{Shamir:2017} does however notice, from analysis of a smaller set of $162516$ SDSS spiral galaxies with the \textsc{ganalyzer} software \citep{Ganalyzer,Shamir:2011}, a slight bias of $82244$ spiral galaxies with clockwise handedness versus $80272$ with counterclockwise handedness.} Ergo, for our purposes in this paper, we only consider the absolute value of pitch angle in regards to any derived relationship.

\section{Data and Methodology}\label{DM}

Our sample of galaxies was chosen from the ever-growing list of galaxies with directly measured black hole masses, by which we mean their sphere of gravitational influence has supposedly been spatially resolved. This includes measurements via proper stellar motion, stellar dynamics, gas dynamics and stimulated astrophysical masers (we do not include SMBH masses estimated via reverberation mapping methods). Additionally, we only consider measurements offering a specific mass rather than an upper or lower limit. This criterion yielded a sample 44 galaxies (see Table~\ref{Sample}).

We carefully considered the implications of variable pitch angles for the same galaxy when viewed in different wavelengths of light. \citet{Pour-Imani:2016} found that pitch angle is statistically more tightly wound (i.e. smaller) when viewed in the light from the older stellar populations. We sought to preferentially measure images that more strongly exhibit the young stellar populations. In doing so, we are able to glimpse the current location of the spiral density wave that is enhancing star formation in the spiral arm. Whereas, older stellar populations were born long ago within the density wave pattern, but have since drifted away from the wave after multiple orbits around their galaxy, making the pitch angle smaller \citep[see fig.~1 from][]{Pour-Imani:2016}.

Our preferred images are those of ultraviolet light (e.g. \textit{GALEX} \textit{FUV} and \textit{NUV}), which reveals young bright stars still in their stellar nurseries, or 8.0 $\micron$ infrared light (e.g. \textit{Spitzer} \textit{IRAC4}), which is sensitive to light from the warmed dust of star-forming regions. Above all, we sought high-resolution imaging that adequately revealed the spiral structure, regardless of the wavelength of light. For instance, near-IR images often (though not always) reveal smoother spiral structure that is more likely to appear grand design in nature (and is easier from which to measure pitch angle). This can be seen in the work of \citet{Thornley:1996}, who demonstrates that spirals appearing flocculent in visible wavelengths of light may appear as grand design spirals if viewed in near-IR wavelengths.

Previous papers exploring the $M_{\rm BH}$--$\phi$ relation have used a single method to measure the value of $\phi$. \citet{Seigar:2008} and \citet{Berrier:2013} both exclusively used two-dimensional fast Fourier transform (\textsc{2dfft}) analysis. Here, we have employed multiple methods to ensure the most reliable pitch angle measurements. Pitch angles were measured using a new template fitting software called \textsc{spirality} \citep{Shields:2015,Spirality} and \textsc{2dfft} software \citep{Davis:2012,2DFFT}. Additionally, computer vision software \citep{Davis:Hayes:2014} was utilized to corroborate the pitch angle measurements.

All of these methods first compensate for the random inclination angle of a galaxy's disc by de-projecting it to a face-on orientation.\footnote{It is interesting to note that the act of measuring pitch angle itself also yields a good indication of the true inclination angle of a galaxy \citep{Poltorak:2007}.} Inclination angles were estimated from each galaxy's outer isophote ellipticity, and these inclination angles were subsequently used to de-project the galaxies via the method of \citet{Davis:2012}. Even with the use of multiple software routines that invoke varied methods of measuring pitch angle, it sometimes remains difficult to clearly analyse flocculent spiral structure. To overcome this, one can apply multiple image processing techniques such as `symmetric component isolation' (see \citealt{Davis:2012}, their section 5.1 and \citealt{Shields:2015}, their section 3.2 and 3.3) to enhance the spiral structure for an adequate measurement. Even so, when presented with images of poor quality, one needs to be mindful that all methods are unfavourably contaminated with spuriously high-pitch angle signals in the presence of low signal to noise.

\begin{landscape}
\begin{table}
\caption{Sample of 44 spiral galaxies with directly measured black hole masses.
\textbf{Columns:}
(1) Galaxy name.
(2) Morphological type, mostly from HyperLeda and NED.
(3) Bulge morphology (`C' = classical bulge, `P' = pseudobulge and `N' = bulge-less).
(4) Bulge morphology reference.
(5) Spiral arm classification \citep{Elmegreen:1987}.
(6) Luminosity distance, mostly from HyperLeda and NED.
(7) Distance reference.
(8) Major diameter, mostly from NED.
(9) Black hole mass, adjusted to the distances in Column 7.
(10) Measurement method for black hole mass (`g' = gas dynamics, `m' = maser, `p' = stellar proper motion and `s' = stellar dynamics).
(11) Black hole mass reference.
(12) Harmonic mode (i.e. number of dominant spiral arms) measured by \textsc{spirality}.
(13) Logarithmic spiral arm pitch angle.
(14) Inclination angle used for de-projection.
(15) Telescope used for pitch angle measurement (images acquired primarily from NED or MAST).
(16) Photometric filter used for pitch angle measurement.
(17) Resolution of pitch angle measurement (i.e. Gaussian PSF FWHM).
(18) Pitch angle reference.
(19) Measurement method for pitch angle measurement (`T' = template fitting, `F' = \textsc{2dfft} and `V' = computer vision).
\textbf{References:}
(1) \citet{Kormendy:Ho:2013}.
(2) \citet{Hu:2009}.
(3) \citet{Greene:2016}.
(4) \citet{Zoccali:2014}.
(5) \citet{Kormendy:2013}.
(6) \citet{Saglia:2016}.
(7) \citet{Fisher:Drory:2010}.
(8) \citet{Hu:2008}.
(9) \citet{Nowak:2010}.
(10) \citet{Sandage:1981}. 
(11) \citet{Sani:2011}.
(12) \citet{Gadotti:2012}.
(13) \citet{Berrier:2013}.
(14) \citet{Tully:2008}.
(15) Luminosity distance computed using redshift (usually from NED) and the cosmographic parameters of \citet{Planck:2015}.
(16) \citet{Yamauchi:2012}.
(17) \citet{Boehle:2016}.
(18) \citet{Riess:2012}.
(19) \citet{Radburn-Smith:2011}.
(20) \citet{Pudge}.
(21) \citet{Tully:1988}.
(22) \citet{Terry:2002}.
(23) \citet{Sorce:2014}.
(24) \citet{Lagattuta:2013}.
(25) \citet{Kudritzki:2012}.
(26) \citet{Lee:2013}.
(27) \citet{Honig:2014}.
(28) \citet{Humphreys:2013}.
(29) \citet{Bose:2014}.
(30) \citet{Jacobs:2009}.
(31) \citet{Silverman:2012}.
(32) \citet{McQuinn:2016}.
(33) \citet{Tully:2015}.
(34) \citet{Gao:2016}.
(35) \citet{McQuinn:2017}.
(36) \citet{Kuo:2013}.
(37) \citet{Greenhill:2003}.
(38) \citet{Reid:2013}.
(39) \citet{Greenhill:2003a}.
(40) \citet{Tadhunter:2003}.
(41) \citet{Gao:2017}.
(42) \citet{Bender:2005}.
(43) \citet{Rodriguez-Rico:2006}.
(44) \citet{Lodato:2003}.
(45) \citet{Onishi:2015}.
(46) \citet{Atkinson:2005}.
(47) \citet{Cappellari:2008}.
(48) \citet{Devereux:2003}.
(49) \citet{Yamauchi:2004}.
(50) \citet{Hicks:2008}.
(51) \citet{Davies:2006}.
(52) \citet{Onken:2014}.
(53) \citet{Pastorini:2007}.
(54) \citet{Brok:2015}.
(55) \citet{Jardel:2011}.
(56) Private value from K. Gebhardt \citep{Kormendy:2011}.
(57) \citet{Greenhill:1997}.
(58) \citet{Blais-Ouellette:2004}.
(59) \citet{Wold:2006}.
(60) This work.
(61) \citet{Vallee:2015}.
(62) \citet{Pour-Imani:2016}.
(63) \citet{Davis:2014}.
}
\label{Sample}
\begin{tabular}{llccccccccccrcllccc}
\hline
Galaxy name & \multicolumn{1}{c}{Type} & Bulge & Rf. & AC & Dist. & Rf. & Size & $\log({M_{\rm BH}/{\rm M_{\sun}}})$ & Met. & Rf. & $m$ & \multicolumn{1}{c}{$|\phi|$} & $i$ & \multicolumn{1}{c}{Telescope} & \multicolumn{1}{c}{Filter} & Res. & Rf. & Met. \\
 & & & & & (Mpc) & & ($\arcmin$) & & & & & \multicolumn{1}{c}{($\degr$)} & ($\degr$) & & & ($\arcsec$) & \\
(1) & \multicolumn{1}{c}{(2)} & (3) & (4) & (5) & (6) & (7) & (8) & (9) & (10) & (11) & (12) & \multicolumn{1}{c}{(13)} & (14) & \multicolumn{1}{c}{(15)} & \multicolumn{1}{c}{(16)} & (17) & (18) & (19) \\
\hline
Circinus & SABb & P & 1 &  & $4.21$ & 14 & $6.9$ & $6.25^{+0.07}_{-0.08}$ & m & 39 & 2 & $17.0\pm3.9$ & $48.8$ & \textit{HST} & \textit{F215N} & $0.19$ & 60 & T \\
Cygnus A & SB$^a$ & C & 2 &  & $258.4$ & 15 & $0.45$ & $9.44^{+0.11}_{-0.14}$ & g & 40 & 2 & $2.7\pm0.2$ & $0$ & \textit{HST} & \textit{F450W} & $0.08$ & 60 & T \\
ESO558-G009 & Sbc & P & 3 &  & $115.4$ & 15 & $1.6$ & $7.26^{+0.03}_{-0.04}$ & m & 41 & 2 & $16.5\pm1.3$ & $75.2$ & \textit{HST} & \textit{F814W} & $0.08$ & 60 & F \\
IC 2560 & SBb & P,C & 1,3 &  & $31.0$ & 16 & $3.2$ & $6.49^{+0.08}_{-0.10}$ & m & 41 & 2 & $22.4\pm1.7$ & $66.4$ & \textit{HST} & \textit{F814W} & $0.08$ & 60 & T \\
J0437+2456$^b$ & SB & P & 3 &  & $72.2$ & 15 & $0.8$ & $6.51^{+0.04}_{-0.05}$ & m & 41 & 2 & $16.9\pm4.1$ & $65.2$ & \textit{HST} & \textit{F814W} & $0.06$ & 60 & T \\
Milky Way & SBbc & P,C & 1,4 &  & $0.008$ & 17 &  & $6.60\pm0.02$ & p & 17 & $4^c$ & $13.1\pm0.6$ &  &  &  &  & 61 &  \\
Mrk 1029 & S & P & 3 &  & $136.9$ & 15 & $0.8$ & $6.33^{+0.10}_{-0.13}$ & m & 41 & 2 & $17.9\pm2.1$ & $0$ & \textit{HST} & \textit{F160W} & $0.09$ & 60 & T \\
NGC 0224 & SBb & C & 1 &  & $0.75$ & 18 & $190$ & $8.15^{+0.22}_{-0.10}$ & s & 42 & 1 & $8.5\pm1.3$ & $78.9$ & \textit{GALEX} & \textit{NUV} & $4.85$ & 13 & F \\
NGC 0253 & SABc & P & 5 &  & $3.47$ & 19 & $27.5$ & $7.00\pm0.30^d$ & g & 43 & 2 & $13.8\pm2.3$ & $73.0$ & \textit{Spitzer} & \textit{IRAC4} & $1.91$ & 60 & F \\
NGC 1068 & SBb & P,C & 1,6 & 3 & $10.1$ & 14 & $7.1$ & $6.75\pm0.02$ & m & 44 & 3 & $17.3\pm1.9$ & $42.2$ & SDSS & \textit{u} & $1.06$ & 60 & F \\
NGC 1097 & SBb & P & 7 & 12 & $24.9$ & 20 & $9.3$ & $8.38\pm0.03$ & g & 45 & 2 & $9.5\pm1.3$ & $48.4$ & \textit{Spitzer} & \textit{IRAC4} & $1.97$ & 62 & F \\
NGC 1300 & SBbc & P & 1 & 12 & $14.5$ & 14 & $6.2$ & $7.71^{+0.17}_{-0.12}$ & g & 46 & 2 & $12.7\pm2.0$ & $30.2$ & du Pont & \textit{B} & $0.69$ & 63 & F \\
NGC 1320 & Sa & P & 3 &  & $37.7$ & 21 & $1.9$ & $6.78^{+0.16}_{-0.26}$ & m & 41 & 1 & $19.3\pm2.0$ & $35.7$ & \textit{HST} & \textit{F330W} & $0.03$ & 60 & F \\
NGC 1398 & SBab & C & 6 & 6 & $24.8$ & 14 & $7.1$ & $8.03\pm0.08$ & s & 6 & 1 & $9.7\pm0.7$ & $42.3$ & \textit{GALEX} & \textit{FUV} & $4.20$ & 60 & T \\
NGC 2273 & SBa & P & 1 &  & $31.6$ & 22 & $3.2$ & $6.97\pm0.03$ & m & 41 & 2 & $15.2\pm3.9$ & $42.1$ & \textit{HST} & \textit{\textit{F336W}} & $0.11$ & 60 & T \\
NGC 2748 & Sbc & P & 1 &  & $18.2$ & 23 & $3.0$ & $7.54^{+0.15}_{-0.23}$ & g & 46 & 1 & $6.8\pm2.2$ & $74.3$ & \textit{Spitzer} & \textit{IRAC1} & $1.89$ & 60 & T \\
NGC 2960 & Sa & P & 6 &  & $71.1$ & 24 & $1.8$ & $7.06\pm0.03$ & m & 41 & 1 & $14.9\pm1.9$ & $58.3$ & \textit{HST} & \textit{\textit{F336W}} & $0.05$ & 60 & F \\
NGC 2974 & SB & C & 8 &  & $21.5$ & 14 & $3.5$ & $8.23\pm0.05$ & s & 8,47 & 3 & $10.5\pm2.9$ & $69.0$ & \textit{GALEX} & \textit{FUV} & $4.17$ & 60 & T \\
NGC 3031 & SBab & C & 1 & 12 & $3.48$ & 25 & $26.9$ & $7.83^{+0.11}_{-0.07}$ & g & 48 & 2 & $13.4\pm2.3$ & $53.4$ & \textit{Spitzer} & \textit{IRAC4} & $1.74$ & 60 & T \\
NGC 3079 & SBcd & P & 6 &  & $16.5$ & 14 & $7.9$ & $6.38^{+0.08}_{-0.10}$ & m & 49 & 2 & $20.6\pm3.8$ & $78.9$ & \textit{Spitzer} & \textit{IRAC2} & $1.76$ & 60 & F \\
NGC 3227 & SABa & P & 1 & 7 & $21.1$ & 22 & $5.4$ & $7.86^{+0.17}_{-0.25}$ & g,s & 50,51 & 2 & $7.7\pm1.4$ & $70.3$ & JKT & H$\alpha$ & $1.87$ & 60 & F \\
NGC 3368 & SABa & P\&C & 9 & 8 & $10.7$ & 26 & $7.6$ & $6.89^{+0.08}_{-0.10}$ & g,s & 9 & 2 & $14.0\pm1.4$ & $0$ & VATT 1.8m & \textit{R} & $0.62$ & 13 & F \\
NGC 3393 & SBa & P & 1 &  & $55.8$ & 15 & $2.2$ & $7.49^{+0.05}_{-0.06}$ & m & 41 & 3 & $13.1\pm2.5$ & $34.5$ & CTIO 0.9m & \textit{B} & $0.99$ & 13 & F \\
NGC 3627 & SBb & P & 6 & 7 & $10.6$ & 26 & $9.1$ & $6.95\pm0.05$ & s & 6 & 2 & $18.6\pm2.9$ & $53.9$ & \textit{Spitzer} & \textit{IRAC4} & $1.90$ & 62 & F \\
NGC 4151 & SABa & C & 6 & 5 & $19.0$ & 27 & $6.3$ & $7.68^{+0.15}_{-0.60}$ & g,s & 50,52 & 3 & $11.8\pm1.8$ & $0$ & VLA & 21 cm & $4.71$ & 13 & F \\
NGC 4258 & SABb & P,C & 7\&10,1 &  & $7.60$ & 28 & $18.6$ & $7.60\pm0.01$ & m & 28 & 1 & $13.2\pm2.5$ & $67.2$ & \textit{GALEX} & \textit{NUV} & $4.70$ & 60 & T \\
\hline
\end{tabular}
\end{table}
\end{landscape}

\begin{landscape}
\begin{table}
\contcaption{}
\begin{tabular}{llccccccccccrcllccc}
\hline
Galaxy name & \multicolumn{1}{c}{Type} & Bulge & Rf. & AC & Dist. & Rf. & Size & $\log({M_{\rm BH}/{\rm M_{\sun}}})$ & Met. & Rf. & $m$ & \multicolumn{1}{c}{$|\phi|$} & $i$ & \multicolumn{1}{c}{Telescope} & \multicolumn{1}{c}{Filter} & Res. & Rf. & Met. \\
 & & & & & (Mpc) & & ($\arcmin$) & & & & & \multicolumn{1}{c}{($\degr$)} & ($\degr$) & & & ($\arcsec$) & \\
(1) & \multicolumn{1}{c}{(2)} & (3) & (4) & (5) & (6) & (7) & (8) & (9) & (10) & (11) & (12) & \multicolumn{1}{c}{(13)} & (14) & \multicolumn{1}{c}{(15)} & \multicolumn{1}{c}{(16)} & (17) & (18) & (19) \\
\hline
NGC 4303 & SBbc & P & 7 & 9 & $12.3$ & 29 & $6.5$ & $6.58^{+0.07}_{-0.26}$ & g & 53 & 2 & $14.7\pm0.9$ & $0$ & \textit{GALEX} & \textit{NUV} & $4.38$ & 60 & T \\
NGC 4388 & SBcd & P & 1 &  & $17.8$ & 23 & $4.84$ & $6.90^{+0.04}_{-0.05}$ & m & 40 & 2 & $18.6\pm2.6$ & $74.0$ & KPNO 2.3m & $K_S$ & $1.29$ & 60 & F \\
NGC 4395 & SBm & N & 10 & 1 & $4.76$ & 30 & $13.2$ & $5.64^{+0.22}_{-0.12}$ & g & 54 & 2 & $22.7\pm3.6$ & $36.2$ & \textit{GALEX} & \textit{FUV} & $4.20$ & 60 & T \\
NGC 4501 & Sb & P & 6 & 9 & $11.2$ & 31 & $6.9$ & $7.13\pm0.08$ & s & 6 & 2 & $12.2\pm3.4$ & $35.7$ & \textit{GALEX} & \textit{NUV} & $4.10$ & 60 & T \\
NGC 4594 & Sa & P,C & 11\&12,1 &  & $9.55$ & 32 & $8.7$ & $8.81\pm0.03$ & s & 55 & 1 & $5.2\pm0.4$ & $80.7$ & \textit{Spitzer} & \textit{IRAC4} & $2.23$ & 60 & T \\
NGC 4699 & SABb & P\&C & 6 & 3 & $23.7$ & 14 & $3.8$ & $8.34\pm0.05$ & s & 6 & 1 & $5.1\pm0.4$ & $50.7$ & \textit{GALEX} & \textit{NUV} & $3.86$ & 60 & T \\
NGC 4736 & SBab & P & 1 & 3 & $4.41$ & 30 & $11.2$ & $6.78^{+0.09}_{-0.11}$ & s & 56 & 1 & $15.0\pm2.3$ & $32.9$ & \textit{GALEX} & \textit{FUV} & $3.90$ & 62 & F \\
NGC 4826 & Sab & P & 1 & 6 & $5.55$ & 23 & $10.0$ & $6.07^{+0.10}_{-0.12}$ & s & 56 & 3 & $24.3\pm1.5$ & $62.5$ & VLA & $21$ cm & $7.26$ & 60 & F \\
NGC 4945 & SBc & P & 1 &  & $3.72$ & 33 & $20.0$ & $6.15\pm0.30^d$ & m & 57 & 2 & $22.2\pm3.0$ & $80.5$ & 2MASS & $K_S$ & $2.77$ & 60 & F \\
NGC 5055 & Sbc & P & 7 & 3 & $8.87$ & 35 & $12.6$ & $8.94^{+0.09}_{-0.11}$ & g & 58 & 1 & $4.1\pm0.4$ & $67.9$ & \textit{GALEX} & \textit{FUV} & $3.80$ & 60 & T \\
NGC 5495 & SBc & P & 3 &  & $101.1$ & 15 & $1.4$ & $7.04^{+0.08}_{-0.09}$ & m & 41 & 2 & $13.3\pm1.4$ & $38.2$ & \textit{HST} & \textit{F814W} & $0.08$ & 60 & F \\
NGC 5765b & SABb & P & 3 &  & $133.9$ & 34 & $0.7$ & $7.72\pm0.03$ & m & 41 & 2 & $13.5\pm3.9$ & $0$ & \textit{HST} & \textit{F814W} & $0.08$ & 60 & T \\
NGC 6264 & SBb & P & 1 &  & $153.9$ & 36 & $0.81$ & $7.51\pm0.02$ & m & 41 & 2 & $7.5\pm2.7$ & $49.8$ & \textit{HST} & \textit{F110W} & $0.57$ & 60 & V \\
NGC 6323 & SBab & P & 1 &  & $116.9$ & 15 & $1.1$ & $7.02\pm0.02$ & m & 41 & 1 & $11.2\pm1.3$ & $68.2$ & \textit{HST} & \textit{\textit{F336W}} & $0.08$ & 60 & T \\
NGC 6926 & SBc & P & 13 & 5 & $87.6$ & 37 & $1.9$ & $7.74^{+0.26}_{-0.74}$ & m & 37 & 2 & $9.1\pm0.7$ & $73.3$ & 2MASS & \textit{J} & $2.97$ & 60 & T \\
NGC 7582 & SBab & P & 1 &  & $19.9$ & 20 & $5.0$ & $7.67^{+0.09}_{-0.08}$ & g & 59 & 1 & $10.9\pm1.6$ & $59.4$ & \textit{GALEX} & \textit{FUV} & $3.92$ & 60 & T \\
UGC 3789 & SABa & P & 1 &  & $49.6$ & 38 & $1.6$ & $7.06^{+0.02}_{-0.03}$ & m & 41 & 2 & $10.4\pm1.9$ & $55.5$ & \textit{HST} & \textit{F438W} & $0.08$ & 60 & T \\
UGC 6093 & SBbc & P & 3 &  & $164.1$ & 15 & $0.94$ & $7.45\pm0.04$ & m & 41 & 2 & $10.2\pm0.9$ & $24.7$ & \textit{HST} & \textit{F814W} & $0.08$ & 60 & T \\
\hline
\multicolumn{19}{l}{$^a$ Cygnus A displays a nuclear bar and spiral arms at a galactocentric radius < $2\farcs25$ ($2.82$ kpc).} \\
\multicolumn{19}{l}{$^b$ SDSS J043703.67+245606.8} \\
\multicolumn{19}{l}{$^c$ Meta-analysis by \citet{Vallee:2015}.} \\
\multicolumn{19}{l}{$^d$ A factor of 2 uncertainty has been assigned here.} \\
\end{tabular}
\end{table}
\end{landscape}

Fully aware of the inherent bias of algorithms to be confused by high-pitch angle noise,\footnote{This is akin to the persistence of low-frequency noise in FFT analysis. Noise abounds in frequencies that correspond to wavelengths of the order of the sampling range. For pitch angle analysis, low frequencies are high-pitch angle patterns with wavelengths of the order of the radial width of the annulus of a galactic disc. They experience less azimuthal winding than low-pitch angle, high-frequency patterns with shorter wavelengths, which wrap around a greater azimuthal range of the galaxy and potentially repeat their spiral pattern across the annulus of a galactic disc.} we took care to identify the fundamental pitch angle for each individual galaxy. This involved analysis that did not blindly quote the strongest Fourier pitch angle frequency, but rather sought to identify secondary and perhaps tertiary Fourier pitch angle frequencies that might represent the true, fundamental pitch angle. By applying multiple, independent software routines (see Appendix~\ref{demo}), we were confident in our ability to identify and rule out false pitch angle measurements. Collectively, this approach represents an improvement over past efforts as we have utilized the most appropriate method and avoided instances where things can go wrong.

Additionally --- unless care is taken --- we note that barred galaxies tend to be biased towards higher pitch angle values due to the presence of large central bars. This was, however, readily checked by varying the innermost radius of the region fit for spiral structure. The measured pitch angle starts to spike once the bar begins to influence the result. Even if such careful steps are taken to remove the influence of bars, the fact remains that much of the inner radial range of the galaxy is unusable for pitch angle measurement.

In contrast, unbarred galaxies can have spiral patterns that encompass the entire radial range of a galaxy except for the bulge (if present). Therefore, the main necessity for accurate pitch angle measurement is the presence of spiral arms that encompass large azimuthal ranges around the galaxy. The easiest galaxies to measure have spiral arm patterns that wrap around a significant fraction of the galaxy. Spiral patterns that wrap around $2{\rm \pi}$~radians become quite simple to measure.

Admittedly, it is challenging to model all the varying morphologies of spiral galaxies as possessing perfectly logarithmic and constant pitch angle spiral arms. To mediate this difficulty, one focuses on identifying the spiral arm \textit{segments} that are brightest and closest to the galactic centre, but beyond any central bars. Specifically, we have higher regard for stable stretches of constant pitch angle that are not at the outermost radial edge of a galaxy. In doing so, our pitch angles avoid, as best as possible, potential external tidal influence on the spiral arm geometry.

\section{Analysis and Results}\label{AR}

\begin{figure}
\includegraphics[clip=true,trim=0mm 0mm 0mm 0mm,width=\columnwidth]{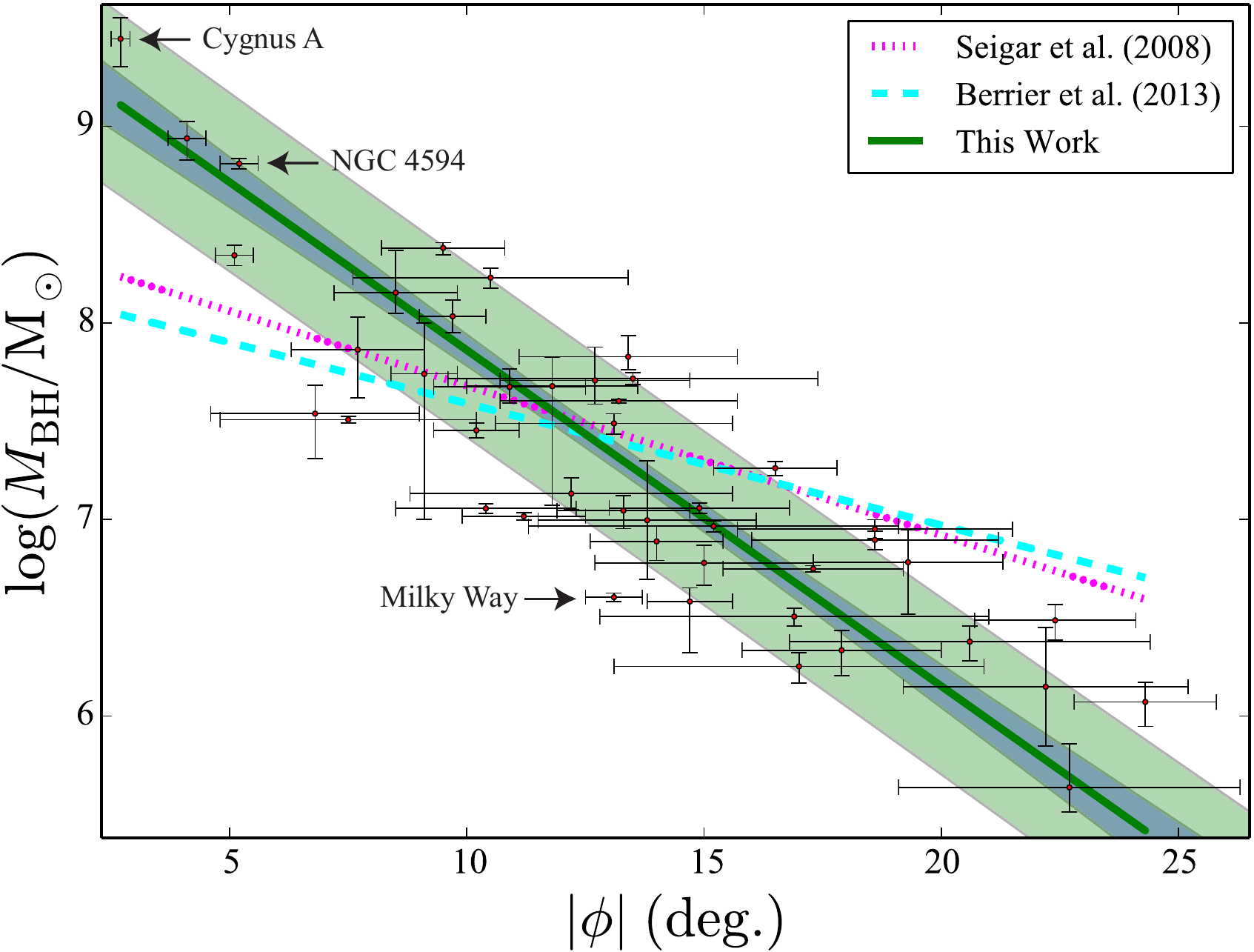}
\caption{Black hole mass (Table~\ref{Sample}, Column 9) versus the absolute value of the pitch angle in degrees (Table~\ref{Sample}, Column 13), represented as red dots bounded by black error bars. Equation~(\ref{M-phi}) is the solid green line (which represents the result of the error-weighted \textsc{bces} (Y|X) regression of $\log{M_{\rm BH}}$ on $|\phi|$). The $1\sigma$ confidence band (smaller dark shaded region) and the $1\sigma$ total rms scatter band (larger light shaded region) depict the error associated with the fit parameters (slope and intercept) and the rms scatter about the best fit of equation~(\ref{M-phi}), respectively. The three galaxies with questionable measurements (see Section \ref{Outliers}) are labelled. For comparison, we have also plotted the ordinary least squares (Y|X) linear regression from \citet{Seigar:2008} and from \citet{Berrier:2013}, represented by a dotted magenta and a dashed cyan line, respectively.}
\label{plot}
\end{figure}

We performed linear fits using the \textsc{bces} (bivariate correlated errors and intrinsic scatter) regression method \citep{BCES},\footnote{We used a version of the \textsc{bces} software translated into the \textsc{python} programming language by R. S. Nemmen for use in astronomical applications \citep{Nemmen:2012}.} which takes into account measurement error in both coordinates and intrinsic scatter in data. For our ($\phi$, $M_{\rm BH}$) data, we use the \textsc{bces} (Y|X) fitting method, which minimizes the residuals in the ${\rm Y}=\log{M_{\rm BH}}$ direction, and the \textsc{bces} Bisector fitting method, which bisects the angle between the \textsc{bces} (Y|X) and the \textsc{bces} (X|Y)\footnote{The \textsc{bces} (X|Y) regression minimizes the residuals in the ${\rm X} = |\phi|$ direction.} slopes. We find from analysis of the full sample of 44 galaxies (see Fig.~\ref{plot}) that the \textsc{bces} (Y|X) regression yields a slope and intercept\footnote{To reduce the uncertainty on the intercept, we performed a regression of $\log({M_{\rm BH}/{\rm M_{\sun}}})$ on ($|\phi| - |\phi|_{\rm median}$), with $|\phi|_{\rm median}\equiv15\degr$ being the approximate median integer value of $|\phi|$.} such that
\begin{equation}
\log({M_{\rm BH}/{\rm M_{\sun}}}) = (7.01\pm0.07) - (0.171\pm0.017)[|\phi|-15\degr],
\label{M-phi}
\end{equation}
with intrinsic scatter $\epsilon=0.33\pm0.08$~dex and a total root mean square (rms) scatter $\Delta = 0.43$~dex in the $\log{M_{\rm BH}}$ direction.\footnote{The intrinsic scatter is the quadratic difference between the total rms scatter and the measurement uncertainties.} The quality of the fit can be described with a Pearson correlation coefficient of $r = -0.88$ and a $p$-value probability of $5.77\times10^{-15}$ that the null hypothesis is true.

As pointed out in \citet{Novak:2006}, since there is no natural division of the variables into `dependant' and `independent' variables in black hole scaling relations, we prefer to represent the $M_{\rm BH}$--$\phi$ relation with a \textit{symmetric} treatment of the variables, as is the case in the \textsc{bces} Bisector regression. However, given that the error bars on the logarithm of the black hole masses are much smaller than the error bars on the pitch angles (see Table~\ref{Sample}), the \textsc{bces} (X|Y) regression, and thus also the symmetric treatment of our data, results in the same relation (equation~\ref{M-phi}) as the asymmetric regression performed above. We additionally used the modified \textsc{fitexy} routine from \citet{Tremaine:2002} and obtained consistent results.

\begin{table*}
\caption{\textsc{bces} (Y|X) linear regressions for the expression $\log({M_{\rm BH}/{\rm M_{\sun}}}) = A[|\phi|-15\degr]+B$.
\textbf{Columns:}
(1) Fit number.
(2) Sample description.
(3) Sample size.
(4) Slope.
(5) $\log({M_{\rm BH}/{\rm M_{\sun}}})$--intercept at $|\phi|=15\degr$.
(6) Intrinsic scatter in the $\log{M_{\rm BH}}$ direction.
(7) Total rms scatter in the $\log{M_{\rm BH}}$ direction.
(8) Pearson correlation coefficient.
(9) $p$-value probability that the null hypothesis is true.
}
\label{fits}
\begin{tabular}{clccccccl}
\hline
Fit & \multicolumn{1}{c}{Sample} & $N$ & $A$ & $B$ & $\epsilon$ & $\Delta$ & $r$ & \multicolumn{1}{c}{$p$-value} \\
 & & & (dex/deg) & (dex) & (dex) & (dex) & & \\
(1) & \multicolumn{1}{c}{(2)} & (3) & (4) & (5) & (6) & (7) & (8) & \multicolumn{1}{c}{(9)} \\
\hline
1 & All & 44 & $-0.171\pm0.017$ & $7.01\pm0.07$ & $0.33\pm0.08$ & 0.43 & $-0.88$ & $5.77\times10^{-15}$ \\
2 & Pseudobulges + hybrids & $37^a$ & $-0.153\pm0.018$ & $6.99\pm0.07$ & $0.31\pm0.08$ & 0.41 & $-0.85$ & $1.68\times10^{-11}$ \\
3 & Classical bulges + hybrids & $13^a$ & $-0.169\pm0.025$ & $7.13\pm0.16$ & $0.31\pm0.08$ & 0.41 & $-0.90$ & $2.31\times10^{-5}$ \\
4 & Barred & 35 & $-0.188\pm0.024$ & $6.96\pm0.09$ & $0.35\pm0.09$ & 0.46 & $-0.86$ & $2.66\times10^{-11}$ \\
5 & Unbarred & 9 & $-0.143\pm0.020$ & $7.11\pm0.12$ & $0.33\pm0.08$ & 0.43 & $-0.92$ & $4.92\times10^{-4}$ \\
6 & $m=2$ & 26 & $-0.188\pm0.028$ & $7.00\pm0.09$ & $0.41\pm0.11$ & 0.49 & $-0.86$ & $1.79\times10^{-8}$ \\
7 & $m\neq2$ & 18 & $-0.153\pm0.019$ & $7.05\pm0.10$ & $0.28\pm0.09$ & 0.40 & $-0.88$ & $1.20\times10^{-6}$ \\
\hline
\multicolumn{9}{l}{$^a$ Seven galaxies (IC 2560, the Milky Way, NGC 1068, NGC 3368, NGC 4258, NGC 4594 and NGC 4699)} \\
\multicolumn{9}{l}{potentially have both types of bulge morphology. The bulge-less galaxy NGC 4395 is excluded.}
\end{tabular}
\end{table*}

\subsection{Sub-samples}

\begin{figure}
\includegraphics[clip=true,trim= 0mm 0mm 0mm 0mm,width=\columnwidth]{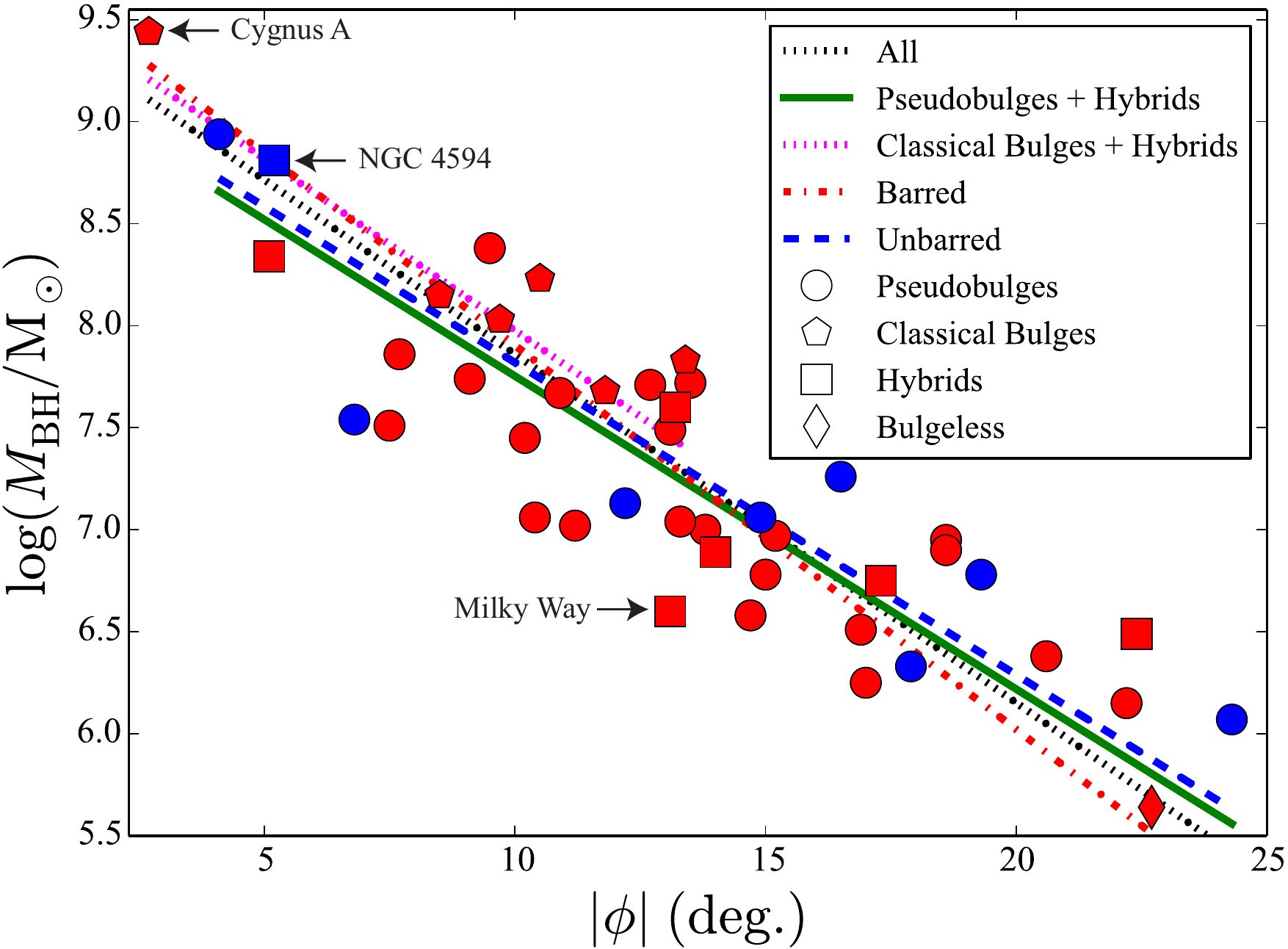}
\caption{Similar to Fig.~\ref{plot}. Galaxies with bulges only classified as classical are indicated by pentagons. Galaxies with bulges only classified as pseudobulges are indicated by circles. Galaxies with bulges ambiguously classified as having either classical, pseudo or both types of bulges are labelled as having hybrid bulges and are marked with squares. NGC 4395, being the only bulge-less galaxy in our sample, is marked with a diamond. Markers filled with the colour red represent galaxies with barred morphologies and markers filled with the colour blue represent galaxies with unbarred morphologies. The full and sub-sample \textsc{bces} (Y|X) linear regressions are plotted as lines with various styles and colours. Fits 1--5 from Table~\ref{fits} are depicted as a dotted black line, a solid green line, a dotted magenta line, an alternating dash--dotted red line and a dashed blue line, respectively. The three galaxies with questionable measurements (see Section \ref{Outliers}) are labelled. Error bars and confidence regions have not been included for clarity.}
\label{plot2}
\end{figure}

We have explored the $M_{\rm BH}$--$\phi$ relation for various subsets that segregate different types of bulges and overall morphologies: pseudobulges, classical bulges, barred and unbarred galaxies. The results of this analysis are presented in Table~\ref{fits} and Fig.~\ref{plot2}. \citet{Graham:2008} \& \citet{Hu:2008} presented evidence that barred/pseudobulge galaxies do not follow the same $M_{\rm BH}$--$\sigma$ scaling relation as unbarred/classical bulges and several authors have speculated that SMBHs do not correlate with galaxy discs \citep[e.g.][]{Kormendy:2011}. However, recent work by \citet{Simmons:2017} indicate that disc-dominated galaxies do indeed co-evolve with their SMBHs. We present evidence that SMBHs clearly correlate well with galactic discs in as much as the existence of an $M_{\rm BH}$--$\phi$ relation demands such a correlation \citep{Treuthardt:2012}. Furthermore, we reveal in Table~\ref{Sample} that most of the galaxies in our sample are alleged in the literature to contain pseudobulges. That is, SMBHs in alleged pseudobulges correlate with their galaxy's discs (Table~\ref{fits}).

We further acknowledge that the label `pseudobulge' is often an ambiguous moniker. The qualifications that differentiate pseudobulges from classical bulges are extensive \citep{Fisher:Drory:2016} and are often difficult to determine definitively \citep{Savorgnan:2016:II,Graham:2016b}. This can be observed from the seven hybrid galaxies in our sample (see Table~\ref{Sample}, Column 3) that either have conflicting pseudobulge versus classical bulge classifications in the literature or are stated as possessing both a pseudobulge and classical bulge, simultaneously.

Observing Fig.~\ref{plot2}, we do note that the six galaxies (Cygnus A, NGC 224, NGC 1398, NGC 2974, NGC 3031 and NGC 4151) classified unambiguously as possessing classical bulges, all lie above the best-fitting linear regression for the entire sample. Since the linear regression naturally acts to divide half of the sample above the line of best fit, the individual chance of a particular galaxy lying above the line of best fit is 50~per~cent, making the probability of these specific six galaxies all lying above the line of best fit 1 chance out of 64. This informs us that the classical bulges tend to have higher black hole masses for a given pitch angle. However, it is important to note the diminished statistical significance of the classical bulge sample due to it having a small sample size of only 13 galaxies (with seven of those being hybrid bulge morphologies). Furthermore, this sample includes all three galaxies with questionable measurements (see Section \ref{Outliers}). However, what is of interest is that the galaxies alleged to have pseudobulges define a tight relation; they are not randomly distributed in the $M_{\rm BH}$--$\phi$ diagram.

We also find a majority of our sample consisting of barred morphologies. All of the relations in Table~\ref{fits} are similar (within the quoted margins of error) concerning both their slopes and intercepts, except for the slopes of the barred and unbarred sample. The barred sample has a statistically dissimilar (i.e. error bars that do not overlap), steeper slope than the unbarred sample. Again, it is important to point out that this observation is derived from small number statistics for the unbarred sample, but the error bars should capture this. This observed dissimilarity is such that barred galaxies tend to have more massive black holes than unbarred galaxies with equivalent pitch angles for $|\phi| \loa 11\fdg8$, and vice versa.

Finally, we investigate whether the number of spiral arms affects the determination of the $M_{\rm BH}$--$\phi$ relation. We compare galaxies with two dominant spiral arms ($m=2$) to those with any other count of dominant spiral arms ($m\neq2$). For all galaxies (except the Milky Way), we use the \textsc{spirality} software to count the number of spiral arms (see the middle and right-hand panels of Fig.~\ref{spirality}. In the end, we find that the two samples are statistically equivalent (see Table~\ref{fits}, Fits 6 and 7).

\subsection{Questionable measurements}\label{Outliers}

Cygnus A has the most massive SMBH in our spiral galaxy sample; it is the only spiral galaxy with an SMBH mass greater than one billion solar masses. While numerous early-type galaxies are known to exceed the billion solar mass mark, it is uncommon for spiral galaxies to achieve this mass. Cygnus A is also the most distant galaxy in our sample, with the most ambiguous morphological classification. Furthermore, the bar and spiral arms in Cygnus A are nuclear features unlike the large-scale features in the discs of all other galaxies in our sample. It may be that Cygnus A is an early-type galaxy with an intermediate-scale disc hosting a spiral, cf. CG 611 \citep{Graham:2017}. Recently, \citet{Perley:2017} discovered what is potentially a secondary SMBH near the central SMBH in Cygnus A, further complicating our understanding of this galaxy.

NGC 4594, the `Sombrero' galaxy, is notorious for being simultaneously elliptical and spiral, behaving like two galaxies, one inside the other \citep{Gadotti:2012}.

As for the Milky Way, our Galaxy has been determined to have an uncommonly tight spiral structure (albeit measured with difficulty by astronomers living inside of it) for its relatively low-mass SMBH. It is worth noting that published values of the Milky Way's pitch angle have varied wildly, ranging from $3\degr\leq|\phi|\leq28\degr$. Meta-analysis of these various published values yields a best-fitting absolute value of $13\fdg1\pm0\fdg6$ \citep{Vallee:2015}, which is what we used here. In fact, the most recent measurement of the Milky Way's pitch angle by \citet{Rastorguev:2017} is even smaller ($|\phi| = 10\fdg4\pm0\fdg3$), and thus even more of an outlier if applied to the $M_{\rm BH}$--$\phi$ relation.

Removing these three galaxies does not change any of the results in Table~\ref{fits} by more than the 1$\sigma$ level.

\subsection{$M_{\rm BH}$--$\sigma$ relation for spiral galaxies}

Here, we analyse the $M_{\rm BH}$--$\sigma$ relationship for our sample of 44 spiral galaxies. We obtained the majority of our central velocity dispersion ($\sigma$) measurements from the HyperLeda data base (see Table~\ref{sigma_table}). Literature values were available for all galaxies except for NGC 6926. Unlike the $M_{\rm BH}$--$\phi$ relation, a marked difference arises between the various \textsc{bces} regressions of our $M_{\rm BH}$--$\sigma$ data. We therefore present both the results of the \textsc{bces} (Y|X) and the \textsc{bces} Bisector regressions.

\begin{figure}
\includegraphics[clip=true,trim= 0mm 0mm 0mm 0mm,width=\columnwidth]{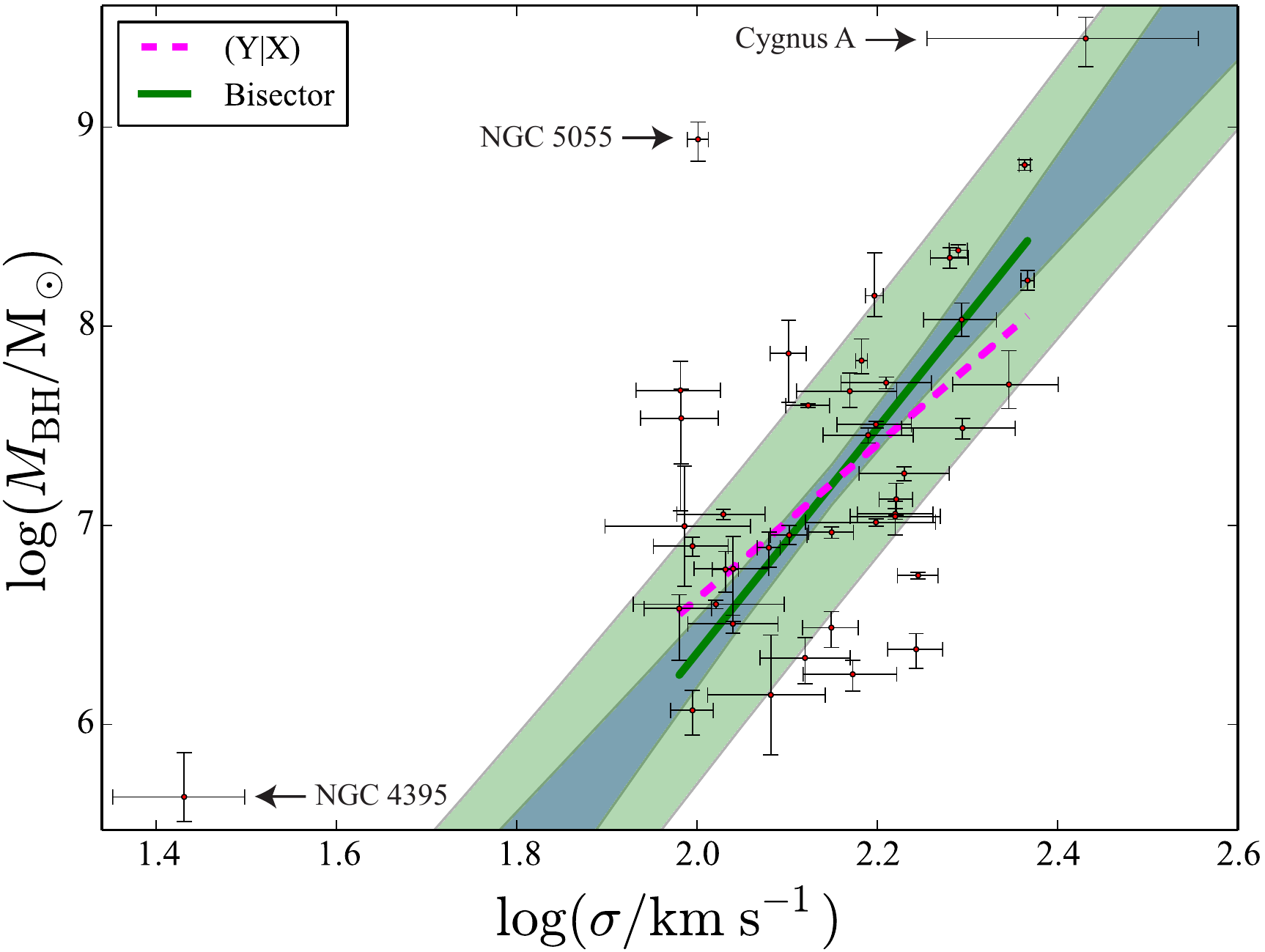}
\caption{Black hole mass (Table~\ref{Sample}, Column 9) versus central velocity dispersion (Table~\ref{sigma_table}, Column 3), represented as red dots bounded by black error bars. The \textsc{bces} (Y|X) and the \textsc{bces} Bisector regressions of Fit \# 1 from Table~\ref{sigma_fits} are depicted by a dashed magenta and a solid green line, respectively. The $1\sigma$ confidence band (smaller grey shaded region) depicts the error associated with the fit parameters (slope and intercept). The $1\sigma$ total rms scatter about the best-fitting \textsc{bces} Bisector regression is shown by the larger green shaded region. We consider the three labelled galaxies as outliers and they are not included in any of our linear regressions involving central velocity dispersion.}
\label{sigma_morph}
\end{figure}

\begin{figure}
\includegraphics[clip=true,trim= 0mm 0mm 0mm 0mm,width=\columnwidth]{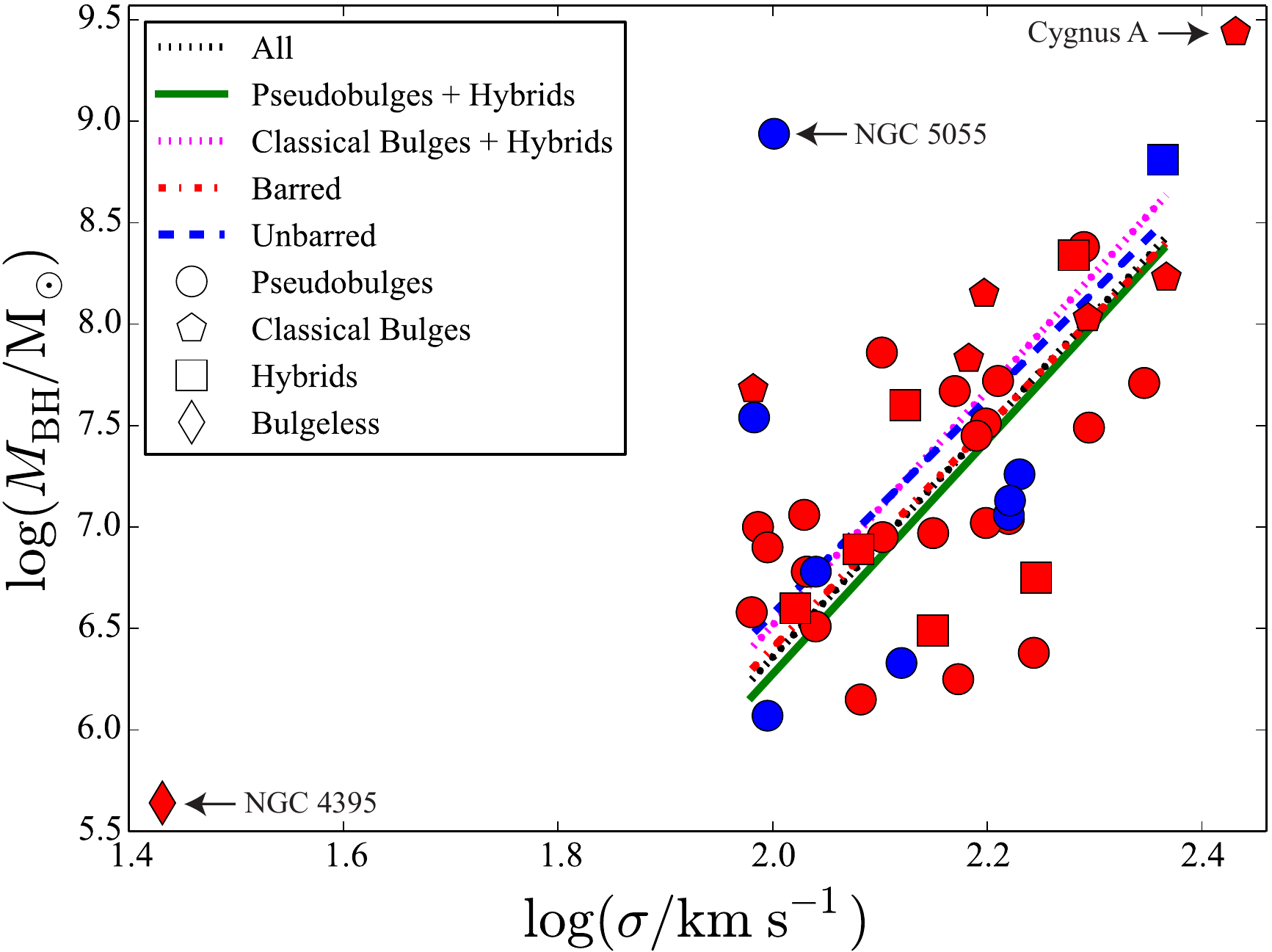}
\caption{Similar to Fig.~\ref{sigma_morph}. Symbols are the same as in Fig.~\ref{plot2}. \textsc{bces} Bisector linear regressions to the full and sub-samples are plotted as lines with various styles and colours. Bisector Fits 1--5 from Table~\ref{sigma_fits} are depicted as a dotted black line, a solid green line, a dotted magenta line, an alternating dash--dotted red line and a dashed blue line, respectively.}
\label{sigma_bulge_morph}
\end{figure}

Plots depicting the overall fit, and delineating barred/unbarred morphologies, as well as different bulge morphologies, can be seen in Figs.~\ref{sigma_morph}~\&~\ref{sigma_bulge_morph}, respectively. It is evident from these figures that three galaxies standout as noticeable outliers: NGC 4395, NGC 5055 and Cygnus A. NGC 4395 has an extremely low velocity dispersion, Cygnus A has the largest uncertainty on $\sigma$ of any galaxy in our sample, and NGC 5055 appears to have an uncharacteristically low velocity dispersion for a galaxy hosting such a large black hole. We have excluded these three galaxies from all linear regressions (see Table~\ref{sigma_fits}), leaving us with a sample of 40 spiral galaxies.

\begin{table}
\caption{Central velocity dispersions.
\textbf{Columns:}
(1) Galaxy name.
(2) Checkmark indicates the galaxy has been identified to possess a bar.
(3) Central velocity dispersion (km~s$^{-1}$).
(4) Central velocity dispersion reference.
}
\label{sigma_table}
\begin{tabular}{lccl}
\hline
Galaxy name & Bar? & $\sigma$ & Reference \\
 & & (km~s$^{-1}$) & \\
(1) & (2) & (3) & (4) \\
\hline
Circinus & \checkmark & $149\pm18$ & HyperLeda \\
Cygnus A & \checkmark & $270\pm90$ & \citet{Kormendy:Ho:2013} \\
ESO558-G009 & & $170\pm^{+21}_{-19}$ & \citet{Greene:2016} \\
IC 2560 & \checkmark & $141\pm10$ & \citet{Kormendy:Ho:2013} \\
J0437+2456 & \checkmark & $110^{+13}_{-12}$ & \citet{Greene:2016} \\
Milky Way & \checkmark & $105\pm20$ & \citet{Kormendy:Ho:2013} \\
Mrk 1029 & & $132^{+16}_{-14}$ & \citet{Greene:2016} \\
NGC 0224 & \checkmark & $157\pm4$ & HyperLeda \\
NGC 0253 & \checkmark & $97\pm18$ & HyperLeda \\
NGC 1068 & \checkmark & $176\pm9$ & HyperLeda \\
NGC 1097 & \checkmark & $195^{+5}_{-4}$ & \citet{Bosch:2016} \\
NGC 1300 & \checkmark & $222\pm30$ & HyperLeda \\
NGC 1320 & & $110\pm10$ & HyperLeda \\
NGC 1398 & \checkmark & $197\pm18$ & HyperLeda \\
NGC 2273 & \checkmark & $141\pm8$ & HyperLeda \\
NGC 2748 & & $96\pm10$ & HyperLeda \\
NGC 2960 & & $166^{+17}_{-15}$ & \citet{Saglia:2016} \\
NGC 2974 & \checkmark & $233\pm4$ & HyperLeda \\
NGC 3031 & \checkmark & $152\pm2$ & HyperLeda \\
NGC 3079 & \checkmark & $175\pm12$ & HyperLeda \\
NGC 3227 & \checkmark & $126\pm6$ & HyperLeda \\
NGC 3368 & \checkmark & $120\pm4$ & HyperLeda \\
NGC 3393 & \checkmark & $197\pm28$ & HyperLeda \\
NGC 3627 & \checkmark & $127\pm6$ & HyperLeda \\
NGC 4151 & \checkmark & $96\pm10$ & HyperLeda \\
NGC 4258 & \checkmark & $133\pm7$ & HyperLeda \\
NGC 4303 & \checkmark & $96\pm8$ & HyperLeda \\
NGC 4388 & \checkmark & $99\pm9$ & HyperLeda \\
NGC 4395 & \checkmark & $27\pm5$ & HyperLeda \\
NGC 4501 & & $166\pm7$ & HyperLeda \\
NGC 4594 & & $231\pm3$ & HyperLeda \\
NGC 4699 & \checkmark & $191\pm9$ & HyperLeda \\
NGC 4736 & \checkmark & $108\pm4$ & HyperLeda \\
NGC 4826 & & $99\pm5$ & HyperLeda \\
NGC 4945 & \checkmark & $121\pm18$ & HyperLeda \\
NGC 5055 & & $100\pm3$ & HyperLeda \\
NGC 5495 & \checkmark & $166^{+20}_{-18}$ & \citet{Greene:2016} \\
NGC 5765b & \checkmark & $162^{+20}_{-18}$ & \citet{Greene:2016} \\
NGC 6264 & \checkmark & $158\pm15$ & \citet{Kormendy:Ho:2013} \\
NGC 6323 & \checkmark & $158\pm26$ & \citet{Kormendy:Ho:2013} \\
NGC 6926 & \checkmark &  &  \\
NGC 7582 & \checkmark & $148\pm19$ & HyperLeda \\
UGC 3789 & \checkmark & $107\pm12$ & \citet{Kormendy:Ho:2013} \\
UGC 6093 & \checkmark & $155^{+19}_{-17}$ & \citet{Greene:2016} \\
\hline
\end{tabular}
\end{table}

\begin{table*}
\caption{\textsc{bces} linear regressions for the expression $\log({M_{\rm BH}/{\rm M_{\sun}}}) = A\log[\sigma/200$~$\rm{km}$~$\rm{s}$$^{-1}]+B$. Similar to Table~\ref{fits}, except that a different expression has been fit, and two types of regression are used.
}
\label{sigma_fits}
\begin{tabular}{cclccccccc}
\hline
Fit & Regression & \multicolumn{1}{c}{Sample} & $N$ & $A$ & $B$ & $\epsilon$ & $\Delta$ & $r$ & $p$-value \\
 & & & & & (dex) & (dex) & (dex) & & \\
(1) & (2) & \multicolumn{1}{c}{(3)} & (4) & (5) & (6) & (7) & (8) & (9) & (10) \\
\hline
\multirow{2}{*}{1} & (Y|X) & \multirow{2}{*}{All} & \multirow{2}{*}{$40^a$} & $3.88\pm0.89$ & $7.80\pm0.16$ & $0.54\pm0.02$ & 0.57 & \multirow{2}{*}{0.56} & \multirow{2}{*}{$1.72\times10^{-4}$} \\
 & Bisector & & & $5.65\pm0.79$ & $8.06\pm0.13$ & $0.58\pm0.03$ & 0.63 & &  \\
\hline
\multirow{2}{*}{2} & (Y|X) & \multirow{2}{*}{Pseudobulges + hybrids} & \multirow{2}{*}{$35^{a,b}$} & $3.97\pm1.03$ & $7.73\pm0.19$ & $0.51\pm0.02$ & 0.55 & \multirow{2}{*}{0.56} & \multirow{2}{*}{$5.03\times10^{-4}$} \\
 & Bisector & & & $5.76\pm0.91$ & $8.01\pm0.17$ & $0.55\pm0.03$ & 0.61 & & \\
\hline
\multirow{2}{*}{3} & (Y|X) & \multirow{2}{*}{Classical bulges + hybrids} & \multirow{2}{*}{$12^{a,b}$} & $4.15\pm1.47$ & $8.08\pm0.20$ & $0.62\pm0.01$ & 0.62 & \multirow{2}{*}{0.63} & \multirow{2}{*}{$2.85\times10^{-2}$} \\
 & Bisector & & & $5.78\pm1.34$ & $8.26\pm0.15$ & $0.66\pm0.01$ & 0.67 & & \\
 \hline
\multirow{2}{*}{4} & (Y|X) & \multirow{2}{*}{Barred} & \multirow{2}{*}{$32^a$} & $3.63\pm0.92$ & $7.78\pm0.17$ & $0.53\pm0.02$ & 0.56 & \multirow{2}{*}{0.52} & \multirow{2}{*}{$2.13\times10^{-3}$} \\
 & Bisector & & & $5.45\pm0.86$ & $8.04\pm0.15$ & $0.57\pm0.03$ & 0.62 & & \\
\hline
\multirow{2}{*}{5} & (Y|X) & \multirow{2}{*}{Unbarred} & \multirow{2}{*}{$8^a$} & $4.52\pm2.15$ & $7.82\pm0.33$ & $0.64\pm0.02$ & 0.68 & \multirow{2}{*}{0.66} & \multirow{2}{*}{$7.31\times10^{-2}$} \\
 & Bisector & & & $6.06\pm1.54$ & $8.06\pm0.24$ & $0.68\pm0.02$ & 0.73 & & \\
 \hline
\multirow{2}{*}{6} & (Y|X) & \multirow{2}{*}{$m=2$} & \multirow{2}{*}{$23^a$} & $3.49\pm1.18$ & $7.60\pm0.23$ & $0.51\pm0.01$ & 0.54 & \multirow{2}{*}{0.46} & \multirow{2}{*}{$2.55\times10^{-2}$} \\
 & Bisector & & & $5.50\pm1.27$ & $7.92\pm0.23$ & $0.54\pm0.02$ & 0.60 & & \\
 \hline
\multirow{2}{*}{7} & (Y|X) & \multirow{2}{*}{$m\neq2$} & \multirow{2}{*}{$17^a$} & $3.88\pm1.18$ & $7.98\pm0.18$ & $0.56\pm0.02$ & 0.58 & \multirow{2}{*}{0.64} & \multirow{2}{*}{$5.83\times10^{-3}$} \\
 & Bisector & & & $5.30\pm0.96$ & $8.17\pm0.15$ & $0.59\pm0.03$ & 0.64 & & \\
\hline
\multicolumn{10}{l}{$^a$ Excluding NGC 6926 (for lack of a velocity dispersion measurement) and the outliers: Cygnus A, NGC 4395 and NGC 5055.} \\
\multicolumn{10}{l}{$^b$ Seven galaxies (IC 2560, the Milky Way, NGC 1068, NGC 3368, NGC 4258, NGC 4594 and NGC 4699) potentially have} \\
\multicolumn{10}{l}{both types of bulge morphology. The bulge-less galaxy NGC 4395 is excluded.}
\end{tabular}
\end{table*}

Many works \citep[e.g.][]{Graham:2008,Hu:2008,Gultekin:2009,Graham:Scott:2013} have identified an offset ($\approx 0.3$~dex) in SMBH mass between galaxies with barred and unbarred morphologies in the $M_{\rm BH}$--$\sigma$ diagram. However, we do not observe any such offset in our $M_{\rm BH}$--$\sigma$ diagram. This could be attributed to numerous reasons. Since \citet{Graham:Scott:2013}, five of the unbarred spiral galaxies in their sample have been identified as possessing bars: NGC 224, NGC 3031, NGC 4388, NGC 4736 \citep[][and references therein]{Savorgnan:2016:II}, plus NGC 6264 \citep[][and references therein]{Saglia:2016}. As such, our sample of 40 only contains 8 unbarred galaxies. Also, many of the black hole mass estimates and distances have been revised. Moreover, \citet{Graham:Scott:2013} observe the offset using a sample that includes elliptical, lenticular and spiral morphologies. Since we do not include any elliptical nor lenticular morphologies, it becomes difficult to directly compare our results.

Two  of the `unbarred' galaxies are NGC 4826 (the `Black Eye Galaxy') and Mrk 1029, both alleged to have pseudobulges. Unexpectedly, all of the eight galaxies without bars have been claimed to host pseudobulges, structures thought to be associated with bars.

Concerning the linear regressions for the various subsamples (see Table~\ref{sigma_fits}), we find no statistical difference between the slope or intercept for all but one of the fits when using the same type of regression. With the symmetric bisector regression, the `Classical Bulges + Hybrids' sample (Fit \#3 from Table~\ref{sigma_fits}) is noticeably offset above the other fits (see Fig. \ref{sigma_bulge_morph}). The vertical offset is 0.20~dex above the `All' sample and 0.25~dex above the `Pseudobulges + Hybrids' sample. However, the intercept values do have overlapping error bars. We additionally used the modified \textsc{fitexy} routine from \citet{Tremaine:2002} and obtained consistent results.

\section{Discussion and Implications}\label{DI}

The spiral density wave theory has been cited for approximately six decades \citep{Shu:2016} as the agent for `grand design' spiral genesis in disc galaxies. \citet{Lin:Shu:1966} specify that the density wave theory predicts that a relationship should exist between spiral arm pitch angle and the central enclosed mass of a galaxy.\footnote{In their chapter~6, \citet{Binney:Tremaine:1987} provide an extensive discussion of the implications and potential limitations of the spiral wave dispersion relation.} They calculated the pitch angle to be a ratio of the density of material in the galaxy's disc relative to a certain quantity made up of the frequencies of orbital motions in the discs, which itself is dependent on the central gravitational mass. Specifically, that the pitch angle at a given radius is determined by the density of the medium in that region of the disc and the enclosed gravitational mass central to that orbital galactocentric radius.\footnote{See equation~4.1 from \citet{Lin:Shu:1966}. This formula is also represented in equation~1 from \citet{Davis:2015} in terms of pitch angle and later simplified in their equation~2. However, please note that due to a corrigendum, their equation~1 is not dimensionally correct and needs to be divided by the galactocentric radius on the righthand side of the formula.}

\subsection{Variable pitch angle}

So far this decade, there have been several studies in the literature concerning the variances in pitch angle measurements caused by numerous factors such as the wavelength of light and galactocentric radius. \citet{Martinez-Garcia:2014} find, from a study of five galaxies across the optical spectrum, that the absolute value of pitch angle gradually increases at longer wavelengths for three galaxies. This result can be contrasted with the larger study of \citet{Pour-Imani:2016}, who use a sample of 41 galaxies imaged from \textit{FUV} to 8.0~$\micron$ wavelengths of light. They find that the absolute pitch angle of a galaxy is statistically smaller\footnote{The analysis of \citet{Pour-Imani:2016} indicates that the most prominent observed difference is between the 3.6~$\micron$ pitch angle ($|\phi_{3.6\,\mu{\rm m}}|$) and the 8.0~$\micron$ pitch angle ($|\phi_{8.0\,\mu{\rm m}}|$). The typical difference is: $|\phi_{8.0\,\mu{\rm m}}| - |\phi_{3.6\,\mu{\rm m}}| = 3\fdg75\pm1\fdg25$.} (tighter winding) when measured using light that highlights old stellar populations, and larger (looser winding) when using light that highlights young star forming regions. As noted in Section \ref{DM}, to account for this variation, we strived to identify the most fundamental pitch angle for each galaxy in light that preferably indicates star forming regions. In doing so, we should be glimpsing the true location of the spiral density wave, which itself should be related to the central mass and ultimately the SMBH mass of a galaxy.

It is also worth considering the disc size--luminosity relations for different morphological types. \citet{Graham:Worley:2008} indicate that disc scalelength is roughly constant with Hubble type, but the disc central surface brightness shows a definite trend of decreasing with Hubble type. For this to occur, the luminosity of the disc must also become fainter with increasing Hubble type. Therefore, the discs become thinner (decreased surface density) as the spiral arms become more open in the late-type spirals. This would indicate that in more open spiral patterns, the overall disc density is small. However, this only implies that the \textit{stellar} density has decreased in the disc. It is likely that the gas density, and thus the gas fraction of the total density, is higher in gas-rich, late-type galaxies. Indeed, \citet{Davis:2015} present evidence (see their equation~2) that the gas density (as compared to the stellar density) in the disc is the primary indicator of spiral tightness since it is primarily within the gas that the density wave propagates.

For the case of variable pitch angle with galactocentric radius, \citet{Savchenko:2013} find that most galaxies cannot be described by a single pitch angle. In those cases, the absolute value of pitch angle decreases with increasing galactocentric radius (i.e. the arms become more tightly wound). This is in agreement with \citet{Davis:2015}, who predict that there should be a natural tendency for pitch angle to decrease with increasing galactocentric radius due to conditions inherent in the density wave theory. Particularly, because as galactocentric radius increases, the enclosed mass must increase and the gas density in the disc typically decreases. Both of these factors tend to tighten the spiral arm pattern (decrease the pitch angle). However, this can be contrasted with the findings of \citet{Davis:Hayes:2014}, whose observations indicate the opposite (i.e. increasing pitch angle with increasing galactocentric radius).

\subsection{Evolution of pitch angle}\label{Evolution}

It is important for the validity of any relationship derived from pitch angle, and for how proposed relations connect to broader galaxy evolution, that spiral patterns not be transient features. Observations of the ubiquity of spiral galaxies, accounting for 56~per~cent of the galaxies in our local Universe \citep{Loveday:1996}, appear to favour the longevity of spiral structure. Over the years, there have been numerous findings from theory and computer simulations of spiral galaxies. \citet{Julian:Toomre:1966} show that spirals can be a transient phenomenon brought on by lumpy perturbers in the disc of a galaxy. Contrarily, \citet{D'Onghia:2013} find that spiral structure can survive long after the original perturbing influence has vanished. However, \citet{Sellwood:Carlberg:2014} argue that if spirals can develop as self-excited instabilities, then the role of heavy clumps in the disc is probably not fundamental to the origin of spiral patterns. Furthermore, they claim that long-lasting spiral structure results from the superposition of several transient spiral modes.\footnote{\citet{Morozov:1991} \& \citet{Morozov:1992} also describe complicated patterns in spiral galaxies as superpositions of unstable hydrodynamic modes.}

\citet{Grand:2013} find from $N$-body simulations that absolute pitch angles tend to decrease with time through a winding-up effect. Contrastingly, \citet{Shields:2014} find from analysing a sample of more than 100 galaxies spanning redshifts up to $z=1.2$, that pitch angle appears to have statistically loosened since $z=0.5$. Although, they admit the possibility of selection effects and biases. One possibility is that the later type spiral galaxies with higher pitch angles might not be observed at great distance due to their lower intrinsic luminosity and surface brightness.

Continuing the discussion of simulations, we draw attention to predictions of pitch angles. Our measured pitch angles (which contain only relatively earlier type spiral galaxies, predominantly Hubble types Sa \& Sb) do not exceed $\approx25\degr$. The pitch angles presented in this work are consistent with the predictions of \citet{Perez-Villegas:2013} that large-scale, long-lasting spiral structure in galaxies should restrict pitch angle values to a maximum of $\approx15\degr$, $\approx18\degr$ and $\approx20\degr$ for Sa, Sb and Sc galaxies, respectively. Furthermore, \citet{Perez-Villegas:2013} show that chaotic behaviour leads to more transient spiral structure for pitch angle values larger than $\approx30\degr$, $\approx40\degr$ and $\approx50\degr$ for Sa, Sb and Sc galaxies, respectively. If these predictions are applicable, this implies that our measured pitch angles should be stable for the vast majority of our sample. Additionally, future studies of black holes in galaxies with later morphological types than exist in our sample (i.e. Sd) should be considered relatively stable for even large pitch angles. Since our sample does not include Sd galaxies, and consists of SMBHs with $M_{\rm BH} \goa 10^6$~${\rm M_{\sun}}$, this implies that future work to identify IMBHs via the $M_{\rm BH}$--$\phi$ relation should be targeting galaxies that have pitch angles $\goa30\degr$.

\subsection{External influences}

Another consideration that could influence pitch angle is from external agents such as tidal interaction, accretion, harassment, cluster environments, etc. Through the process of investigating the pitch angles of 125 galaxies in the Great Observatories Origins Deep Survey South Field \citep{Dickinson:2003}, \citet{Davis:2010} found little to no difference, on average, between the pitch angles of galaxies in and out of overdense regions, nor between the pitch angles of red or blue spiral galaxies. More recently, \citet{Semczuk:2017} studied $N$-body simulations of a Milky Way like galaxy orbiting a Virgo-like cluster. These simulations produce tidally induced logarithmic spirals upon pericentre passage around the cluster. Their findings indicate that, similarly to \citet{Grand:2013}, spiral arms wind up with time (decreasing the absolute value of pitch angle). However, upon successive pericentre passages, the spirals are again tidally stretched out and the pitch angle loosens with this cycle repeating, indefinitely. Concerning our sample, our galaxies are generally local field galaxies and should have little instance of tidal interaction events.

\subsection{Intrinsic scatter}

Our intrinsic scatter in equation~(\ref{M-phi}) is $\approx 77$~per~cent of the total rms scatter, implying that $\approx 77$~per~cent of the scatter comes from intrinsic scatter about the $M_{\rm BH}$--$\phi$ relation and the other $\approx 23$~per~cent arises from measurement error. The median black hole mass measurement error is 19~per~cent or 0.08~dex and the median pitch angle measurement error is 14~per~cent or $1\fdg9$ across our sample of 44 galaxies. Given this much smaller measurement error on $\log{M_{\rm BH}}$ than on $|\phi|$, i.e. 0.08 versus 1.9 in Figs.~\ref{plot}~\&~\ref{plot2}, we note again (see Section \ref{AR}) that the \textsc{bces} (X|Y) regression, and thus the \textsc{bces} Bisector symmetric regression, is almost identical (to two significant figures) to the \textsc{bces} (Y|X) regression.

One major contributing factor to the observed intrinsic scatter in the $M_{\rm BH}$--$\phi$ relation could be attributed to not accounting for the gas fraction in galaxies. Another factor that may be affecting the intrinsic scatter is Toomre's Stability Criterion \citep{Safronov:1960,Toomre:1964}, whose parameter $Q$ is related to gas fraction (since it depends on the gas surface density). Observationally, \citet{Seigar:2005,Seigar:2006,Seigar:2014} find that pitch angle is well-correlated with the galactic shear rate, which depends on the mass contained within a specified galactocentric radius. \citet{Grand:2013} corroborate these results in their $N$-body simulations. Recently, \citet{Kim:2017} have also shown that the pitch angle of nuclear spirals is similarly correlated with the background shear. Therefore, when shear is low, spirals are loosely wound (and vice versa) both on the scale of nuclear spirals in a galactic centre and for spiral arms residing in a galactic disc.

One would expect that late-type galaxies, which tend to have lower shear rates and therefore larger pitch angles, would accordingly have a higher gas fraction. This seems to explain the observational results of \citet{Davis:2015}, who indicate the existence of a Fundamental Plane relationship between bulge mass, disc density and pitch angle. Concerning the $M_{\rm BH}$--$\phi$ relation, if gas fractions were accurately known, it could act as a correcting factor to reduce the intrinsic scatter in the $M_{\rm BH}$--$\phi$ relation by the addition of a third parameter.

\subsection{Comparison to previous studies}

\begin{figure}
\includegraphics[clip=true,trim= 9mm 0mm 20mm 15mm,width=\columnwidth]{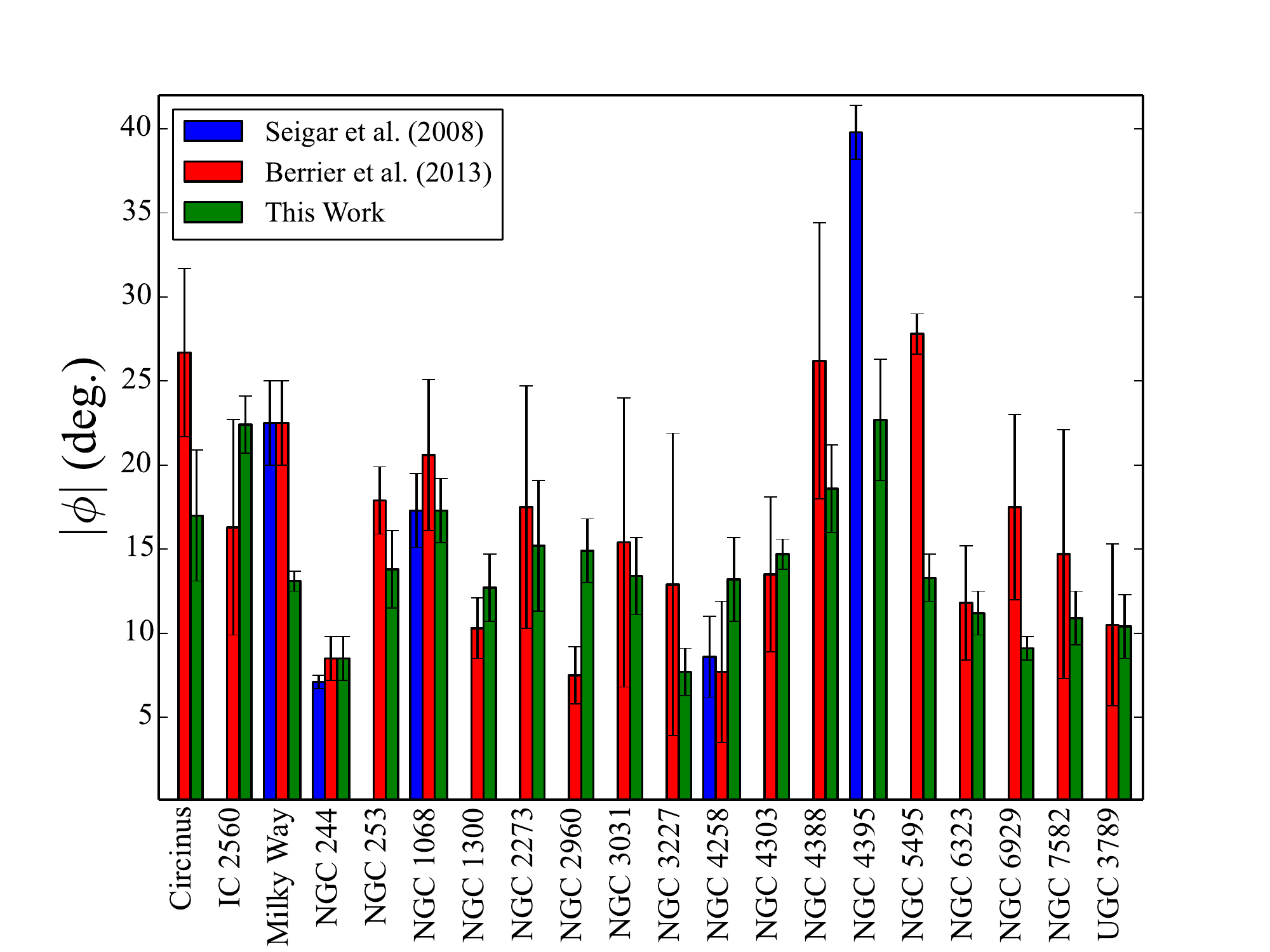}
\caption{Comparison of our pitch angle measurements, in green, for the 20 galaxies that are common in the directly measured SMBH mass samples of \citet{Seigar:2008}, in blue, and/or \citet{Berrier:2013}, in red. Black error bars are provided for individual measurements and reflect that our measurements are in agreement with these previous studies 70~per~cent of the time.}
\label{compare}
\end{figure}

This work updates the previous work of \citet{Seigar:2008} and \citet{Berrier:2013}. In \citet{Seigar:2008}, 5\footnote{In table~2 of \citet{Seigar:2008}, 12 galaxies are listed under the category of `BH Estimates from Direct Measurements'. However, two of those are merely upper limits and five are reverberation mapping estimates.} spiral galaxies with directly measured black hole mass estimates were studied. Five years later, \citet{Berrier:2013} increased this number to 22.\footnote{Of the 22 galaxies in the direct measurement sample of \citet{Berrier:2013}, three measurements were adopted and left unchanged for use in this paper (NGC 224, NGC 3368 and NGC 3393) and not plotted in Fig.~\ref{compare}.} Our current work has doubled the sample size. The median pitch angle of these respective samples has gradually decreased over the years from $17\fdg3$ \citep{Seigar:2008} to $14\fdg4$ \citet{Berrier:2013}, and finally $13\fdg3$ (this work).

As seen in Fig.~\ref{plot}, the slope of our linear fit is noticeably steeper than that described by the ordinary least-squares (Y|X) linear regression of \citet{Berrier:2013}; slope = $-0.062\pm0.009$~dex~deg$^{-1}$ (or compared to $-0.076\pm0.005$~dex~deg$^{-1}$ found by \citealt{Seigar:2008}). This can be attributed to many possible reasons. During the intervening 4 yr, many of the distances to the SMBH host galaxies have been revised. All redshift-dependent distances have been revised with newer cosmographical parameters. Any change in distance will have a proportional change on the estimated black hole mass. Additionally, many of these mass measurements themselves have been updated.

Pitch angle estimates have also evolved since these previous studies (see Fig.~\ref{compare}). Pitch angle measurements conducted by \citet{Seigar:2008} and \citet{Berrier:2013} were exclusively measured via \textsc{2dfft} algorithms. Our work has continued to use this trusted method of pitch angle measurement, but we have also incorporated newer template fitting and computer vision methods. Access to better resolution imaging has also affected our pitch angle measurements even for identical galaxies from previous studies, which employed primarily ground-based imaging. Our work has benefited from an availability of space-based imaging for the majority of our sample. These combined effects enable us to improve the accuracy and precision of pitch angle measurements for these well-studied galaxies.

Access to better resolution imaging also has likely statistically tightened (decreased $|\phi|$) our measured spirals. As previously mentioned, poor-resolution has a tendency to bias pitch angle measurements to looser (increased $|\phi|$) values. Therefore, better spatial resolution should reduce high-pitch angle noise and increase the chance of measuring the fundamental pitch angle. This effect could potentially explain why our linear regression has yielded a steeper slope than previous studies, by preferentially `tightening' the low-surface brightness, low-mass SMBH host galaxies that would benefit the most from better imaging (galaxies towards the bottom, righthand quadrant of Fig.~\ref{plot}). Additionally, the samples of \citet{Seigar:2008} and \citet{Berrier:2013} did not include the four lowest pitch angle galaxies (seen in the extreme upper-left corners of Figs.~\ref{plot}~\&~\ref{plot2}). These four extreme points will also contribute to a comparative steepening of the slope of the $M_{\rm BH}$--$\phi$ relation presented in this work.

Six of our galaxies have $>1\sigma$ discrepancies between our pitch angle measurement and our previously published values. Concerning Fourier methods of pitch angle measurement, it sometimes occurs that signals will be present at the true pitch angle and at multiples of 2 of that value. That is, the fundamental pitch angle can be overlooked by such codes, which report a pitch angle different by a factor of 2, especially exacerbated in noisy and/or flocculent images. As can be seen in Fig.~\ref{compare}, in all cases where the pitch angle measurements disagree beyond one standard deviation, the discrepancy is approximately a factor of 2. We feel confident over-ruling previous measurements with values differing by a factor of two because we have been able to use multiple software methods and analyse different imaging, such as \textit{GALEX} images, that can better bring out the spiral structure in many cases.

\subsection{Utility of the $M_{\rm BH}$--$\phi$ relation}

The potential to identify galaxies that may host IMBHs will be greatly enhanced via application of the $M_{\rm BH}$--$\phi$ relation (equation~\ref{M-phi}). A sample of candidate late-type galaxies hosting IMBHs could be initially identified from a catalogue of images \citep[e.g.][]{Baillard:2011}. Then, quantitative pitch angle measurements could provide IMBH mass estimates. Finally, a follow-up cross-check with the \textit{Chandra} X-ray archive could validate the existence of these IMBHs by looking for active galactic nuclear emission.

The $M_{\rm BH}$--$\phi$ relation, as defined in equation~(\ref{M-phi}), is capable of interpolating black hole masses in the range $5.42 \leq \log({M_{\rm BH}/{\rm M_{\sun}}}) \leq 9.11$ from their associated pitch angles $24\fdg3 \geq |\phi| \geq 2\fdg7$, respectively. Beyond that, it can be extrapolated for black hole masses down to zero ($|\phi| = 56\fdg0$) or as high as $\log({M_{\rm BH}/{\rm M_{\sun}}}) = 9.57$ ($\phi = 0\degr$). In terms of predicting IMBH masses in the range $10^2 \leq M_{\rm BH}/{\rm M_{\sun}} \leq 10^5$, this would dictate that their host galaxies would be late-type spirals with pitch angles of $44\fdg3 \geq |\phi| \geq 26\fdg7$, respectively. Pitch angles $\goa50\degr$ are very rarely measured in the literature.

Gravitational wave detections can aid future extensions of the $M_{\rm BH}$--$\phi$ relation by providing direct estimates of black hole masses, with better accuracy than current astronomical techniques based on electromagnetic radiation, for which the black hole's sphere of gravitational influence needs to be spatially resolved. The Advanced Laser Interferometer Gravitational-Wave Observatory \citep{aLIGO} is capable of detecting gravitational waves generated by black hole merger events with total masses up to 100 ${\rm M_{\sun}}$ \citep{LIGO}. This is right at the transition between stellar mass black holes ($M_{\rm BH} < 10^2$~${\rm M_{\sun}}$) and IMBHs ($10^2 < M_{\rm BH}/{\rm M_{\sun}} < 10^5$).

As the sensitivity and the localization abilities of gravitational radiation detectors increases (i.e. the proposed Evolved Laser Interferometer Space Antenna; \citealt{eLISA} \& \citealt{Danzmann:2015}), we should be able to conduct follow-up electromagnetic radiation observations and potentially glimpse (galaxy) evolution in action. Upcoming ground-based detectors will be able to probe longer wavelength gravitational radiation than is currently possible. In particular, the Kamioka Gravitational Wave Detector (KAGRA) will be sensitive to IMBH mergers up to 2000~${\rm M_{\sun}}$ \citep{Shinkai:2016}, implying that (late-type)--(late-type) galaxy mergers with IMBHs could potentially generate a detectable signal.

It would be easy to assume that the existence of an $M_{\rm BH}$--$\phi$ relation might simply be a consequence of the well-known (black hole mass)--(host spheroid mass) relation. However, the efficacy of the $M_{\rm BH}$--$\phi$ relation in predicting black hole masses in bulge-less galaxies may imply otherwise. Moreover, the $M_{\rm BH}$--$\phi$ relation is significantly tighter than expectations arising from the consequence of other scaling relations (see equation~\ref{indirect}). This will surely become more evident as the population of bulge-less galaxies with directly measured black holes masses grows to numbers greater than the current tally of one (NGC 4395). Furthermore, the low scatter in the $M_{\rm BH}$--$\phi$ relation makes it at least as accurate at predicting black hole masses in spiral galaxies as the other known mass scaling relations.

\subsection{Galaxies with (possible) low-mass black holes}

\subsubsection{NGC 4395}

NGC 4395 is the only galaxy in our sample that has been classified as a bulge-less galaxy \citep{Sandage:1981}. It additionally stands out as having the lowest mass black hole in our sample, at just under a half-million solar masses. Furthermore, it is the only Magellanic type galaxy in our sample. With the classification, SBm, it also exhibits a noticeable barred structure despite any indications of a central bulge.

\subsubsection{M33}

To test the validity of our relation at the low-mass end, we now analyse M33 (NGC 598); which is classified as a bulge-less galaxy \citep{M33,Minniti:1993} and thought to have one of the smallest, or perhaps no black hole residing at its centre. Since studies \citep{Gebhardt:2001,Merritt:2001} only provide upper limits to its black hole mass, we do not use it to determine our relation, but rather test the $M_{\rm BH}$--$\phi$ relation's extrapolation to the low-mass end. We adopt a luminosity distance of 839~kpc \citep{Gieren:2013} and a distance-adjusted  black hole mass of $\log({M_{\rm BH}/{\rm M_{\sun}}})\leq3.20$ \citep{Gebhardt:2001} and $\log({M_{\rm BH}/{\rm M_{\sun}}})\leq3.47$ \citep{Merritt:2001}. Using a \textit{GALEX} \textit{FUV} image, we measure a pitch angle of $|\phi| = 40\fdg0\pm3\fdg0$. This is in agreement with the measurement of \citet{Seigar:2011}, who reports $|\phi| = 42\fdg2\pm3\fdg0$ from a \textit{Spitzer}/\textit{IRAC1}, 3.6~$\micron$ image. Applying equation~(\ref{M-phi}), we obtain $\log({M_{\rm BH}/{\rm M_{\sun}}})=2.73\pm0.70$. Thus, our mass estimate of the potential black hole in M33 is consistent at the $-0.67\sigma$ and $-1.06\sigma$ level with the published upper limit black hole mass estimates of \citet{Gebhardt:2001} and \citet{Merritt:2001}, respectively.

\subsubsection{Circinus galaxy}

Lastly, we investigated the recent three-dimensional radiation-hydrodynamic simulation of gas around the low-luminosity AGN of the Circinus Galaxy \citep{Wada:2016}. Their results produce prominent spiral arm structure in the geometrically thick disc surrounding the AGN, assuming a central SMBH mass of $2\times10^6$~${\rm M_{\sun}}$ (consistent with our adopted directly measured mass in Table~\ref{Sample}). We analyse the snapshot of their number density distribution of $\rm{H_2O}$ (see their fig.~2a, left-hand panel). We find an interesting eight-arm structure with a pitch angle of $27\fdg7\pm1\fdg7$ and a radius of $\approx9$ pc. This result differs significantly from our measurement of the pitch angle in the galactic disc ($17\fdg0\pm3\fdg9$). Possible explanations could simply be different physical mechanisms in the vicinity of a black hole's sphere of gravitational influence \citep[$r_h\approx1.2$~pc for Circinus's SMBH, for definition of $r_h$, see][]{Peebles:1972} or the different relative densities of this inner $\approx 9$~pc disc radius versus the much larger $\approx48$~kpc \citep{Jones:1999} disc radius of the entire galaxy. According to the predictions of spiral density wave theory \citep[see equation~2 from][]{Davis:2015}, the higher relative local densities in the inner disc should produce a higher pitch angle spiral density wave than in the larger, more tenuous galactic disc.

\subsection{Vortex nature of spiral galaxy structure}

There exists a sizeable amount of material in the literature concerning the study of vortices, cyclones and anticyclones in galaxies and their reproducibility in laboratory fluid dynamic experiments \citep[for a 20 yr review, see][]{Fridman:2007}. Fridman has done a lot of research concerning the origin of spiral structure since \citet{Fridman:1978}. Much of his work investigating the motion of gas in discs of galaxies has revealed strong evidence for the existence and nature of vortices \citep[e.g.][]{Fridman:Khoruzhii:1999,Fridman:1999,Fridman:2000,Fridman:2001c,Fridman:2001b,Fridman:2001}. Furthermore, \citet{Chavanis:2002} present a thorough discussion on the statistical mechanics of vortices in galaxies, while \citet{Vatistas:2010} provide an account of the striking similarity between the rotation curves in galaxies and in terrestrial hurricanes and tornadoes. In addition, \citet{Vorobyov:2006} specifically describe their numerical simulations that indicate anticyclones in the gas flow around the location of galactic corotation increase in intensity with increasing galactic spiral arm pitch angle absolute value.

In meteorology, it is well-known that the source mechanism for the observed rotation in cyclones and anticyclones is the Coriolis effect caused by the rotation of the Earth. Some speculation has been made postulating that the observed rotation in galaxies is analogously derivative of a Coriolis effect originating from an alleged rotation of the Universe \citep{Li:1998,Chaliasos:2006,Chaliasos:2006b}. Other studies theorise that primeval turbulence in the cosmos (rotation of cells/voids/walls of the cosmic web) instilled angular momenta in forming galaxies \citep{Dallaporta:1972,Jones:1973,Casuso:2015}, rather than rotation of the entire Universe.

There also exists significant literature concerning the similarity of spiral galaxies to turbulent eddies. Early research indicates that large-scale irregularities that occur in spiral galaxies can be described as eddies and are in many ways similar to those that occur in our own atmosphere \citep{Dickinson:1963,Dickinson:1964,Dickinson:1964b}. Subsequent hydrodynamic studies of eddies revealed a possible scenario regarding the formation of rotating spiral galaxies based on the concept of the formation of tangential discontinuity and its decay into eddies of the galactic scale, spawned from protogalactic vorticity in the metagalactic medium \citep{Ushakov:1983,Ushakov:1984,Chernin:1993,Chernin:1996}. \citet{Silk:1972} speculate that galactic-scale turbulent eddies originating prior to the era of recombination were frozen out of the turbulent flow at epochs following recombination.

Direct simulations with rotating shallow water experiments have been used to adequately model vortical structures in planetary atmospheres and oceans \citep{Nezlin:1990b,Nezlin:1990,Nezlin:1990c}. The application of rotating shallow water experiments is also applicable for the adequate modelling of spiral structures of gaseous galactic discs. \citet{Nezlin:1991} hypothesize that experimental Rayleigh friction between shallow water and the bottom of a vessel is physically analogous to friction between structure and stars in the disc of a galaxy. Favourable comparisons exist between the rotation of a compressible inviscid fluid disc and the dynamics observed in hurricanes and spiral galaxies and yield vortex wave streamlines that are logarithmic spirals \citep{White:1972}.

Curiously, further evidence in nature for a connection between spiral arms and the central mass concentration manifests itself in tropical cyclones \citep{Dvorak:1975}. Indeed, the Hubble-Jeans Sequence lends itself quite well to tropical cyclones, such that those with high wind speeds have large central dense `overcasts' (CDOs) and tightly wound spiral arms while those with low wind speeds have small CDOs and loosely wound spiral arms. One notable difference between galaxies and tropical cyclones is that larger galactic bulges garner higher (S\'ersic indices and) central mass concentrations while tropical cyclones with larger CDOs possess lower central atmospheric pressures. Both mechanisms, whether gravity or pressure, can be described by the nature and effect of their central potential wells on the surrounding spiral structure.

\section*{Acknowledgements}

AWG was supported under the Australian Research Council's funding scheme DP17012923. This research has made use of NASA's Astrophysics Data System. We acknowledge the usage of the HyperLeda data base \citep{HyperLeda}, \url{http://leda.univ-lyon1.fr}. This research has made use of the NASA/IPAC Extragalactic Database (NED), \url{https://ned.ipac.caltech.edu/}. Some of these data presented in this paper were obtained from the Mikulski Archive for Space Telescopes (MAST), \url{http://archive.stsci.edu/}. We used the \textsc{red idl} cosmology routines written by Leonidas and John Moustakas. Finally, we used the statistical and plotting routines written by R. S. Nemmen that accompany his \textsc{python} translation of the \textsc{bces} software.




\bibliographystyle{mnras}
\bibliography{bibliography}

\begin{thebibliography}{}
\makeatletter
\relax
\def\mn@urlcharsother{\let\do\@makeother \do\$\do\&\do\#\do\^\do\_\do\%\do\~}
\def\mn@doi{\begingroup\mn@urlcharsother \@ifnextchar [ {\mn@doi@}
  {\mn@doi@[]}}
\def\mn@doi@[#1]#2{\def\@tempa{#1}\ifx\@tempa\@empty \href
  {http://dx.doi.org/#2} {doi:#2}\else \href {http://dx.doi.org/#2} {#1}\fi
  \endgroup}
\def\mn@eprint#1#2{\mn@eprint@#1:#2::\@nil}
\def\mn@eprint@arXiv#1{\href {http://arxiv.org/abs/#1} {{\tt arXiv:#1}}}
\def\mn@eprint@dblp#1{\href {http://dblp.uni-trier.de/rec/bibtex/#1.xml}
  {dblp:#1}}
\def\mn@eprint@#1:#2:#3:#4\@nil{\def\@tempa {#1}\def\@tempb {#2}\def\@tempc
  {#3}\ifx \@tempc \@empty \let \@tempc \@tempb \let \@tempb \@tempa \fi \ifx
  \@tempb \@empty \def\@tempb {arXiv}\fi \@ifundefined
  {mn@eprint@\@tempb}{\@tempb:\@tempc}{\expandafter \expandafter \csname
  mn@eprint@\@tempb\endcsname \expandafter{\@tempc}}}

\bibitem[\protect\citeauthoryear{Abbott et~al.,}{Abbott et~al.}{2016}]{LIGO}
Abbott B.~P.,  et~al., 2016, \mn@doi [Phys. Rev. Lett.]
  {10.1103/PhysRevLett.116.061102}, 116, 061102

\bibitem[\protect\citeauthoryear{{Akritas} \& {Bershady}}{{Akritas} \&
  {Bershady}}{1996}]{BCES}
{Akritas} M.~G.,  {Bershady} M.~A.,  1996, \mn@doi [\apj] {10.1086/177901},
  \href {http://adsabs.harvard.edu/abs/1996ApJ...470..706A} {470, 706}

\bibitem[\protect\citeauthoryear{{Amaro-Seoane} et~al.,}{{Amaro-Seoane}
  et~al.}{2012}]{eLISA}
{Amaro-Seoane} P.,  et~al., 2012, \mn@doi [Classical and Quantum Gravity]
  {10.1088/0264-9381/29/12/124016}, \href
  {http://adsabs.harvard.edu/abs/2012CQGra..29l4016A} {29, 124016}

\bibitem[\protect\citeauthoryear{{Atkinson} et~al.,}{{Atkinson}
  et~al.}{2005}]{Atkinson:2005}
{Atkinson} J.~W.,  et~al., 2005, \mn@doi [\mnras]
  {10.1111/j.1365-2966.2005.08904.x}, \href
  {http://adsabs.harvard.edu/abs/2005MNRAS.359..504A} {359, 504}

\bibitem[\protect\citeauthoryear{{Baillard} et~al.,}{{Baillard}
  et~al.}{2011}]{Baillard:2011}
{Baillard} A.,  et~al., 2011, \mn@doi [\aap] {10.1051/0004-6361/201016423},
  \href {http://adsabs.harvard.edu/abs/2011A%26A...532A..74B} {532, A74}

\bibitem[\protect\citeauthoryear{{Bender} et~al.,}{{Bender}
  et~al.}{2005}]{Bender:2005}
{Bender} R.,  et~al., 2005, \mn@doi [\apj] {10.1086/432434}, \href
  {http://adsabs.harvard.edu/abs/2005ApJ...631..280B} {631, 280}

\bibitem[\protect\citeauthoryear{{Berrier} et~al.,}{{Berrier}
  et~al.}{2013}]{Berrier:2013}
{Berrier} J.~C.,  et~al., 2013, \mn@doi [\apj] {10.1088/0004-637X/769/2/132},
  \href {http://adsabs.harvard.edu/abs/2013ApJ...769..132B} {769, 132}

\bibitem[\protect\citeauthoryear{{Binney} \& {Tremaine}}{{Binney} \&
  {Tremaine}}{1987}]{Binney:Tremaine:1987}
{Binney} J.,  {Tremaine} S.,  1987, {Galactic dynamics}

\bibitem[\protect\citeauthoryear{{Blais-Ouellette}, {Amram}, {Carignan}  \&
  {Swaters}}{{Blais-Ouellette} et~al.}{2004}]{Blais-Ouellette:2004}
{Blais-Ouellette} S.,  {Amram} P.,  {Carignan} C.,   {Swaters} R.,  2004,
  \mn@doi [\aap] {10.1051/0004-6361:20034263}, \href
  {http://adsabs.harvard.edu/abs/2004A%26A...420..147B} {420, 147}

\bibitem[\protect\citeauthoryear{{Boehle} et~al.,}{{Boehle}
  et~al.}{2016}]{Boehle:2016}
{Boehle} A.,  et~al., 2016, \mn@doi [\apj] {10.3847/0004-637X/830/1/17}, \href
  {http://adsabs.harvard.edu/abs/2016ApJ...830...17B} {830, 17}

\bibitem[\protect\citeauthoryear{{Bose} \& {Kumar}}{{Bose} \&
  {Kumar}}{2014}]{Bose:2014}
{Bose} S.,  {Kumar} B.,  2014, \mn@doi [\apj] {10.1088/0004-637X/782/2/98},
  \href {http://cdsads.u-strasbg.fr/abs/2014ApJ...782...98B} {782, 98}

\bibitem[\protect\citeauthoryear{Cappellari et~al.,}{Cappellari
  et~al.}{2007}]{Cappellari:2008}
Cappellari M.,  et~al., 2007, \mn@doi [Proceedings of the International
  Astronomical Union] {10.1017/S1743921308017687}, 3, 215

\bibitem[\protect\citeauthoryear{{Casuso} \& {Beckman}}{{Casuso} \&
  {Beckman}}{2015}]{Casuso:2015}
{Casuso} E.,  {Beckman} J.~E.,  2015, \mn@doi [\mnras] {10.1093/mnras/stv549},
  \href {http://adsabs.harvard.edu/abs/2015MNRAS.449.2910C} {449, 2910}

\bibitem[\protect\citeauthoryear{{Chaliasos}}{{Chaliasos}}{2006a}]{Chaliasos:2006}
{Chaliasos} E.,  2006a, preprint, \href
  {http://adsabs.harvard.edu/abs/2006astro.ph..1659C} {} (\mn@eprint {arXiv}
  {astro-ph/0601659})

\bibitem[\protect\citeauthoryear{{Chaliasos}}{{Chaliasos}}{2006b}]{Chaliasos:2006b}
{Chaliasos} E.,  2006b, preprint, \href
  {http://adsabs.harvard.edu/abs/2006physics...2035C} {} (\mn@eprint {arXiv}
  {physics/0602035})

\bibitem[\protect\citeauthoryear{{Chavanis}}{{Chavanis}}{2002}]{Chavanis:2002}
{Chavanis} P.-H.,  2002, in {Dauxois} T.,  {Ruffo} S.,  {Arimondo} E.,
  {Wilkens} M.,  eds,  Lecture Notes in Physics, Berlin Springer Verlag Vol.
  602, Dynamics and Thermodynamics of Systems with Long-Range Interactions. pp
  208--289 (\mn@eprint {} {cond-mat/0212223})

\bibitem[\protect\citeauthoryear{{Chernin}}{{Chernin}}{1993}]{Chernin:1993}
{Chernin} A.~D.,  1993, \aap, \href
  {http://adsabs.harvard.edu/abs/1993A%26A...267..315C} {267, 315}

\bibitem[\protect\citeauthoryear{{Chernin}}{{Chernin}}{1996}]{Chernin:1996}
{Chernin} A.~D.,  1996, \mn@doi [Nuovo Cimento Rivista Serie]
  {10.1007/BF02743057}, \href
  {http://adsabs.harvard.edu/abs/1996NCimR..19f...1C} {19, 1}

\bibitem[\protect\citeauthoryear{{D'Onghia}, {Vogelsberger}  \&
  {Hernquist}}{{D'Onghia} et~al.}{2013}]{D'Onghia:2013}
{D'Onghia} E.,  {Vogelsberger} M.,   {Hernquist} L.,  2013, \mn@doi [\apj]
  {10.1088/0004-637X/766/1/34}, \href
  {http://adsabs.harvard.edu/abs/2013ApJ...766...34D} {766, 34}

\bibitem[\protect\citeauthoryear{{Dallaporta} \& {Lucchin}}{{Dallaporta} \&
  {Lucchin}}{1972}]{Dallaporta:1972}
{Dallaporta} N.,  {Lucchin} F.,  1972, \aap, \href
  {http://adsabs.harvard.edu/abs/1972A%26A....19..123D} {19, 123}

\bibitem[\protect\citeauthoryear{{Danzmann}}{{Danzmann}}{2015}]{Danzmann:2015}
{Danzmann} K.,  2015, IAU General Assembly, \href
  {http://adsabs.harvard.edu/abs/2015IAUGA..2248153D} {22, 2248153}

\bibitem[\protect\citeauthoryear{{Davies} et~al.,}{{Davies}
  et~al.}{2006}]{Davies:2006}
{Davies} R.~I.,  et~al., 2006, \mn@doi [\apj] {10.1086/504963}, \href
  {http://adsabs.harvard.edu/abs/2006ApJ...646..754D} {646, 754}

\bibitem[\protect\citeauthoryear{{Davis} \& {Hayes}}{{Davis} \&
  {Hayes}}{2014}]{Davis:Hayes:2014}
{Davis} D.~R.,  {Hayes} W.~B.,  2014, \mn@doi [\apj]
  {10.1088/0004-637X/790/2/87}, \href
  {http://adsabs.harvard.edu/abs/2014ApJ...790...87D} {790, 87}

\bibitem[\protect\citeauthoryear{{Davis} et~al.,}{{Davis}
  et~al.}{2010}]{Davis:2010}
{Davis} B.~L.,  et~al., 2010, in American Astronomical Society Meeting
  Abstracts \#215. p.~382

\bibitem[\protect\citeauthoryear{{Davis}, {Berrier}, {Shields}, {Kennefick},
  {Kennefick}, {Seigar}, {Lacy}  \& {Puerari}}{{Davis}
  et~al.}{2012}]{Davis:2012}
{Davis} B.~L.,  {Berrier} J.~C.,  {Shields} D.~W.,  {Kennefick} J.,
  {Kennefick} D.,  {Seigar} M.~S.,  {Lacy} C.~H.~S.,   {Puerari} I.,  2012,
  \mn@doi [\apjs] {10.1088/0067-0049/199/2/33}, \href
  {http://adsabs.harvard.edu/abs/2012ApJS..199...33D} {199, 33}

\bibitem[\protect\citeauthoryear{{Davis} et~al.,}{{Davis}
  et~al.}{2014}]{Davis:2014}
{Davis} B.~L.,  et~al., 2014, \mn@doi [\apj] {10.1088/0004-637X/789/2/124},
  \href {http://adsabs.harvard.edu/abs/2014ApJ...789..124D} {789, 124}

\bibitem[\protect\citeauthoryear{{Davis} et~al.,}{{Davis}
  et~al.}{2015}]{Davis:2015}
{Davis} B.~L.,  et~al., 2015, \mn@doi [\apjl] {10.1088/2041-8205/802/1/L13},
  \href {http://adsabs.harvard.edu/abs/2015ApJ...802L..13D} {802, L13}

\bibitem[\protect\citeauthoryear{{Davis}, {Berrier}, {Shields}, {Kennefick},
  {Kennefick}, {Seigar}, {Lacy}  \& {Puerari}}{{Davis} et~al.}{2016}]{2DFFT}
{Davis} B.~L.,  {Berrier} J.~C.,  {Shields} D.~W.,  {Kennefick} J.,
  {Kennefick} D.,  {Seigar} M.~S.,  {Lacy} C.~H.~S.,   {Puerari} I.,  2016,
  {2DFFT: Measuring Galactic Spiral Arm Pitch Angle}, Astrophysics Source Code
  Library (\mn@eprint {ascl} {1608.015})

\bibitem[\protect\citeauthoryear{{Devereux}, {Ford}, {Tsvetanov}  \&
  {Jacoby}}{{Devereux} et~al.}{2003}]{Devereux:2003}
{Devereux} N.,  {Ford} H.,  {Tsvetanov} Z.,   {Jacoby} G.,  2003, \mn@doi [\aj]
  {10.1086/367595}, \href {http://adsabs.harvard.edu/abs/2003AJ....125.1226D}
  {125, 1226}

\bibitem[\protect\citeauthoryear{{Dickinson}}{{Dickinson}}{1963}]{Dickinson:1963}
{Dickinson} R.~E.,  1963, \mn@doi [Geofisica Pura e Applicata]
  {10.1007/BF01993342}, \href
  {http://adsabs.harvard.edu/abs/1963GeoPA..56..174D} {56, 174}

\bibitem[\protect\citeauthoryear{{Dickinson}}{{Dickinson}}{1964a}]{Dickinson:1964}
{Dickinson} R.~E.,  1964a, \mn@doi [Pure and Applied Geophysics]
  {10.1007/BF00880515}, \href
  {http://adsabs.harvard.edu/abs/1964PApGe..59..142D} {59, 142}

\bibitem[\protect\citeauthoryear{{Dickinson}}{{Dickinson}}{1964b}]{Dickinson:1964b}
{Dickinson} R.~E.,  1964b, \mn@doi [Pure and Applied Geophysics]
  {10.1007/BF00880516}, \href
  {http://adsabs.harvard.edu/abs/1964PApGe..59..155D} {59, 155}

\bibitem[\protect\citeauthoryear{{Dickinson}, {Giavalisco}  \& {GOODS
  Team}}{{Dickinson} et~al.}{2003}]{Dickinson:2003}
{Dickinson} M.,  {Giavalisco} M.,   {GOODS Team} 2003, in {Bender} R.,
  {Renzini} A.,  eds, The Mass of Galaxies at Low and High Redshift. p.~324
  (\mn@eprint {} {astro-ph/0204213}), \mn@doi{10.1007/10899892_78}

\bibitem[\protect\citeauthoryear{{Dressler}}{{Dressler}}{1989}]{Dressler:1989}
{Dressler} A.,  1989, in {Osterbrock} D.~E.,  {Miller} J.~S.,  eds,  IAU
  Symposium Vol. 134, Active Galactic Nuclei. p.~217

\bibitem[\protect\citeauthoryear{Dvorak}{Dvorak}{1975}]{Dvorak:1975}
Dvorak V.~F.,  1975, \mn@doi [Monthly Weather Review]
  {10.1175/1520-0493(1975)103<0420:TCIAAF>2.0.CO;2}, 103, 420

\bibitem[\protect\citeauthoryear{{Elmegreen} \& {Elmegreen}}{{Elmegreen} \&
  {Elmegreen}}{1987}]{Elmegreen:1987}
{Elmegreen} D.~M.,  {Elmegreen} B.~G.,  1987, \mn@doi [\apj] {10.1086/165034},
  \href {http://adsabs.harvard.edu/abs/1987ApJ...314....3E} {314, 3}

\bibitem[\protect\citeauthoryear{{Fisher} \& {Drory}}{{Fisher} \&
  {Drory}}{2010}]{Fisher:Drory:2010}
{Fisher} D.~B.,  {Drory} N.,  2010, \mn@doi [\apj]
  {10.1088/0004-637X/716/2/942}, \href
  {http://adsabs.harvard.edu/abs/2010ApJ...716..942F} {716, 942}

\bibitem[\protect\citeauthoryear{Fisher \& Drory}{Fisher \&
  Drory}{2016}]{Fisher:Drory:2016}
Fisher D.~B.,  Drory N.,  2016, An Observational Guide to Identifying
  Pseudobulges and Classical Bulges in Disc Galaxies.
Springer International Publishing, Cham, pp 41--75,
  \mn@doi{10.1007/978-3-319-19378-6_3}, \url
  {http://dx.doi.org/10.1007/978-3-319-19378-6_3}

\bibitem[\protect\citeauthoryear{{Fridman}}{{Fridman}}{1978}]{Fridman:1978}
{Fridman} A.~M.,  1978, \mn@doi [Soviet Physics Uspekhi]
  {10.1070/PU1978v021n06ABEH005562}, \href
  {http://adsabs.harvard.edu/abs/1978SvPhU..21..536F} {21, 536}

\bibitem[\protect\citeauthoryear{{Fridman}}{{Fridman}}{2007}]{Fridman:2007}
{Fridman} A.~M.,  2007, \mn@doi [Physics Uspekhi]
  {10.1070/PU2007v050n02ABEH006210}, \href
  {http://adsabs.harvard.edu/abs/2007PhyU...50..115F} {50, 115}

\bibitem[\protect\citeauthoryear{{Fridman} \& {Khoruzhii}}{{Fridman} \&
  {Khoruzhii}}{1999}]{Fridman:Khoruzhii:1999}
{Fridman} A.~M.,  {Khoruzhii} O.~V.,  1999, in {Sellwood} J.~A.,  {Goodman} J.,
   eds,  Astronomical Society of the Pacific Conference Series Vol. 160,
  Astrophysical Discs - an EC Summer School. p.~341

\bibitem[\protect\citeauthoryear{{Fridman} \& {Khoruzhii}}{{Fridman} \&
  {Khoruzhii}}{2000}]{Fridman:2000}
{Fridman} A.~M.,  {Khoruzhii} O.~V.,  2000, \mn@doi [Physics Letters A]
  {10.1016/S0375-9601(00)00680-0}, \href
  {http://adsabs.harvard.edu/abs/2000PhLA..276..199F} {276, 199}

\bibitem[\protect\citeauthoryear{{Fridman}, {Khoruzhii}, {Polyachenko},
  {Zasov}, {Sil'chenko}, {Afanas'ev}, {Dodonov}  \& {Moiseev}}{{Fridman}
  et~al.}{1999}]{Fridman:1999}
{Fridman} A.~M.,  {Khoruzhii} O.~V.,  {Polyachenko} E.,  {Zasov} A.~V.,
  {Sil'chenko} O.~K.,  {Afanas'ev} V.~L.,  {Dodonov} S.~N.,   {Moiseev} A.~V.,
  1999, \mn@doi [Physics Letters A] {10.1016/S0375-9601(99)00786-0}, \href
  {http://adsabs.harvard.edu/abs/1999PhLA..264...85F} {264, 85}

\bibitem[\protect\citeauthoryear{{Fridman} et~al.,}{{Fridman}
  et~al.}{2001a}]{Fridman:2001c}
{Fridman} A.~M.,  et~al., 2001a, in {Funes} J.~G.,  {Corsini} E.~M.,  eds,
  Astronomical Society of the Pacific Conference Series Vol. 230, Galaxy Disks
  and Disk Galaxies. pp 187--198

\bibitem[\protect\citeauthoryear{{Fridman} et~al.,}{{Fridman}
  et~al.}{2001b}]{Fridman:2001b}
{Fridman} A.~M.,  et~al., 2001b, \mn@doi [\mnras]
  {10.1046/j.1365-8711.2001.04218.x}, \href
  {http://adsabs.harvard.edu/abs/2001MNRAS.323..651F} {323, 651}

\bibitem[\protect\citeauthoryear{{Fridman}, {Khoruzhii, O. V.}, {Lyakhovich, V.
  V.}, {Sil'chenko, O. K.}, {ov, A. V. Za}, {iev, V. L. Afana}, {Dodonov, S.
  N.}  \& {teix, J. Boule}}{{Fridman} et~al.}{2001c}]{Fridman:2001}
{Fridman} A.~M.,  {Khoruzhii, O. V.} {Lyakhovich, V. V.} {Sil'chenko, O. K.}
  {ov, A. V. Za} {iev, V. L. Afana} {Dodonov, S. N.}  {teix, J. Boule} 2001c,
  \mn@doi [A&A] {10.1051/0004-6361:20010392}, 371, 538

\bibitem[\protect\citeauthoryear{{Gadotti} \& {S{\'a}nchez-Janssen}}{{Gadotti}
  \& {S{\'a}nchez-Janssen}}{2012}]{Gadotti:2012}
{Gadotti} D.~A.,  {S{\'a}nchez-Janssen} R.,  2012, \mn@doi [\mnras]
  {10.1111/j.1365-2966.2012.20925.x}, \href
  {http://adsabs.harvard.edu/abs/2012MNRAS.423..877G} {423, 877}

\bibitem[\protect\citeauthoryear{{Gao} et~al.,}{{Gao} et~al.}{2016}]{Gao:2016}
{Gao} F.,  et~al., 2016, \mn@doi [\apj] {10.3847/0004-637X/817/2/128}, \href
  {http://adsabs.harvard.edu/abs/2016ApJ...817..128G} {817, 128}

\bibitem[\protect\citeauthoryear{{Gao} et~al.,}{{Gao} et~al.}{2017}]{Gao:2017}
{Gao} F.,  et~al., 2017, \mn@doi [\apj] {10.3847/1538-4357/834/1/52}, \href
  {http://adsabs.harvard.edu/abs/2017ApJ...834...52G} {834, 52}

\bibitem[\protect\citeauthoryear{{Gebhardt} et~al.,}{{Gebhardt}
  et~al.}{2001}]{Gebhardt:2001}
{Gebhardt} K.,  et~al., 2001, \mn@doi [\aj] {10.1086/323481}, \href
  {http://adsabs.harvard.edu/abs/2001AJ....122.2469G} {122, 2469}

\bibitem[\protect\citeauthoryear{{Giacomazzo}, {Baker}, {Miller}, {Reynolds}
  \& {van Meter}}{{Giacomazzo} et~al.}{2012}]{Giacomazzo:2012}
{Giacomazzo} B.,  {Baker} J.~G.,  {Miller} M.~C.,  {Reynolds} C.~S.,   {van
  Meter} J.~R.,  2012, \mn@doi [\apjl] {10.1088/2041-8205/752/1/L15}, \href
  {http://adsabs.harvard.edu/abs/2012ApJ...752L..15G} {752, L15}

\bibitem[\protect\citeauthoryear{{Gieren} et~al.,}{{Gieren}
  et~al.}{2013}]{Gieren:2013}
{Gieren} W.,  et~al., 2013, \mn@doi [\apj] {10.1088/0004-637X/773/1/69}, \href
  {http://adsabs.harvard.edu/abs/2013ApJ...773...69G} {773, 69}

\bibitem[\protect\citeauthoryear{{Graham}}{{Graham}}{2008}]{Graham:2008}
{Graham} A.~W.,  2008, \mn@doi [\pasa] {10.1071/AS08013}, \href
  {http://adsabs.harvard.edu/abs/2008PASA...25..167G} {25, 167}

\bibitem[\protect\citeauthoryear{Graham}{Graham}{2016}]{Graham:2016b}
Graham A.~W.,  2016, Galaxy Bulges and Their Massive Black Holes: A Review.
Springer International Publishing, Cham, pp 263--313,
  \mn@doi{10.1007/978-3-319-19378-6_11}, \url
  {http://dx.doi.org/10.1007/978-3-319-19378-6_11}

\bibitem[\protect\citeauthoryear{{Graham} \& {Scott}}{{Graham} \&
  {Scott}}{2013}]{Graham:Scott:2013}
{Graham} A.~W.,  {Scott} N.,  2013, \mn@doi [\apj]
  {10.1088/0004-637X/764/2/151}, \href
  {http://adsabs.harvard.edu/abs/2013ApJ...764..151G} {764, 151}

\bibitem[\protect\citeauthoryear{{Graham} \& {Worley}}{{Graham} \&
  {Worley}}{2008}]{Graham:Worley:2008}
{Graham} A.~W.,  {Worley} C.~C.,  2008, \mn@doi [\mnras]
  {10.1111/j.1365-2966.2008.13506.x}, \href
  {http://adsabs.harvard.edu/abs/2008MNRAS.388.1708G} {388, 1708}

\bibitem[\protect\citeauthoryear{{Graham}, {Janz}, {Penny}, {Chilingarian},
  {Ciambur}, {Forbes}  \& {Davies}}{{Graham} et~al.}{2017}]{Graham:2017}
{Graham} A.~W.,  {Janz} J.,  {Penny} S.~J.,  {Chilingarian} I.~V.,  {Ciambur}
  B.~C.,  {Forbes} D.~A.,   {Davies} R.~L.,  2017, \mn@doi [\apj]
  {10.3847/1538-4357/aa6e56}, \href
  {http://adsabs.harvard.edu/abs/2017ApJ...840...68G} {840, 68}

\bibitem[\protect\citeauthoryear{{Grand}, {Kawata}  \& {Cropper}}{{Grand}
  et~al.}{2013}]{Grand:2013}
{Grand} R.~J.~J.,  {Kawata} D.,   {Cropper} M.,  2013, \mn@doi [\aap]
  {10.1051/0004-6361/201321308}, \href
  {http://adsabs.harvard.edu/abs/2013A%26A...553A..77G} {553, A77}

\bibitem[\protect\citeauthoryear{{Greene} et~al.,}{{Greene}
  et~al.}{2016}]{Greene:2016}
{Greene} J.~E.,  et~al., 2016, \mn@doi [\apjl] {10.3847/2041-8205/826/2/L32},
  \href {http://adsabs.harvard.edu/abs/2016ApJ...826L..32G} {826, L32}

\bibitem[\protect\citeauthoryear{{Greenhill}, {Moran}  \&
  {Herrnstein}}{{Greenhill} et~al.}{1997}]{Greenhill:1997}
{Greenhill} L.~J.,  {Moran} J.~M.,   {Herrnstein} J.~R.,  1997, \mn@doi [\apjl]
  {10.1086/310643}, \href {http://adsabs.harvard.edu/abs/1997ApJ...481L..23G}
  {481, L23}

\bibitem[\protect\citeauthoryear{{Greenhill}, {Kondratko}, {Lovell}, {Kuiper},
  {Moran}, {Jauncey}  \& {Baines}}{{Greenhill} et~al.}{2003a}]{Greenhill:2003}
{Greenhill} L.~J.,  {Kondratko} P.~T.,  {Lovell} J.~E.~J.,  {Kuiper} T.~B.~H.,
  {Moran} J.~M.,  {Jauncey} D.~L.,   {Baines} G.~P.,  2003a, \mn@doi [\apjl]
  {10.1086/367602}, \href {http://adsabs.harvard.edu/abs/2003ApJ...582L..11G}
  {582, L11}

\bibitem[\protect\citeauthoryear{{Greenhill} et~al.,}{{Greenhill}
  et~al.}{2003b}]{Greenhill:2003a}
{Greenhill} L.~J.,  et~al., 2003b, \mn@doi [\apj] {10.1086/374862}, \href
  {http://adsabs.harvard.edu/abs/2003ApJ...590..162G} {590, 162}

\bibitem[\protect\citeauthoryear{{G{\"u}ltekin} et~al.,}{{G{\"u}ltekin}
  et~al.}{2009}]{Gultekin:2009}
{G{\"u}ltekin} K.,  et~al., 2009, \mn@doi [\apj] {10.1088/0004-637X/698/1/198},
  \href {http://adsabs.harvard.edu/abs/2009ApJ...698..198G} {698, 198}

\bibitem[\protect\citeauthoryear{{Hayes}, {Davis}  \& {Silva}}{{Hayes}
  et~al.}{2017}]{Hayes:2016}
{Hayes} W.~B.,  {Davis} D.,   {Silva} P.,  2017, \mn@doi [\mnras]
  {10.1093/mnras/stw3290}, \href
  {http://adsabs.harvard.edu/abs/2017MNRAS.466.3928H} {466, 3928}

\bibitem[\protect\citeauthoryear{{Hicks} \& {Malkan}}{{Hicks} \&
  {Malkan}}{2008}]{Hicks:2008}
{Hicks} E.~K.~S.,  {Malkan} M.~A.,  2008, \mn@doi [\apjs] {10.1086/521650},
  \href {http://adsabs.harvard.edu/abs/2008ApJS..174...31H} {174, 31}

\bibitem[\protect\citeauthoryear{{H{\"o}nig}, {Watson}, {Kishimoto}  \&
  {Hjorth}}{{H{\"o}nig} et~al.}{2014}]{Honig:2014}
{H{\"o}nig} S.~F.,  {Watson} D.,  {Kishimoto} M.,   {Hjorth} J.,  2014, \mn@doi
  [\nat] {10.1038/nature13914}, \href
  {http://adsabs.harvard.edu/abs/2014Natur.515..528H} {515, 528}

\bibitem[\protect\citeauthoryear{{Hu}}{{Hu}}{2008}]{Hu:2008}
{Hu} J.,  2008, \mn@doi [\mnras] {10.1111/j.1365-2966.2008.13195.x}, \href
  {http://adsabs.harvard.edu/abs/2008MNRAS.386.2242H} {386, 2242}

\bibitem[\protect\citeauthoryear{{Hu}}{{Hu}}{2009}]{Hu:2009}
{Hu} J.,  2009, preprint, \href
  {http://adsabs.harvard.edu/abs/2009arXiv0908.2028H} {} (\mn@eprint {arXiv}
  {0908.2028})

\bibitem[\protect\citeauthoryear{{Hubble}}{{Hubble}}{1926}]{Hubble:1926}
{Hubble} E.~P.,  1926, \mn@doi [\apj] {10.1086/143018}, \href
  {http://adsabs.harvard.edu/abs/1926ApJ....64..321H} {64}

\bibitem[\protect\citeauthoryear{{Hubble}}{{Hubble}}{1936}]{Hubble:1936}
{Hubble} E.~P.,  1936, {Realm of the Nebulae}

\bibitem[\protect\citeauthoryear{{Humphreys}, {Reid}, {Moran}, {Greenhill}  \&
  {Argon}}{{Humphreys} et~al.}{2013}]{Humphreys:2013}
{Humphreys} E.~M.~L.,  {Reid} M.~J.,  {Moran} J.~M.,  {Greenhill} L.~J.,
  {Argon} A.~L.,  2013, \mn@doi [\apj] {10.1088/0004-637X/775/1/13}, \href
  {http://cdsads.u-strasbg.fr/abs/2013ApJ...775...13H} {775, 13}

\bibitem[\protect\citeauthoryear{{Jacobs}, {Rizzi}, {Tully}, {Shaya}, {Makarov}
   \& {Makarova}}{{Jacobs} et~al.}{2009}]{Jacobs:2009}
{Jacobs} B.~A.,  {Rizzi} L.,  {Tully} R.~B.,  {Shaya} E.~J.,  {Makarov} D.~I.,
   {Makarova} L.,  2009, \mn@doi [\aj] {10.1088/0004-6256/138/2/332}, \href
  {http://cdsads.u-strasbg.fr/abs/2009AJ....138..332J} {138, 332}

\bibitem[\protect\citeauthoryear{{Jardel} et~al.,}{{Jardel}
  et~al.}{2011}]{Jardel:2011}
{Jardel} J.~R.,  et~al., 2011, \mn@doi [\apj] {10.1088/0004-637X/739/1/21},
  \href {http://adsabs.harvard.edu/abs/2011ApJ...739...21J} {739, 21}

\bibitem[\protect\citeauthoryear{{Jeans}}{{Jeans}}{1919}]{Jeans:1919}
{Jeans} J.~H.,  1919, {Problems of cosmogony and stellar dynamics}

\bibitem[\protect\citeauthoryear{{Jeans}}{{Jeans}}{1928}]{Jeans:1928}
{Jeans} J.~H.,  1928, {Astronomy and cosmogony}

\bibitem[\protect\citeauthoryear{{Jones}}{{Jones}}{1973}]{Jones:1973}
{Jones} B.~T.,  1973, \mn@doi [\apj] {10.1086/152048}, \href
  {http://adsabs.harvard.edu/abs/1973ApJ...181..269J} {181, 269}

\bibitem[\protect\citeauthoryear{{Jones}, {Koribalski}, {Elmouttie}  \&
  {Haynes}}{{Jones} et~al.}{1999}]{Jones:1999}
{Jones} K.~L.,  {Koribalski} B.~S.,  {Elmouttie} M.,   {Haynes} R.~F.,  1999,
  \mn@doi [\mnras] {10.1046/j.1365-8711.1999.02057.x}, \href
  {http://adsabs.harvard.edu/abs/1999MNRAS.302..649J} {302, 649}

\bibitem[\protect\citeauthoryear{{Julian} \& {Toomre}}{{Julian} \&
  {Toomre}}{1966}]{Julian:Toomre:1966}
{Julian} W.~H.,  {Toomre} A.,  1966, \mn@doi [\apj] {10.1086/148957}, \href
  {http://adsabs.harvard.edu/abs/1966ApJ...146..810J} {146, 810}

\bibitem[\protect\citeauthoryear{Kim \& Elmegreen}{Kim \&
  Elmegreen}{2017}]{Kim:2017}
Kim W.-T.,  Elmegreen B.~G.,  2017, \mn@doi [\apjl] {10.3847/2041-8213/aa70a1},
  841, L4

\bibitem[\protect\citeauthoryear{{Kormendy}}{{Kormendy}}{2013}]{Kormendy:2013}
{Kormendy} J.,  2013, {Secular Evolution in Disk Galaxies}.
p.~1

\bibitem[\protect\citeauthoryear{{Kormendy} \& {Ho}}{{Kormendy} \&
  {Ho}}{2013}]{Kormendy:Ho:2013}
{Kormendy} J.,  {Ho} L.~C.,  2013, \mn@doi [\araa]
  {10.1146/annurev-astro-082708-101811}, \href
  {http://adsabs.harvard.edu/abs/2013ARA%26A..51..511K} {51, 511}

\bibitem[\protect\citeauthoryear{{Kormendy} \& {Richstone}}{{Kormendy} \&
  {Richstone}}{1995}]{Kormendy:Richstone:1995}
{Kormendy} J.,  {Richstone} D.,  1995, \mn@doi [\araa]
  {10.1146/annurev.aa.33.090195.003053}, \href
  {http://adsabs.harvard.edu/abs/1995ARA%26A..33..581K} {33, 581}

\bibitem[\protect\citeauthoryear{{Kormendy}, {Bender}  \& {Cornell}}{{Kormendy}
  et~al.}{2011}]{Kormendy:2011}
{Kormendy} J.,  {Bender} R.,   {Cornell} M.~E.,  2011, \mn@doi [\nat]
  {10.1038/nature09694}, \href
  {http://adsabs.harvard.edu/abs/2011Natur.469..374K} {469, 374}

\bibitem[\protect\citeauthoryear{{Kudritzki}, {Urbaneja}, {Gazak}, {Bresolin},
  {Przybilla}, {Gieren}  \& {Pietrzy{\'n}ski}}{{Kudritzki}
  et~al.}{2012}]{Kudritzki:2012}
{Kudritzki} R.-P.,  {Urbaneja} M.~A.,  {Gazak} Z.,  {Bresolin} F.,  {Przybilla}
  N.,  {Gieren} W.,   {Pietrzy{\'n}ski} G.,  2012, \mn@doi [\apj]
  {10.1088/0004-637X/747/1/15}, \href
  {http://cdsads.u-strasbg.fr/abs/2012ApJ...747...15K} {747, 15}

\bibitem[\protect\citeauthoryear{{Kuo}, {Braatz}, {Reid}, {Lo}, {Condon},
  {Impellizzeri}  \& {Henkel}}{{Kuo} et~al.}{2013}]{Kuo:2013}
{Kuo} C.~Y.,  {Braatz} J.~A.,  {Reid} M.~J.,  {Lo} K.~Y.,  {Condon} J.~J.,
  {Impellizzeri} C.~M.~V.,   {Henkel} C.,  2013, \mn@doi [\apj]
  {10.1088/0004-637X/767/2/155}, \href
  {http://cdsads.u-strasbg.fr/abs/2013ApJ...767..155K} {767, 155}

\bibitem[\protect\citeauthoryear{{LIGO Scientific Collaboration} et~al.,}{{LIGO
  Scientific Collaboration} et~al.}{2015}]{aLIGO}
{LIGO Scientific Collaboration} et~al., 2015, \mn@doi [Classical and Quantum
  Gravity] {10.1088/0264-9381/32/7/074001}, \href
  {http://adsabs.harvard.edu/abs/2015CQGra..32g4001L} {32, 074001}

\bibitem[\protect\citeauthoryear{{Lagattuta}, {Mould}, {Staveley-Smith},
  {Hong}, {Springob}, {Masters}, {Koribalski}  \& {Jones}}{{Lagattuta}
  et~al.}{2013}]{Lagattuta:2013}
{Lagattuta} D.~J.,  {Mould} J.~R.,  {Staveley-Smith} L.,  {Hong} T.,
  {Springob} C.~M.,  {Masters} K.~L.,  {Koribalski} B.~S.,   {Jones} D.~H.,
  2013, \mn@doi [\apj] {10.1088/0004-637X/771/2/88}, \href
  {http://adsabs.harvard.edu/abs/2013ApJ...771...88L} {771, 88}

\bibitem[\protect\citeauthoryear{{Lee} \& {Jang}}{{Lee} \&
  {Jang}}{2013}]{Lee:2013}
{Lee} M.~G.,  {Jang} I.~S.,  2013, \mn@doi [\apj] {10.1088/0004-637X/773/1/13},
  \href {http://cdsads.u-strasbg.fr/abs/2013ApJ...773...13L} {773, 13}

\bibitem[\protect\citeauthoryear{{Li}}{{Li}}{1998}]{Li:1998}
{Li} L.-X.,  1998, \mn@doi [General Relativity and Gravitation]
  {10.1023/A:1018867011142}, \href
  {http://adsabs.harvard.edu/abs/1998GReGr..30..497L} {30, 497}

\bibitem[\protect\citeauthoryear{{Lin} \& {Shu}}{{Lin} \&
  {Shu}}{1966}]{Lin:Shu:1966}
{Lin} C.~C.,  {Shu} F.~H.,  1966, Proceedings of the National Academy of
  Science, \href {http://adsabs.harvard.edu/abs/1966PNAS...55..229L} {55, 229}

\bibitem[\protect\citeauthoryear{{Lintott} et~al.,}{{Lintott}
  et~al.}{2008}]{Lintott:2008}
{Lintott} C.~J.,  et~al., 2008, \mn@doi [\mnras]
  {10.1111/j.1365-2966.2008.13689.x}, \href
  {http://adsabs.harvard.edu/abs/2008MNRAS.389.1179L} {389, 1179}

\bibitem[\protect\citeauthoryear{{Lintott} et~al.,}{{Lintott}
  et~al.}{2011}]{Lintott:2011}
{Lintott} C.,  et~al., 2011, \mn@doi [\mnras]
  {10.1111/j.1365-2966.2010.17432.x}, \href
  {http://adsabs.harvard.edu/abs/2011MNRAS.410..166L} {410, 166}

\bibitem[\protect\citeauthoryear{{Lodato} \& {Bertin}}{{Lodato} \&
  {Bertin}}{2003}]{Lodato:2003}
{Lodato} G.,  {Bertin} G.,  2003, \mn@doi [\aap] {10.1051/0004-6361:20021672},
  \href {http://adsabs.harvard.edu/abs/2003A%26A...398..517L} {398, 517}

\bibitem[\protect\citeauthoryear{{Loveday}}{{Loveday}}{1996}]{Loveday:1996}
{Loveday} J.,  1996, \mn@doi [\mnras] {10.1093/mnras/278.4.1025}, \href
  {http://adsabs.harvard.edu/abs/1996MNRAS.278.1025L} {278, 1025}

\bibitem[\protect\citeauthoryear{{Magorrian} et~al.,}{{Magorrian}
  et~al.}{1998}]{Magorrian:1998}
{Magorrian} J.,  et~al., 1998, \mn@doi [\aj] {10.1086/300353}, \href
  {http://ads.nao.ac.jp/abs/1998AJ....115.2285M} {115, 2285}

\bibitem[\protect\citeauthoryear{{Makarov}, {Prugniel}, {Terekhova}, {Courtois}
   \& {Vauglin}}{{Makarov} et~al.}{2014}]{HyperLeda}
{Makarov} D.,  {Prugniel} P.,  {Terekhova} N.,  {Courtois} H.,   {Vauglin} I.,
  2014, \mn@doi [\aap] {10.1051/0004-6361/201423496}, \href
  {http://adsabs.harvard.edu/abs/2014A%26A...570A..13M} {570, A13}

\bibitem[\protect\citeauthoryear{{Marconi} \& {Hunt}}{{Marconi} \&
  {Hunt}}{2003}]{Marconi:Hunt:2003}
{Marconi} A.,  {Hunt} L.~K.,  2003, \mn@doi [\apjl] {10.1086/375804}, \href
  {http://ads.nao.ac.jp/abs/2003ApJ...589L..21M} {589, L21}

\bibitem[\protect\citeauthoryear{{Mart{\'{\i}}nez-Garc{\'{\i}}a}, {Puerari},
  {Rosales-Ortega}, {Gonz{\'a}lez-L{\'o}pezlira}, {Fuentes-Carrera}  \&
  {Luna}}{{Mart{\'{\i}}nez-Garc{\'{\i}}a} et~al.}{2014}]{Martinez-Garcia:2014}
{Mart{\'{\i}}nez-Garc{\'{\i}}a} E.~E.,  {Puerari} I.,  {Rosales-Ortega} F.~F.,
  {Gonz{\'a}lez-L{\'o}pezlira} R.~A.,  {Fuentes-Carrera} I.,   {Luna} A.,
  2014, \mn@doi [\apjl] {10.1088/2041-8205/793/1/L19}, \href
  {http://adsabs.harvard.edu/abs/2014ApJ...793L..19M} {793, L19}

\bibitem[\protect\citeauthoryear{{McQuinn}, {Skillman}, {Dolphin}, {Berg}  \&
  {Kennicutt}}{{McQuinn} et~al.}{2016}]{McQuinn:2016}
{McQuinn} K.~B.~W.,  {Skillman} E.~D.,  {Dolphin} A.~E.,  {Berg} D.,
  {Kennicutt} R.,  2016, \mn@doi [\aj] {10.3847/0004-6256/152/5/144}, \href
  {http://adsabs.harvard.edu/abs/2016AJ....152..144M} {152, 144}

\bibitem[\protect\citeauthoryear{{McQuinn}, {Skillman}, {Dolphin}, {Berg}  \&
  {Kennicutt}}{{McQuinn} et~al.}{2017}]{McQuinn:2017}
{McQuinn} K.~B.~W.,  {Skillman} E.~D.,  {Dolphin} A.~E.,  {Berg} D.,
  {Kennicutt} R.,  2017, \mn@doi [\aj] {10.3847/1538-3881/aa7aad}, \href
  {http://adsabs.harvard.edu/abs/2017AJ....154...51M} {154, 51}

\bibitem[\protect\citeauthoryear{{Merritt}, {Ferrarese}  \& {Joseph}}{{Merritt}
  et~al.}{2001}]{Merritt:2001}
{Merritt} D.,  {Ferrarese} L.,   {Joseph} C.~L.,  2001, \mn@doi [Science]
  {10.1126/science.1063896}, \href
  {http://adsabs.harvard.edu/abs/2001Sci...293.1116M} {293, 1116}

\bibitem[\protect\citeauthoryear{{Minniti}, {Olszewski}  \& {Rieke}}{{Minniti}
  et~al.}{1993}]{Minniti:1993}
{Minniti} D.,  {Olszewski} E.~W.,   {Rieke} M.,  1993, \mn@doi [\apjl]
  {10.1086/186884}, \href {http://adsabs.harvard.edu/abs/1993ApJ...410L..79M}
  {410, L79}

\bibitem[\protect\citeauthoryear{{Morozov} \& {Mustsevoj}}{{Morozov} \&
  {Mustsevoj}}{1991}]{Morozov:1991}
{Morozov} A.~G.,  {Mustsevoj} V.~V.,  1991, Astronomicheskij Tsirkulyar, \href
  {http://adsabs.harvard.edu/abs/1991ATsir1550....1M} {1550, 1}

\bibitem[\protect\citeauthoryear{{Morozov}, {Mustsevoj}  \&
  {Prosvirov}}{{Morozov} et~al.}{1992}]{Morozov:1992}
{Morozov} A.~G.,  {Mustsevoj} V.~V.,   {Prosvirov} E.,  1992, Soviet Astronomy
  Letters, \href {http://adsabs.harvard.edu/abs/1992SvAL...18...20M} {18, 20}

\bibitem[\protect\citeauthoryear{{Nemmen}, {Georganopoulos}, {Guiriec},
  {Meyer}, {Gehrels}  \& {Sambruna}}{{Nemmen} et~al.}{2012}]{Nemmen:2012}
{Nemmen} R.~S.,  {Georganopoulos} M.,  {Guiriec} S.,  {Meyer} E.~T.,  {Gehrels}
  N.,   {Sambruna} R.~M.,  2012, \mn@doi [Science] {10.1126/science.1227416},
  \href {http://adsabs.harvard.edu/abs/2012Sci...338.1445N} {338, 1445}

\bibitem[\protect\citeauthoryear{{Nezlin}}{{Nezlin}}{1990}]{Nezlin:1990b}
{Nezlin} M.~V.,  1990, in Experimental study and characterization of chaos.
  World Scientific, pp 43--110, \mn@doi{10.1142/9789812832573_0002}

\bibitem[\protect\citeauthoryear{{Nezlin}}{{Nezlin}}{1991}]{Nezlin:1991}
{Nezlin} M.~V.,  1991, Akademiia Nauk SSSR Fizika Atmosfery i Okeana, \href
  {http://adsabs.harvard.edu/abs/1991FizAO..27...32N} {27, 32}

\bibitem[\protect\citeauthoryear{{Nezlin} \& {Snezhkin}}{{Nezlin} \&
  {Snezhkin}}{1990}]{Nezlin:1990}
{Nezlin} M.~V.,  {Snezhkin} E.~N.,  1990, Moscow Izdatel Nauka, \href
  {http://adsabs.harvard.edu/abs/1990MoIzN.........N} {}

\bibitem[\protect\citeauthoryear{{Nezlin}, {Rylov}, {Trubnikov}  \&
  {Khutoretski}}{{Nezlin} et~al.}{1990}]{Nezlin:1990c}
{Nezlin} M.~V.,  {Rylov} A.~Y.,  {Trubnikov} A.~S.,   {Khutoretski} A.~V.,
  1990, \mn@doi [Geophysical and Astrophysical Fluid Dynamics]
  {10.1080/03091929008219505}, \href
  {http://adsabs.harvard.edu/abs/1990GApFD..52..211N} {52, 211}

\bibitem[\protect\citeauthoryear{{Novak}, {Faber}  \& {Dekel}}{{Novak}
  et~al.}{2006}]{Novak:2006}
{Novak} G.~S.,  {Faber} S.~M.,   {Dekel} A.,  2006, \mn@doi [\apj]
  {10.1086/498333}, \href {http://adsabs.harvard.edu/abs/2006ApJ...637...96N}
  {637, 96}

\bibitem[\protect\citeauthoryear{{Nowak}, {Thomas}, {Erwin}, {Saglia}, {Bender}
   \& {Davies}}{{Nowak} et~al.}{2010}]{Nowak:2010}
{Nowak} N.,  {Thomas} J.,  {Erwin} P.,  {Saglia} R.~P.,  {Bender} R.,
  {Davies} R.~I.,  2010, \mn@doi [\mnras] {10.1111/j.1365-2966.2009.16167.x},
  \href {http://adsabs.harvard.edu/abs/2010MNRAS.403..646N} {403, 646}

\bibitem[\protect\citeauthoryear{{Onishi}, {Iguchi}, {Sheth}  \&
  {Kohno}}{{Onishi} et~al.}{2015}]{Onishi:2015}
{Onishi} K.,  {Iguchi} S.,  {Sheth} K.,   {Kohno} K.,  2015, \mn@doi [\apj]
  {10.1088/0004-637X/806/1/39}, \href
  {http://adsabs.harvard.edu/abs/2015ApJ...806...39O} {806, 39}

\bibitem[\protect\citeauthoryear{{Onken} et~al.,}{{Onken}
  et~al.}{2014}]{Onken:2014}
{Onken} C.~A.,  et~al., 2014, \mn@doi [\apj] {10.1088/0004-637X/791/1/37},
  \href {http://adsabs.harvard.edu/abs/2014ApJ...791...37O} {791, 37}

\bibitem[\protect\citeauthoryear{{Pastorini} et~al.,}{{Pastorini}
  et~al.}{2007}]{Pastorini:2007}
{Pastorini} G.,  et~al., 2007, \mn@doi [\aap] {10.1051/0004-6361:20066784},
  \href {http://adsabs.harvard.edu/abs/2007A%26A...469..405P} {469, 405}

\bibitem[\protect\citeauthoryear{{Peebles}}{{Peebles}}{1972}]{Peebles:1972}
{Peebles} P.~J.~E.,  1972, \mn@doi [\apj] {10.1086/151797}, \href
  {http://adsabs.harvard.edu/abs/1972ApJ...178..371P} {178, 371}

\bibitem[\protect\citeauthoryear{{P{\'e}rez-Villegas}, {Pichardo}  \&
  {Moreno}}{{P{\'e}rez-Villegas} et~al.}{2013}]{Perez-Villegas:2013}
{P{\'e}rez-Villegas} A.,  {Pichardo} B.,   {Moreno} E.,  2013, \mn@doi [\apj]
  {10.1088/0004-637X/772/2/91}, \href
  {http://adsabs.harvard.edu/abs/2013ApJ...772...91P} {772, 91}

\bibitem[\protect\citeauthoryear{P{\'e}rez et~al.,}{P{\'e}rez
  et~al.}{2016}]{Perez:2016}
P{\'e}rez L.~M.,  et~al., 2016, \mn@doi [Science] {10.1126/science.aaf8296},
  353, 1519

\bibitem[\protect\citeauthoryear{{Perley}, {Perley}, {Dhawan}  \&
  {Carilli}}{{Perley} et~al.}{2017}]{Perley:2017}
{Perley} D.~A.,  {Perley} R.~A.,  {Dhawan} V.,   {Carilli} C.~L.,  2017,
  \mn@doi [\apj] {10.3847/1538-4357/aa725b}, \href
  {http://adsabs.harvard.edu/abs/2017ApJ...841..117P} {841, 117}

\bibitem[\protect\citeauthoryear{{Planck Collaboration} et~al.,}{{Planck
  Collaboration} et~al.}{2016}]{Planck:2015}
{Planck Collaboration} et~al., 2016, \mn@doi [\aap]
  {10.1051/0004-6361/201525830}, \href
  {http://adsabs.harvard.edu/abs/2016A%26A...594A..13P} {594, A13}

\bibitem[\protect\citeauthoryear{{Poltorak} \& {Fridman}}{{Poltorak} \&
  {Fridman}}{2007}]{Poltorak:2007}
{Poltorak} S.~G.,  {Fridman} A.~M.,  2007, \mn@doi [\arep]
  {10.1134/S1063772907060042}, \href
  {http://adsabs.harvard.edu/abs/2007ARep...51..460P} {51, 460}

\bibitem[\protect\citeauthoryear{{Pour-Imani}, {Kennefick}, {Kennefick},
  {Davis}, {Shields}  \& {Shameer Abdeen}}{{Pour-Imani}
  et~al.}{2016}]{Pour-Imani:2016}
{Pour-Imani} H.,  {Kennefick} D.,  {Kennefick} J.,  {Davis} B.~L.,  {Shields}
  D.~W.,   {Shameer Abdeen} M.,  2016, \mn@doi [\apjl]
  {10.3847/2041-8205/827/1/L2}, \href
  {http://adsabs.harvard.edu/abs/2016ApJ...827L...2P} {827, L2}

\bibitem[\protect\citeauthoryear{{Radburn-Smith} et~al.,}{{Radburn-Smith}
  et~al.}{2011}]{Radburn-Smith:2011}
{Radburn-Smith} D.~J.,  et~al., 2011, \mn@doi [\apjs]
  {10.1088/0067-0049/195/2/18}, \href
  {http://cdsads.u-strasbg.fr/abs/2011ApJS..195...18R} {195, 18}

\bibitem[\protect\citeauthoryear{{Rafikov}}{{Rafikov}}{2016}]{Rafikov:2016}
{Rafikov} R.~R.,  2016, \mn@doi [\apj] {10.3847/0004-637X/831/2/122}, \href
  {http://adsabs.harvard.edu/abs/2016ApJ...831..122R} {831, 122}

\bibitem[\protect\citeauthoryear{{Rastorguev}, {Utkin}, {Zabolotskikh},
  {Dambis}, {Bajkova}  \& {Bobylev}}{{Rastorguev}
  et~al.}{2017}]{Rastorguev:2017}
{Rastorguev} A.~S.,  {Utkin} N.~D.,  {Zabolotskikh} M.~V.,  {Dambis} A.~K.,
  {Bajkova} A.~T.,   {Bobylev} V.~V.,  2017, \mn@doi [Astrophysical Bulletin]
  {10.1134/S1990341317020043}, \href
  {http://adsabs.harvard.edu/abs/2017AstBu..72..122R} {72, 122}

\bibitem[\protect\citeauthoryear{{Reid}, {Braatz}, {Condon}, {Lo}, {Kuo},
  {Impellizzeri}  \& {Henkel}}{{Reid} et~al.}{2013}]{Reid:2013}
{Reid} M.~J.,  {Braatz} J.~A.,  {Condon} J.~J.,  {Lo} K.~Y.,  {Kuo} C.~Y.,
  {Impellizzeri} C.~M.~V.,   {Henkel} C.,  2013, \mn@doi [\apj]
  {10.1088/0004-637X/767/2/154}, \href
  {http://cdsads.u-strasbg.fr/abs/2013ApJ...767..154R} {767, 154}

\bibitem[\protect\citeauthoryear{{Riess}, {Fliri}  \& {Valls-Gabaud}}{{Riess}
  et~al.}{2012}]{Riess:2012}
{Riess} A.~G.,  {Fliri} J.,   {Valls-Gabaud} D.,  2012, \mn@doi [\apj]
  {10.1088/0004-637X/745/2/156}, \href
  {http://cdsads.u-strasbg.fr/abs/2012ApJ...745..156R} {745, 156}

\bibitem[\protect\citeauthoryear{{Ringermacher} \& {Mead}}{{Ringermacher} \&
  {Mead}}{2010}]{Ringermacher:2010}
{Ringermacher} H.~I.,  {Mead} L.~R.,  2010, in American Astronomical Society
  Meeting Abstracts \#215. p.~592

\bibitem[\protect\citeauthoryear{{Rodr{\'{\i}}guez-Rico}, {Goss}, {Zhao},
  {G{\'o}mez}  \& {Anantharamaiah}}{{Rodr{\'{\i}}guez-Rico}
  et~al.}{2006}]{Rodriguez-Rico:2006}
{Rodr{\'{\i}}guez-Rico} C.~A.,  {Goss} W.~M.,  {Zhao} J.-H.,  {G{\'o}mez} Y.,
  {Anantharamaiah} K.~R.,  2006, \mn@doi [\apj] {10.1086/503796}, \href
  {http://adsabs.harvard.edu/abs/2006ApJ...644..914R} {644, 914}

\bibitem[\protect\citeauthoryear{{Rodr{\'{\i}}guez}, {Clocchiatti}  \&
  {Hamuy}}{{Rodr{\'{\i}}guez} et~al.}{2014}]{Pudge}
{Rodr{\'{\i}}guez} {\'O}.,  {Clocchiatti} A.,   {Hamuy} M.,  2014, \mn@doi
  [\aj] {10.1088/0004-6256/148/6/107}, \href
  {http://adsabs.harvard.edu/abs/2014AJ....148..107R} {148, 107}

\bibitem[\protect\citeauthoryear{{Safronov}}{{Safronov}}{1960}]{Safronov:1960}
{Safronov} V.~S.,  1960, Annales d'Astrophysique, \href
  {http://adsabs.harvard.edu/abs/1960AnAp...23..979S} {23, 979}

\bibitem[\protect\citeauthoryear{{Saglia} et~al.,}{{Saglia}
  et~al.}{2016}]{Saglia:2016}
{Saglia} R.~P.,  et~al., 2016, \mn@doi [\apj] {10.3847/0004-637X/818/1/47},
  \href {http://adsabs.harvard.edu/abs/2016ApJ...818...47S} {818, 47}

\bibitem[\protect\citeauthoryear{{Sandage} \& {Tammann}}{{Sandage} \&
  {Tammann}}{1981}]{Sandage:1981}
{Sandage} A.,  {Tammann} G.~A.,  1981, {A revised Shapley-Ames Catalog of
  bright galaxies}

\bibitem[\protect\citeauthoryear{{Sani}, {Marconi}, {Hunt}  \&
  {Risaliti}}{{Sani} et~al.}{2011}]{Sani:2011}
{Sani} E.,  {Marconi} A.,  {Hunt} L.~K.,   {Risaliti} G.,  2011, \mn@doi
  [\mnras] {10.1111/j.1365-2966.2011.18229.x}, \href
  {http://adsabs.harvard.edu/abs/2011MNRAS.413.1479S} {413, 1479}

\bibitem[\protect\citeauthoryear{{Savchenko} \& {Reshetnikov}}{{Savchenko} \&
  {Reshetnikov}}{2013}]{Savchenko:2013}
{Savchenko} S.~S.,  {Reshetnikov} V.~P.,  2013, \mn@doi [\mnras]
  {10.1093/mnras/stt1627}, \href
  {http://adsabs.harvard.edu/abs/2013MNRAS.436.1074S} {436, 1074}

\bibitem[\protect\citeauthoryear{{Savorgnan}, {Graham}, {Marconi}  \&
  {Sani}}{{Savorgnan} et~al.}{2016}]{Savorgnan:2016:II}
{Savorgnan} G.~A.~D.,  {Graham} A.~W.,  {Marconi} A.,   {Sani} E.,  2016,
  \mn@doi [\apj] {10.3847/0004-637X/817/1/21}, \href
  {http://adsabs.harvard.edu/abs/2016ApJ...817...21S} {817, 21}

\bibitem[\protect\citeauthoryear{{Seigar}}{{Seigar}}{2011}]{Seigar:2011}
{Seigar} M.~S.,  2011, \mn@doi [ISRN Astronomy and Astrophysics]
  {10.5402/2011/725697}, \href
  {http://adsabs.harvard.edu/abs/2011ISRAA2011E...4S} {2011, 725697}

\bibitem[\protect\citeauthoryear{{Seigar}, {Block}, {Puerari}, {Chorney}  \&
  {James}}{{Seigar} et~al.}{2005}]{Seigar:2005}
{Seigar} M.~S.,  {Block} D.~L.,  {Puerari} I.,  {Chorney} N.~E.,   {James}
  P.~A.,  2005, \mn@doi [\mnras] {10.1111/j.1365-2966.2005.08970.x}, \href
  {http://adsabs.harvard.edu/abs/2005MNRAS.359.1065S} {359, 1065}

\bibitem[\protect\citeauthoryear{{Seigar}, {Bullock}, {Barth}  \&
  {Ho}}{{Seigar} et~al.}{2006}]{Seigar:2006}
{Seigar} M.~S.,  {Bullock} J.~S.,  {Barth} A.~J.,   {Ho} L.~C.,  2006, \mn@doi
  [\apj] {10.1086/504463}, \href
  {http://adsabs.harvard.edu/abs/2006ApJ...645.1012S} {645, 1012}

\bibitem[\protect\citeauthoryear{{Seigar}, {Kennefick}, {Kennefick}  \&
  {Lacy}}{{Seigar} et~al.}{2008}]{Seigar:2008}
{Seigar} M.~S.,  {Kennefick} D.,  {Kennefick} J.,   {Lacy} C.~H.~S.,  2008,
  \mn@doi [\apjl] {10.1086/588727}, \href
  {http://adsabs.harvard.edu/abs/2008ApJ...678L..93S} {678, L93}

\bibitem[\protect\citeauthoryear{{Seigar}, {Davis}, {Berrier}  \&
  {Kennefick}}{{Seigar} et~al.}{2014}]{Seigar:2014}
{Seigar} M.~S.,  {Davis} B.~L.,  {Berrier} J.,   {Kennefick} D.,  2014, \mn@doi
  [\apj] {10.1088/0004-637X/795/1/90}, \href
  {http://adsabs.harvard.edu/abs/2014ApJ...795...90S} {795, 90}

\bibitem[\protect\citeauthoryear{{Sellwood} \& {Carlberg}}{{Sellwood} \&
  {Carlberg}}{2014}]{Sellwood:Carlberg:2014}
{Sellwood} J.~A.,  {Carlberg} R.~G.,  2014, \mn@doi [\apj]
  {10.1088/0004-637X/785/2/137}, \href
  {http://adsabs.harvard.edu/abs/2014ApJ...785..137S} {785, 137}

\bibitem[\protect\citeauthoryear{{Semczuk}, {{\L}okas}  \& {del
  Pino}}{{Semczuk} et~al.}{2017}]{Semczuk:2017}
{Semczuk} M.,  {{\L}okas} E.~L.,   {del Pino} A.,  2017, \mn@doi [\apj]
  {10.3847/1538-4357/834/1/7}, \href
  {http://adsabs.harvard.edu/abs/2017ApJ...834....7S} {834, 7}

\bibitem[\protect\citeauthoryear{{Shamir}}{{Shamir}}{2011a}]{Ganalyzer}
{Shamir} L.,  2011a, {Ganalyzer: A tool for automatic galaxy image analysis},
  Astrophysics Source Code Library (\mn@eprint {ascl} {1105.011})

\bibitem[\protect\citeauthoryear{{Shamir}}{{Shamir}}{2011b}]{Shamir:2011}
{Shamir} L.,  2011b, \mn@doi [\apj] {10.1088/0004-637X/736/2/141}, \href
  {http://adsabs.harvard.edu/abs/2011ApJ...736..141S} {736, 141}

\bibitem[\protect\citeauthoryear{{Shamir}}{{Shamir}}{2017}]{Shamir:2017}
{Shamir} L.,  2017, \mn@doi [\pasa] {10.1017/pasa.2017.4}, \href
  {http://adsabs.harvard.edu/abs/2017PASA...34...11S} {34, e011}

\bibitem[\protect\citeauthoryear{{Shields} et~al.,}{{Shields}
  et~al.}{2014}]{Shields:2014}
{Shields} D.~W.,  et~al., 2014, in {Seigar} M.~S.,  {Treuthardt} P.,  eds,
  Astronomical Society of the Pacific Conference Series Vol. 480, Structure and
  Dynamics of Disk Galaxies. p.~130

\bibitem[\protect\citeauthoryear{{Shields} et~al.,}{{Shields}
  et~al.}{2015a}]{Shields:2015}
{Shields} D.~W.,  et~al., 2015a, preprint, \href
  {http://adsabs.harvard.edu/abs/2015arXiv151106365S} {} (\mn@eprint {arXiv}
  {1511.06365})

\bibitem[\protect\citeauthoryear{{Shields} et~al.,}{{Shields}
  et~al.}{2015b}]{Spirality}
{Shields} D.~W.,  et~al., 2015b, {Spirality: Spiral arm pitch angle
  measurement}, Astrophysics Source Code Library (\mn@eprint {ascl} {1512.015})

\bibitem[\protect\citeauthoryear{{Shinkai}, {Kanda}  \& {Ebisuzaki}}{{Shinkai}
  et~al.}{2017}]{Shinkai:2016}
{Shinkai} H.-a.,  {Kanda} N.,   {Ebisuzaki} T.,  2017, \mn@doi [\apj]
  {10.3847/1538-4357/835/2/276}, \href
  {http://adsabs.harvard.edu/abs/2017ApJ...835..276S} {835, 276}

\bibitem[\protect\citeauthoryear{{Shu}}{{Shu}}{2016}]{Shu:2016}
{Shu} F.~H.,  2016, \mn@doi [\araa] {10.1146/annurev-astro-081915-023426},
  \href {http://adsabs.harvard.edu/abs/2016ARA%26A..54..667S} {54, 667}

\bibitem[\protect\citeauthoryear{{Silk} \& {Ames}}{{Silk} \&
  {Ames}}{1972}]{Silk:1972}
{Silk} J.,  {Ames} S.,  1972, \mn@doi [\apj] {10.1086/151767}, \href
  {http://adsabs.harvard.edu/abs/1972ApJ...178...77S} {178, 77}

\bibitem[\protect\citeauthoryear{{Silverman} et~al.,}{{Silverman}
  et~al.}{2012}]{Silverman:2012}
{Silverman} J.~M.,  et~al., 2012, \mn@doi [\mnras]
  {10.1111/j.1365-2966.2012.21270.x}, \href
  {http://cdsads.u-strasbg.fr/abs/2012MNRAS.425.1789S} {425, 1789}

\bibitem[\protect\citeauthoryear{{Simmons}, {Smethurst}  \&
  {Lintott}}{{Simmons} et~al.}{2017}]{Simmons:2017}
{Simmons} B.~D.,  {Smethurst} R.~J.,   {Lintott} C.,  2017, \mn@doi [\mnras]
  {10.1093/mnras/stx1340}, \href
  {http://adsabs.harvard.edu/abs/2017MNRAS.470.1559S} {470, 1559}

\bibitem[\protect\citeauthoryear{{Sorce}, {Tully}, {Courtois}, {Jarrett},
  {Neill}  \& {Shaya}}{{Sorce} et~al.}{2014}]{Sorce:2014}
{Sorce} J.~G.,  {Tully} R.~B.,  {Courtois} H.~M.,  {Jarrett} T.~H.,  {Neill}
  J.~D.,   {Shaya} E.~J.,  2014, \mn@doi [\mnras] {10.1093/mnras/stu1450},
  \href {http://cdsads.u-strasbg.fr/abs/2014MNRAS.444..527S} {444, 527}

\bibitem[\protect\citeauthoryear{{Tadhunter}, {Marconi}, {Axon}, {Wills},
  {Robinson}  \& {Jackson}}{{Tadhunter} et~al.}{2003}]{Tadhunter:2003}
{Tadhunter} C.,  {Marconi} A.,  {Axon} D.,  {Wills} K.,  {Robinson} T.~G.,
  {Jackson} N.,  2003, \mn@doi [\mnras] {10.1046/j.1365-8711.2003.06588.x},
  \href {http://adsabs.harvard.edu/abs/2003MNRAS.342..861T} {342, 861}

\bibitem[\protect\citeauthoryear{{Terry}, {Paturel}  \& {Ekholm}}{{Terry}
  et~al.}{2002}]{Terry:2002}
{Terry} J.~N.,  {Paturel} G.,   {Ekholm} T.,  2002, \mn@doi [\aap]
  {10.1051/0004-6361:20021018}, \href
  {http://cdsads.u-strasbg.fr/abs/2002A%26A...393...57T} {393, 57}

\bibitem[\protect\citeauthoryear{{Thornley}}{{Thornley}}{1996}]{Thornley:1996}
{Thornley} M.~D.,  1996, \mn@doi [\apjl] {10.1086/310250}, \href
  {http://adsabs.harvard.edu/abs/1996ApJ...469L..45T} {469, L45}

\bibitem[\protect\citeauthoryear{{Toomre}}{{Toomre}}{1964}]{Toomre:1964}
{Toomre} A.,  1964, \mn@doi [\apj] {10.1086/147861}, \href
  {http://adsabs.harvard.edu/abs/1964ApJ...139.1217T} {139, 1217}

\bibitem[\protect\citeauthoryear{{Tremaine} et~al.,}{{Tremaine}
  et~al.}{2002}]{Tremaine:2002}
{Tremaine} S.,  et~al., 2002, \mn@doi [\apj] {10.1086/341002}, \href
  {http://adsabs.harvard.edu/abs/2002ApJ...574..740T} {574, 740}

\bibitem[\protect\citeauthoryear{{Treuthardt}, {Seigar}, {Sierra},
  {Al-Baidhany}, {Salo}, {Kennefick}, {Kennefick}  \& {Lacy}}{{Treuthardt}
  et~al.}{2012}]{Treuthardt:2012}
{Treuthardt} P.,  {Seigar} M.~S.,  {Sierra} A.~D.,  {Al-Baidhany} I.,  {Salo}
  H.,  {Kennefick} D.,  {Kennefick} J.,   {Lacy} C.~H.~S.,  2012, \mn@doi
  [\mnras] {10.1111/j.1365-2966.2012.21118.x}, \href
  {http://adsabs.harvard.edu/abs/2012MNRAS.423.3118T} {423, 3118}

\bibitem[\protect\citeauthoryear{{Tully}}{{Tully}}{1988}]{Tully:1988}
{Tully} R.~B.,  1988, {Nearby galaxies catalog}

\bibitem[\protect\citeauthoryear{{Tully}, {Shaya}, {Karachentsev}, {Courtois},
  {Kocevski}, {Rizzi}  \& {Peel}}{{Tully} et~al.}{2008}]{Tully:2008}
{Tully} R.~B.,  {Shaya} E.~J.,  {Karachentsev} I.~D.,  {Courtois} H.~M.,
  {Kocevski} D.~D.,  {Rizzi} L.,   {Peel} A.,  2008, \mn@doi [\apj]
  {10.1086/527428}, \href {http://cdsads.u-strasbg.fr/abs/2008ApJ...676..184T}
  {676, 184}

\bibitem[\protect\citeauthoryear{{Tully}, {Libeskind}, {Karachentsev},
  {Karachentseva}, {Rizzi}  \& {Shaya}}{{Tully} et~al.}{2015}]{Tully:2015}
{Tully} R.~B.,  {Libeskind} N.~I.,  {Karachentsev} I.~D.,  {Karachentseva}
  V.~E.,  {Rizzi} L.,   {Shaya} E.~J.,  2015, \mn@doi [\apjl]
  {10.1088/2041-8205/802/2/L25}, \href
  {http://cdsads.u-strasbg.fr/abs/2015ApJ...802L..25T} {802, L25}

\bibitem[\protect\citeauthoryear{{Ushakov} \& {Chernin}}{{Ushakov} \&
  {Chernin}}{1983}]{Ushakov:1983}
{Ushakov} A.~Y.,  {Chernin} A.~D.,  1983, \azh, \href
  {http://adsabs.harvard.edu/abs/1983AZh....60..625U} {60, 625}

\bibitem[\protect\citeauthoryear{{Ushakov} \& {Chernin}}{{Ushakov} \&
  {Chernin}}{1984}]{Ushakov:1984}
{Ushakov} A.~Y.,  {Chernin} A.~D.,  1984, \azh, \href
  {http://adsabs.harvard.edu/abs/1984AZh....61...18U} {61, 18}

\bibitem[\protect\citeauthoryear{{Vall{\'e}e}}{{Vall{\'e}e}}{2015}]{Vallee:2015}
{Vall{\'e}e} J.~P.,  2015, \mn@doi [\mnras] {10.1093/mnras/stv862}, \href
  {http://adsabs.harvard.edu/abs/2015MNRAS.450.4277V} {450, 4277}

\bibitem[\protect\citeauthoryear{{Vatistas}}{{Vatistas}}{2010}]{Vatistas:2010}
{Vatistas} G.~H.,  2010, preprint, \href
  {http://adsabs.harvard.edu/abs/2010arXiv1012.1384V} {} (\mn@eprint {arXiv}
  {1012.1384})

\bibitem[\protect\citeauthoryear{{Vorobyov}}{{Vorobyov}}{2006}]{Vorobyov:2006}
{Vorobyov} E.~I.,  2006, \mn@doi [\mnras] {10.1111/j.1365-2966.2006.10550.x},
  \href {http://adsabs.harvard.edu/abs/2006MNRAS.370.1046V} {370, 1046}

\bibitem[\protect\citeauthoryear{{Wada}, {Schartmann}  \& {Meijerink}}{{Wada}
  et~al.}{2016}]{Wada:2016}
{Wada} K.,  {Schartmann} M.,   {Meijerink} R.,  2016, \mn@doi [\apjl]
  {10.3847/2041-8205/828/2/L19}, \href
  {http://adsabs.harvard.edu/abs/2016ApJ...828L..19W} {828, L19}

\bibitem[\protect\citeauthoryear{{White}}{{White}}{1972}]{White:1972}
{White} M.~L.,  1972, \mn@doi [\apss] {10.1007/BF00642741}, \href
  {http://adsabs.harvard.edu/abs/1972Ap%26SS..16..295W} {16, 295}

\bibitem[\protect\citeauthoryear{{Wold}, {Lacy}, {K{\"a}ufl}  \&
  {Siebenmorgen}}{{Wold} et~al.}{2006}]{Wold:2006}
{Wold} M.,  {Lacy} M.,  {K{\"a}ufl} H.~U.,   {Siebenmorgen} R.,  2006, \mn@doi
  [\aap] {10.1051/0004-6361:20053385}, \href
  {http://adsabs.harvard.edu/abs/2006A%26A...460..449W} {460, 449}

\bibitem[\protect\citeauthoryear{{Yamauchi}, {Nakai}, {Sato}  \&
  {Diamond}}{{Yamauchi} et~al.}{2004}]{Yamauchi:2004}
{Yamauchi} A.,  {Nakai} N.,  {Sato} N.,   {Diamond} P.,  2004, \mn@doi [\pasj]
  {10.1093/pasj/56.4.605}, \href
  {http://adsabs.harvard.edu/abs/2004PASJ...56..605Y} {56, 605}

\bibitem[\protect\citeauthoryear{{Yamauchi}, {Nakai}, {Ishihara}, {Diamond}  \&
  {Sato}}{{Yamauchi} et~al.}{2012}]{Yamauchi:2012}
{Yamauchi} A.,  {Nakai} N.,  {Ishihara} Y.,  {Diamond} P.,   {Sato} N.,  2012,
  \mn@doi [\pasj] {10.1093/pasj/64.5.103}, \href
  {http://cdsads.u-strasbg.fr/abs/2012PASJ...64..103Y} {64}

\bibitem[\protect\citeauthoryear{{York} et~al.,}{{York}
  et~al.}{2000}]{York:2000}
{York} D.~G.,  et~al., 2000, \mn@doi [\aj] {10.1086/301513}, \href
  {http://adsabs.harvard.edu/abs/2000AJ....120.1579Y} {120, 1579}

\bibitem[\protect\citeauthoryear{{Yoshizawa} \& {Wakamatsu}}{{Yoshizawa} \&
  {Wakamatsu}}{1975}]{Yoshizawa:1975}
{Yoshizawa} M.,  {Wakamatsu} K.,  1975, \aap, \href
  {http://adsabs.harvard.edu/abs/1975A%26A....44..363Y} {44, 363}

\bibitem[\protect\citeauthoryear{{Zanotti}, {Rezzolla}, {Del Zanna}  \&
  {Palenzuela}}{{Zanotti} et~al.}{2010}]{Zanotti:2010}
{Zanotti} O.,  {Rezzolla} L.,  {Del Zanna} L.,   {Palenzuela} C.,  2010,
  \mn@doi [\aap] {10.1051/0004-6361/201014969}, \href
  {http://adsabs.harvard.edu/abs/2010A%26A...523A...8Z} {523, A8}

\bibitem[\protect\citeauthoryear{{Zoccali} et~al.,}{{Zoccali}
  et~al.}{2014}]{Zoccali:2014}
{Zoccali} M.,  et~al., 2014, \mn@doi [\aap] {10.1051/0004-6361/201323120},
  \href {http://adsabs.harvard.edu/abs/2014A%26A...562A..66Z} {562, A66}

\bibitem[\protect\citeauthoryear{{den Brok} et~al.,}{{den Brok}
  et~al.}{2015}]{Brok:2015}
{den Brok} M.,  et~al., 2015, \mn@doi [\apj] {10.1088/0004-637X/809/1/101},
  \href {http://adsabs.harvard.edu/abs/2015ApJ...809..101D} {809, 101}

\bibitem[\protect\citeauthoryear{{van den Bergh}}{{van den Bergh}}{1991}]{M33}
{van den Bergh} S.,  1991, \mn@doi [\pasp] {10.1086/132860}, \href
  {http://adsabs.harvard.edu/abs/1991PASP..103..609V} {103, 609}

\bibitem[\protect\citeauthoryear{{van den Bosch}}{{van den
  Bosch}}{2016}]{Bosch:2016}
{van den Bosch} R.~C.~E.,  2016, \mn@doi [\apj] {10.3847/0004-637X/831/2/134},
  \href {http://adsabs.harvard.edu/abs/2016ApJ...831..134V} {831, 134}

\makeatother
\end{thebibliography}




\appendix

\section{Demonstration of Independent Measurement Methods}\label{demo}

Here, we demonstrate and compare the results from three independent pitch angle measurement methods used for this study: \textsc{2dfft}, template fitting and computer vision. For the purposes of our demonstration, we focus on UGC 6093. This is a good example case because it demonstrates that accurate pitch angles can be measured for distant galaxies. UGC 6093 is the second most distant galaxy in our sample at $z = 0.036118$.

\subsection{Computer vision}

\begin{figure}
\begin{center}
\includegraphics[clip=true,trim= 0mm 0mm 0mm 0mm,width=\columnwidth]{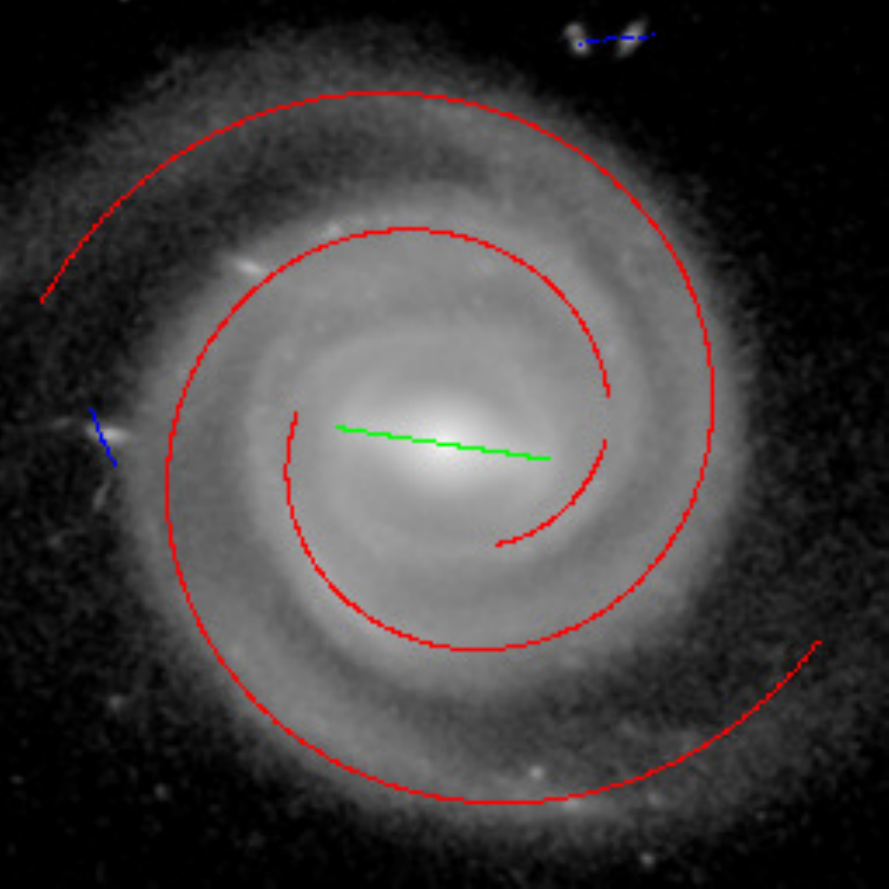}
\caption{\textit{HST} WFC3 \textit{F814W} image of UGC 6093, pre-processed (rotated so the position angle of the resulting $y$-axis is $-145\fdg2$ E of N and de-projected for disc inclination by $24\fdg7$ to a face-on orientation) and overlaid with the spiral arcs detected by the \textsc{sparcfire} computer vision software \citep{Davis:Hayes:2014}. Red spiral arcs represent spiral segments with `Z-wise' chirality and blue spiral arcs represent spiral segments with `S-wise' chirality. The green line represents the detection of a central bar. For a sense of scale, the central bar subtends an angle of $9\farcs19$, equivalent to a physical distance of 6.81 kpc at the distance of UGC 6093.}
\label{sparcfire}
\end{center}
\end{figure}

In Fig.~\ref{sparcfire}, we present the output of the \textsc{sparcfire}\footnote{\url{http://sparcfire.ics.uci.edu}} computer vision software \citep{Davis:Hayes:2014}. It is evident from the overlaid arcs that the dominant chirality in the image is `Z-wise' and we can ignore the spurious short `S-wise' arcs. There are three `Z-wise' arcs that we will analyse to determine a single pitch angle value for the galaxy. From longest to shortest, their arc lengths ($s$) and pitch angles are: $s_1 = 3050$~pixels (120\farcs8 = 89.53~kpc) with $\phi_1 = -9\fdg65$, $s_2 = 2875$~pixels (113\farcs9 = 84.39~kpc) with $\phi_2 = -9\fdg35$ and $s_3 = 296$~pixels (11\farcs7 = 8.69~kpc) with $\phi_3 = -18\fdg03$. This yields a weighted (by arc length) mean pitch angle of $-9\fdg91\pm1\fdg82$ for UGC 6093. \textsc{sparcfire} also detects the presence of a central bar with a total length of 232~pixels ($9\farcs19$ = 6.81~kpc).

\subsection{Template fitting}

The spiral template fitting software, \textsc{spirality} \citep{Shields:2015,Spirality}, operates by examining pixel values on an image along logarithmic spiral coordinate axes with an origin at the centre of a galaxy. It then computes the median pixel value along every axis. By repeating this process for numerous templates, each with different values of pitch angle, it computes the variation of median spiral axis pixel values. Spiral axes with pitch angles that are drastically different from the true pitch angle of a galaxy (or with the wrong chirality) will have little or no variance amongst different axes across the phase angle space of the spiral coordinate system. However, spiral axes that closely resemble the true pitch angle of the galaxy will have increased variance as some axes will lie upon bright spiral arms and others will lie upon the darker gaps in between spiral arms. The true pitch angle of the galaxy will be the set of axes with a pitch angle that maximises the variance.

\begin{figure*}
\includegraphics[clip=true,trim= 0mm 0mm 0mm 0mm,height=4.77cm]{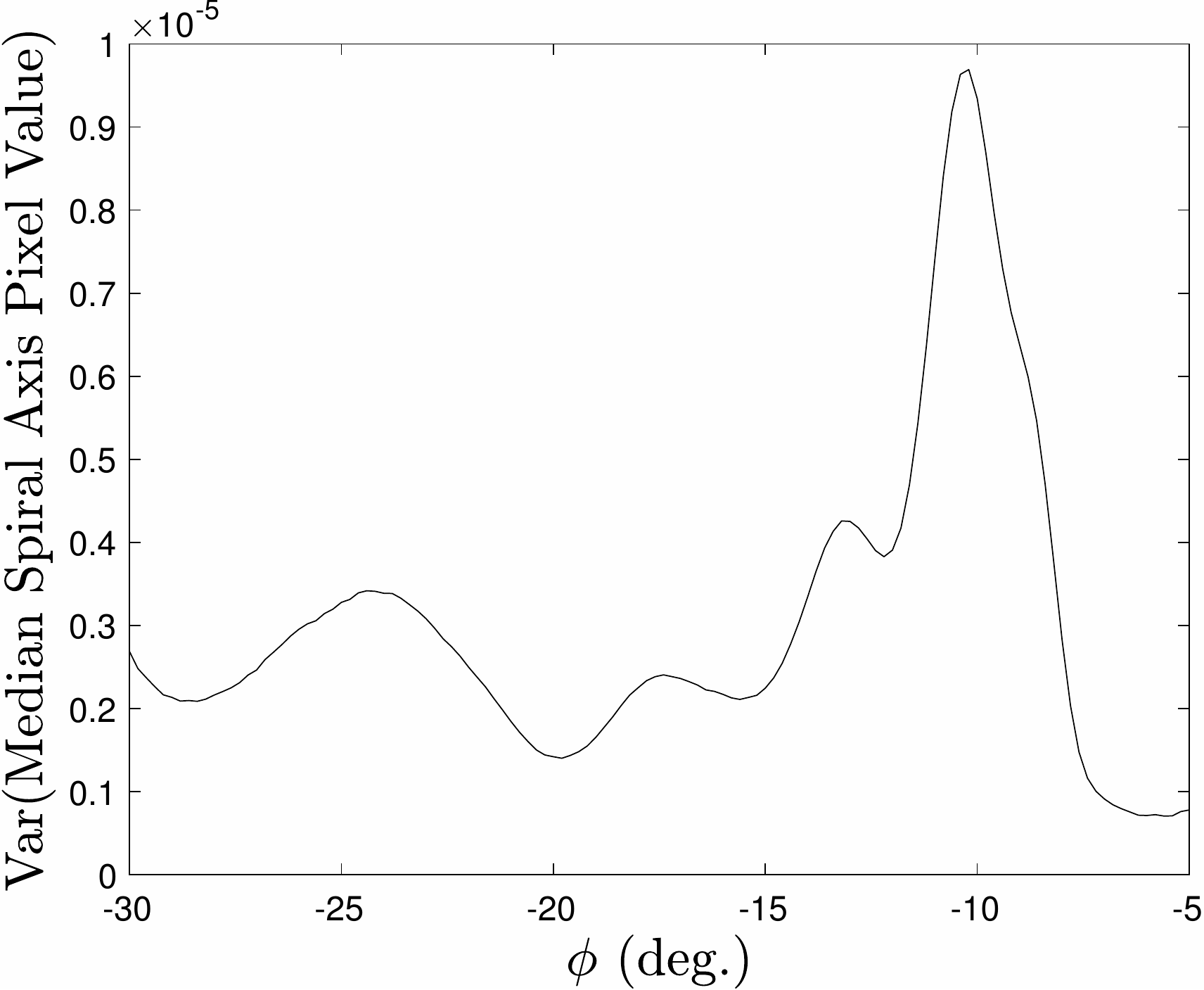}
\includegraphics[clip=true,trim= 0mm 0mm 0mm 0mm,height=4.62cm]{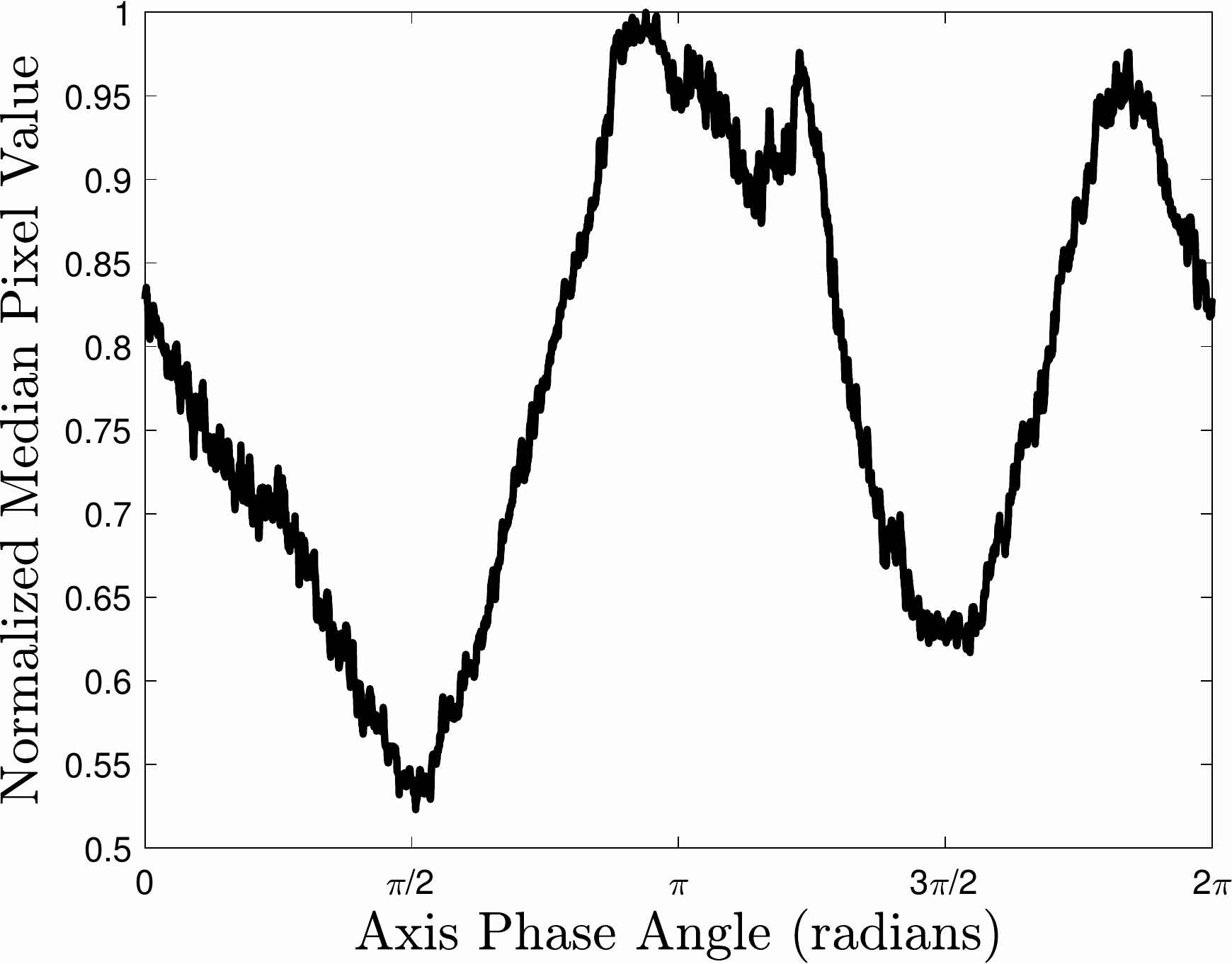}
\includegraphics[clip=true,trim= 0mm 0mm 0mm 0mm,height=4.62cm]{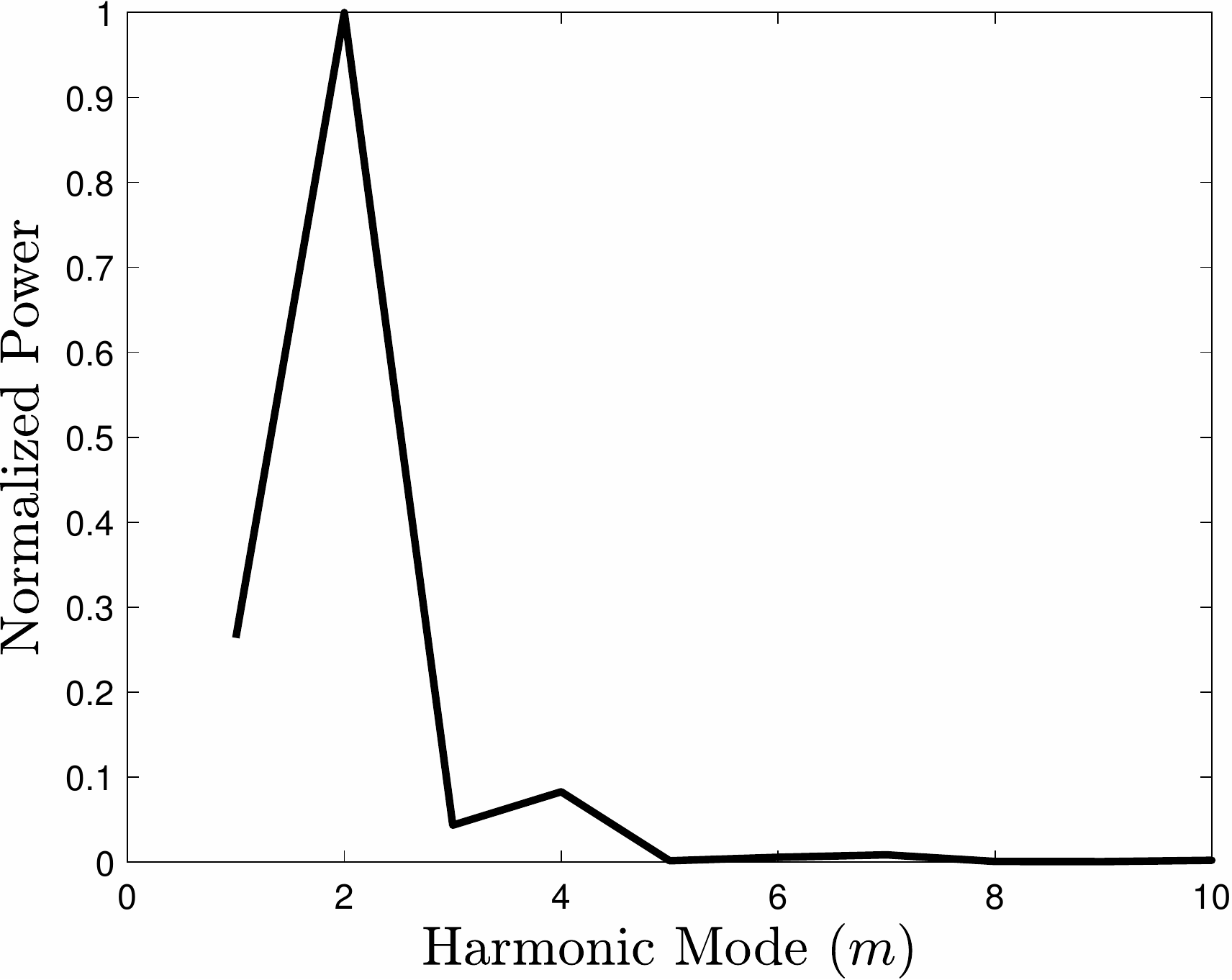} 
\caption{Template fitting results for UGC 6093 by the \textsc{spirality} software \citep{Shields:2015,Spirality}. \textbf{Left} -- variance of the median pixel value across all logarithmic spiral axes as a function of the pitch angle of the logarithmic spiral coordinate system. The maximal variance is found at $\phi = -10\fdg22\pm0\fdg94$. \textbf{Middle} -- the median pixel value along each spiral axis, where the logarithmic spiral coordinate system has a pitch angle equal to that of the maximal variance, $\phi = -10\fdg22$. Each local maximum represents a spiral arm of the galaxy. \textbf{Right} -- fast Fourier transform of the middle panel, identifying UGC 6093 to have two spiral arms (as is apparent to the eye in Fig.~\ref{sparcfire}).}
\label{spirality}
\end{figure*}

For UGC 6093, we define the outer visual radius to be 908 pixels (36\farcs0 = 26.7~kpc). In order to achieve optimum precision, every pixel on the face of the galaxy needs to be sampled at least once. This requires that we instruct \textsc{spirality} to construct at least $2*{\rm \pi}*908$ = 5706 spiral axes (rounded up to the nearest integer). This equates to a phase angle separation of 5.51 $\times$ $10^{-4}$~rad between each spiral axis. Fig.~\ref{spirality} illustrates the output of \textsc{spirality} for UGC 6093. The left panel shows a peak in the variance of medians at $\phi = -10\fdg22\pm0\fdg94$. The middle panel illustrates the maximal variance in median pixel value when $\phi = -10\fdg22$. The right panel is a fast Fourier transform (FFT) of the middle panel, which effectively `counts' the number of spiral arms (two in this case). The FFT is useful in checking whether or not the number of spiral arms that the code `sees' matches the number seen by the user's eyes and thus confirming that the code is sampling the observed structure of the target galaxy. It can also aid in determining the number of spiral arms in more ambiguous flocculent spiral galaxies.

\subsection{2DFFT}

\begin{figure}
\begin{center}
\includegraphics[clip=true,trim= 6mm 1mm 18mm 9mm,width=\columnwidth]{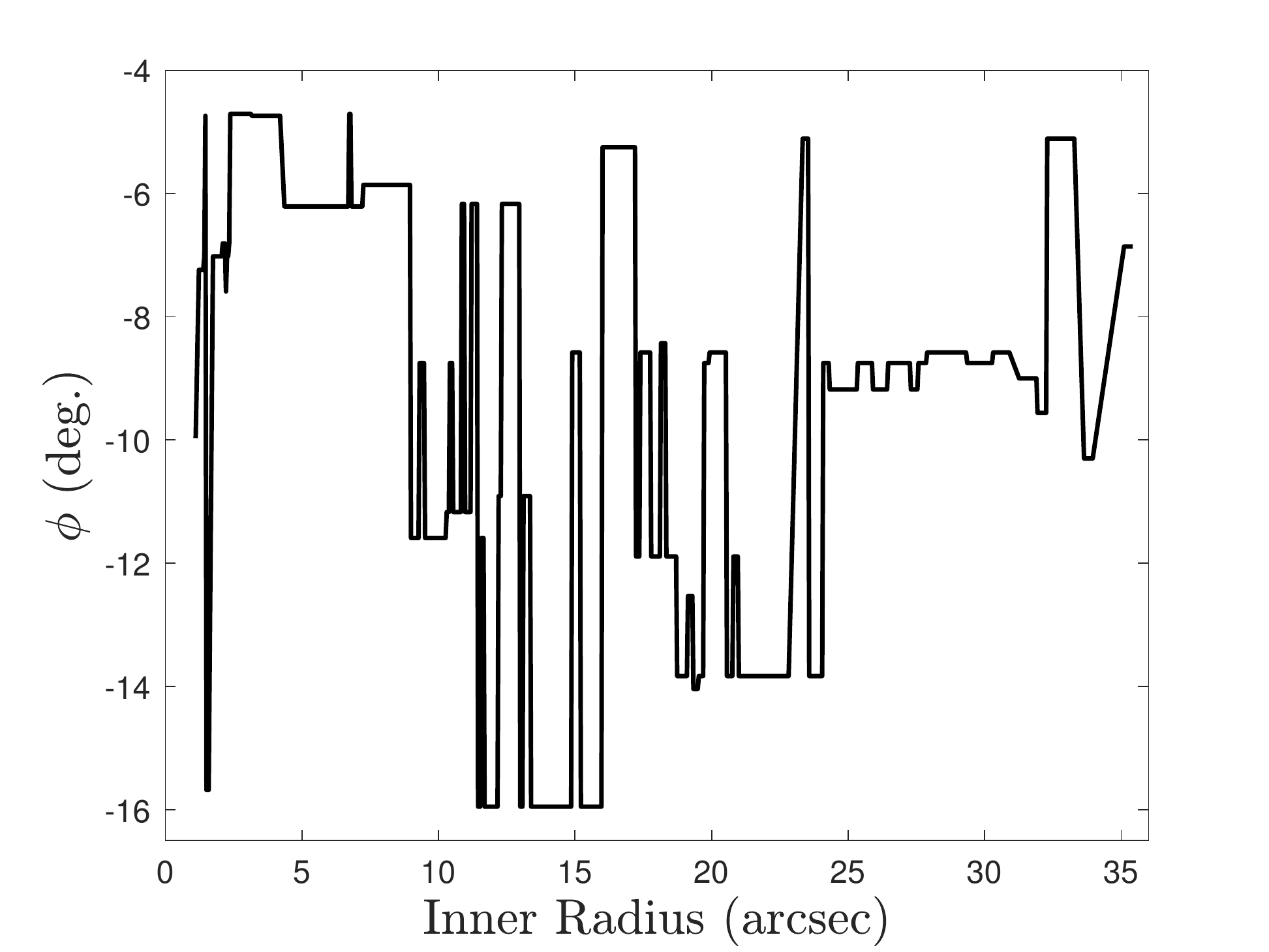}
\caption{\textsc{2dfft} result for the $m = 4$ harmonic mode of UGC 6093. The pitch angle (in degrees) is plotted as a function of the inner radius of the annulus that is fit (in arcsec). The mean pitch angle in this plot is $\phi = -9\fdg50 \pm 3\fdg61$.}
\label{2dfft}
\end{center}
\end{figure}

The \textsc{2dfft} software \citep{Davis:2012,2DFFT} operates by decomposing galaxy images into logarithmic spirals and determining for each number of harmonic modes ($1\leq m \leq6$) the pitch angle that maximizes the Fourier amplitude. \textsc{2dfft} works by analysing a measurement annulus with a fixed outer radius at the outer visible edge of the galaxy ($36\farcs0$). The inner radius is allowed to vary across the entire radius of the galaxy. For UGC 6093, the \textsc{2dfft} software identifies the most stable pitch angle for the dominant chirality as $\phi = -9\fdg50 \pm 3\fdg61$ for the $m = 4$ harmonic mode (see Fig.~\ref{2dfft}). Note, this harmonic mode is not in agreement with that determined by \textsc{spirality} (see the right-hand panel of Fig.~\ref{spirality}). Because \textsc{2dfft} is constrained to see only symmetric patterns in spiral galaxies (unlike \textsc{sparcfire} and \textsc{spirality}), it sometimes misidentifies the true number of arms. However, in this case, since $m = 4$ is a multiple of $m = 2$, the code is still accounting for the two primary arms and is simply trying to account for spiral arm spurs and asymmetries in the gaps between the primary arms.\footnote{\citet{Davis:2012} demonstrate that when signal to noise is high (as it is in this image of UGC 6093), the harmonic mode has less physical meaning as multiple harmonic modes are more likely to be in agreement.}

\subsection{Final pitch angle}

For UGC 6093, we can be confident that its pitch angle is approximately $-10\degr$. This determination is in agreement with all three codes and visual inspection. For this galaxy, we select the measurement made by \textsc{spirality}, $|\phi| = 10\fdg2 \pm 0\fdg9$, as our preferred measurement. Although, all three codes are in agreement, within error bars,\footnote{Error bars are the standard deviation, weighted by the radial extent of visible spirals across the disc of a galaxy, added in quadrature with the computational precision of the software (see respective source papers for complete details).} the measurement of \textsc{2dfft} has the highest associated error, likely because of its confusion with the number of arms. \textsc{sparcfire} provides a more accurate measurement, but still displays a conflict with its identification of a higher (approximately factor of 2) pitch angle for the shortest `Z-wise' arc segment. For these reasons, and because the measurement of \textsc{spirality} has the lowest associated error, we select the measurement from \textsc{spirality} as our preferred measurement for UGC 6093.

For difficult galaxies, we are mindful of the pros and cons of each method. Specifically, computer vision is very fast and fully automated so it is easy to use. However, it does occasionally struggle with highly flocculent galaxies by identifying many spurious short arm segments. \textsc{2dfft} is the oldest and perhaps the most robust method, but it is restricted to measuring only symmetric harmonic modes and it can be prone to high-pitch angle (low-frequency) noise in images with low signal to noise. Finally, template fitting is not hindered by harmonic modes and the software is highly customizable with many optional parameters, making it the least automated of the three methods.

Therefore, we have been able to improve upon past studies of the $M_{\rm BH}$--$\phi$ relation by using the most appropriate method and avoiding situations where things can go wrong. The final column in Table~\ref{Sample} indicates which method was used for each galaxy. As can be seen, the template fitting method was used for 23 of the 44 galaxies, the \textsc{2dfft} method was used for 19, computer vision was used for one (NGC 6264) and the Milky Way's pitch angle was independently determined from various published values.


\bsp	
\label{lastpage}
\end{document}